\documentclass{article}
\usepackage{amssymb}
\usepackage{epubtk}
\usepackage{epsf}
\usepackage{graphicx}

\showlistoftablesfalse

\begin{document}

\title{Characteristic Evolution and Matching}

\author{\epubtkAuthorData{Jeffrey Winicour}
        {Max Planck Institute for Gravitational Physics \\
        (Albert Einstein Institute) \\
        Am M\"uhlenberg 1 \\
        14476 Potsdam-Golm, Germany \\
        and \\
        Department of Physics and Astronomy \\
        University of Pittsburgh \\
        Pittsburgh, PA 15260, U.S.A.}
       {winicour@pitt.edu}
       {http://www.physicsandastronomy.pitt.edu/people/jeffrey_winicour}}

\date{}
\maketitle


\begin{abstract}
  I review the development of numerical evolution codes for general
  relativity based upon the characteristic initial value
  problem. Progress in characteristic evolution is traced from the early
  stage of 1D feasibility studies to 2D axisymmetric codes that accurately
  simulate the oscillations and gravitational collapse of relativistic stars
  and to current 3D codes that provide pieces of a binary black hole
  spacetime. Cauchy codes have now been successful
  at simulating all aspects of the binary black hole problem inside an
  artificially constructed outer boundary. A prime application
  of characteristic evolution is to extend such simulations to null infinity
  where the waveform from the binary inspiral and merger
  can be unambiguously computed. This has now been
  accomplished by Cauchy-characteristic extraction, where data for
  the characteristic evolution is supplied by Cauchy data on an
  extraction worldtube inside the artificial outer boundary. The ultimate
  application of characteristic evolution is to eliminate the role of
  this outer boundary by constructing a global
  solution via Cauchy-characteristic matching. 
  Progress in this direction is discussed.
  \end{abstract}

\epubtkKeywords{Numerical relativity, Numerical methods,
  Characteristic initial value problem}

\newpage

\epubtkUpdate
    [Id=A,
     ApprovedBy=subjecteditor,
     AcceptDate={8 August 2011},
     PublishDate={8 August 2011},
     Type=major]{%
Besides various small changes, Sections~\ref{sec:worldtube-nullcone},
\ref{sec:cosmology}, \ref{sec:worldtube-conservation} and
\ref{sec:ccebh} to \ref{sec:initial} are new, with 37 new references.
}

\newpage


\section{Introduction}
\label{intro}

It is my pleasure to review progress in numerical relativity based upon
characteristic evolution. In the spirit of \emph{Living Reviews in
Relativity}, I invite my colleagues to continue to send me contributions and
comments at winicour@pitt.edu.

We are now in an era in which Einstein's equations can effectively be considered
solved at the local level. Several groups, as reported here and in
other \emph{Living Reviews}, have developed 3D Cauchy evolution codes
which are stable and accurate in some sufficiently bounded
domain. The pioneering works~\cite{fP05} (based upon a harmonic formulation)
and \cite{mCcLpMyZ06, jBjCdCmKjvM06b} (based upon BSSN
formulations~\cite{mStN95,tBsS99}) have initiated dramatic progress in the
ability of these codes to simulate the inspiral and merger of binary black
holes, the premier problem in classical relativity. Global solutions of
binary black holes are another matter. Characteristic evolution codes
have been successful in treating the exterior region of asymptotically flat
spacetimes extending to future null infinity. Just as several coordinate
patches are necessary to describe a spacetime with nontrivial topology,
the most effective attack on the binary black hole waveform might
involve a global solution patched together from pieces of spacetime
handled by a combination of different codes and techniques.

Most of the effort in numerical relativity has centered about Cauchy
codes based upon the \{3\,+\,1\} formalism~\cite{york}, which evolve the
spacetime inside an artificially constructed outer boundary. It has been
common practice in Cauchy simulations of binary black holes to
compute the waveform from data on a finite extraction worldtube
inside the outer boundary, using perturbative methods based upon
introducing a Schwarzschild background in the exterior
region~\cite{ab1,ab2,ab3,all1,rupright,rezzmatz,nagar}. In order to properly approximate
the waveform at null infinity the extraction worldtube must be sufficiently large
but at the same time causally and numerically isolated from errors
propagating in from the outer boundary. Considerable improvement in this approach
has resulted from efficient methods for dealing with a very large outer boundary
and from techniques to extrapolate the extracted waveform to infinity.
However, this is not an ideally efficient approach and is especially
impractical to apply to simulations of stellar collapse.
A different approach which is specifically tailored to
study radiation at null infinity can be based upon the characteristic
initial value problem. This eliminates error due to
asymptotic approximations and the gauge effects introduced
by the choice of a finite extraction worldtube.

In the 1960s, Bondi~\cite{1bondi,bondi} and Penrose~\cite{Penrose}
pioneered the use of null hypersurfaces to describe gravitational waves.
The characteristic initial value problem did not receive much attention before
its importance in general relativity was recognized.  Historically, the
development of computational physics has focused on hydrodynamics, where the
characteristics typically do not define useful coordinate surfaces and there
is no generic outer boundary behavior comparable to null infinity. 
But this new approach has flourished in general relativity. It has
led to the first unambiguous description of gravitational radiation in a fully
nonlinear context. By formulating asymptotic flatness in terms of characteristic
hypersurfaces extending to infinity, it was possible to reconstruct, in a
nonlinear geometric setting, the basic properties of gravitational waves which
had been developed in linearized theory on a Minkowski background. The major new
nonlinear features were the Bondi mass and news function, and the mass loss
formula relating them. The Bondi news function is an invariantly defined complex
radiation amplitude $N= N_{\oplus}+i N_{\otimes}$, whose real and imaginary parts
correspond to the time derivatives $\partial_t h_{\oplus}$ and $\partial_t
h_{\otimes}$ of the ``plus'' and ``cross'' polarization modes of the strain $h$
incident on a gravitational wave antenna. The corresponding waveforms are
important both for the design of detection templates for a binary black hole
inspiral and merger and for the determination of the resulting recoil velocity.

The recent success of Cauchy evolutions in simulating binary black holes
emphasizes the need to apply global techniques to accurate waveform
extraction. This has stimulated several attempts to increase the accuracy of
characteristic evolution. The Cauchy simulations
have incorporated increasingly sophisticated numerical techniques, such as
mesh refinement, multi-domain decomposition, pseudo-spectral collocation
and high order (in some cases eighth order) finite
difference approximations. The initial characteristic codes were developed with
unigrid second order accuracy. One of the prime factors affecting the accuracy
of any characteristic code is the introduction of a smooth coordinate system
covering the sphere, which labels the null directions on the outgoing light
cones. This is also an underlying problem in meteorology and oceanography. In a
pioneering paper on large-scale numerical weather prediction,
Phillips~\cite{phillips} put forward a list of desirable features for a mapping
of the sphere to be useful for global forecasting. The first requirement was the
freedom from singularities. This led to two distinct choices which had been
developed earlier in purely geometrical studies:  stereographic coordinates (two
coordinate patches) and cubed-sphere coordinates (six patches). Both coordinate
systems have been rediscovered in the context of numerical relativity (see
Section~\ref{sec:sphercoor}). The cubed-sphere method has stimulated two new
attempts at improved codes for characteristic evolution (see
Section~\ref{sec:nummeth}). An ingenious third treatment, based upon a toroidal map
of the sphere, was devised in developing a characteristic code
for Einstein equations~\cite{bartnumeth} (see Section~ \ref{sec:toroidal}).

Another issue affecting code accuracy is the choice between a second or first
differential order reduction of the evolution system. Historically, the
predominant importance of computational fluid dynamics has  favored first order
systems, in particular the reduction to symmetric hyperbolic form. However, in
acoustics and elasticity theory, where the natural treatment is in terms of
second order wave equations, an effective argument for the second order form has
been made~\cite{krort, kreissortpet}. In general relativity, the
question of whether first or
second order formulations are more natural depends on how  Einstein's equations
are reduced to a hyperbolic system by some choice of coordinates and variables.
The second order form is more natural in the harmonic formulation, where the
Einstein equations reduce to quasilinear wave equations. The first order form is
more natural in the Friedrich--Nagy formulation~\cite{friednag},
which includes the Weyl tensor
among the evolution variables, and was used in the first demonstration of a
well-posed initial-boundary value problem for Einstein's equations.
Investigations of first order formulations of the characteristic initial value
problem are discussed in Section~\ref{sec:fvssec}.

The major drawback of a stand-alone characteristic
approach arises from the formation of caustics in the light rays generating the
null hypersurfaces. In the most ambitious scheme proposed at the theoretical
level such caustics would be treated ``head-on'' as part of the evolution
problem~\cite{friedst1}. This is a profoundly attractive idea. Only a few
structural stable caustics can arise in numerical evolution, and their
geometrical properties are well enough understood to model their singular
behavior numerically~\cite{friedst2}, although a computational implementation
has not yet been attempted. 

In the typical setting for the characteristic initial value problem, the domain
of dependence of a single smooth null hypersurface is empty. In order to obtain a
nontrivial evolution problem, the null hypersurface must either be completed to a
caustic-crossover region where it pinches off, or an additional inner boundary
must be introduced. So far, the only caustics that have been successfully evolved
numerically in general relativity are pure point caustics (the complete null cone
problem). When spherical symmetry is not present, the stability conditions near
the vertex of a light cone place a strong restriction on the allowed time
step~\cite{igw}. Nevertheless, point caustics in general relativity have been
successfully handled for axisymmetric vacuum spacetimes~\cite{papa}. Progress
toward extending these results to realistic astrophysical sources has been made
by coupling an axisymmetric characteristic gravitational-hydro code with a high
resolution shock capturing code for the relativistic hydrodynamics, as
initiated in the thesis of F.~Siebel~\cite{Siebel}. This
has enabled the global characteristic simulation of
the oscillation and collapse of a relativistic star in which the emitted
gravitational waves are computed at null infinity (see Sections~\ref{sec:shydro}
and~\ref{sec:ahydro}). Nevertheless, computational demands to extend these
results to 3D evolution would be prohibitive using current generation
supercomputers, due to the small timestep required at the vertex of the null
cone (see Section~\ref{sec:bondiprob}). This is an unfortunate feature of
the present day finite difference codes, which might be eliminated by the use,
say, of a spectral approach. Away from
the caustics, characteristic evolution offers myriad computational and
geometrical advantages. Vacuum simulations of black hole spacetimes, where the
inner boundary can be taken to be the white hole horizon, offer a scenario where
both the timestep and caustic problems can be avoided and three-dimensional
simulations are practical (as discussed in Section~\ref{sec:mode}).
An early example was the study of gravitational radiation from the
post-merger phase of a binary black black hole using a fully
nonlinear three-dimensional characteristic code~\cite{Zlochower,zlochmode}.

At least in the near future, fully three-dimensional computational
applications of characteristic evolution are likely to be restricted to some
mixed form, in which data is prescribed on a non-singular but incomplete
initial null hypersurface $N$ and on a second inner boundary $B$, which
together with the initial null hypersurface determines a nontrivial domain of
dependence. The hypersurface $B$ may be either (i)~null, (ii)~timelike or
(iii)~spacelike, as schematically depicted in Figure~\ref{fig:civp}. The first
two possibilities give rise to (i)~the double null problem and (ii)~the
nullcone-worldtube problem. Possibility (iii) has more than one interpretation.
It may be regarded as a Cauchy initial-boundary value problem where the outer
boundary is null. An alternative interpretation is the Cauchy-characteristic
matching (CCM) problem, in which the Cauchy and characteristic evolutions are
matched transparently across a worldtube W, as indicated in
Figure~\ref{fig:civp}.

\epubtkImage{civp.png}{%
  \begin{figure}[htbp]
    \centerline{\includegraphics[width=0.8\textwidth]{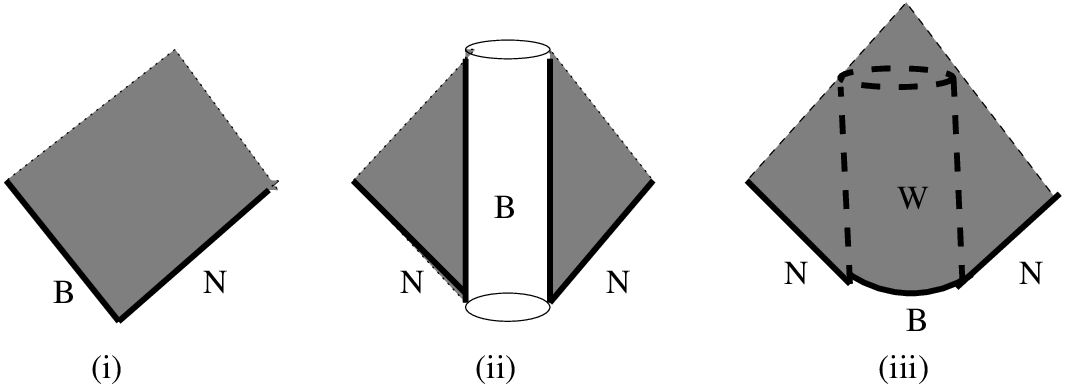}}
    \caption{The three applications of characteristic evolution
      with data given on an initial null hypersurface \emph{N} and boundary
      \emph{B}. The shaded regions indicate the corresponding domains of
      dependence.}
    \label{fig:civp}
  \end{figure}}

In CCM, it is possible to choose the matching interface between the Cauchy and
characteristic regions to be a null hypersurface, but it is more practical to
match across a timelike worldtube. CCM combines the advantages of
characteristic evolution in treating the outer radiation zone in spherical
coordinates which are naturally adapted to the topology of the worldtube with
the advantages of Cauchy evolution in treating the inner region in Cartesian
coordinates, where spherical coordinates would break down.

In this review, we trace the development of characteristic algorithms from
model 1D problems to a 2D axisymmetric code which computes the gravitational
radiation from the oscillation and gravitational collapse of a relativistic
star and to a 3D code designed to calculate the waveform emitted in the merger
to ringdown phase of a binary black hole. And we trace the development of CCM
from early feasibility studies to successful implementation in the linear
regime and through current attempts to treat the binary black hole problem.
 
CCM eliminates the need of outer boundary data for the Cauchy
evolution and  supplies the waveform at null infinity via a characteristic evolution.
At present, the only successful 3D application of CCM in general relativity
has been to to the linearized
matching problem between a 3D characteristic code and a 3D Cauchy code based
upon harmonic coordinates~\cite{harm} (see Section~\ref{sec:linccm}). Here the
linearized Cauchy code satisfies a well-posed initial-boundary value problem,
which seems to be a critical missing ingredient in previous attempts at CCM in
general relativity. Recently a well-posed initial-boundary value problem has
been established for fully nonlinear harmonic evolution~\cite{hKjW06} (see
Section~\ref{sec:outercb}), which should facilitate the extension of CCM to the
nonlinear case.

Cauchy-characteristic extraction (CCE), which
is one of the pieces of the CCM strategy, also supplies the waveform at null infinity
by means of a characteristic evolution. However, in this case the artificial outer
Cauchy boundary is left unchanged and the data for
the characteristic evolution is extracted
from Cauchy data on an interior worldtube. Since my last review, the most important
development has been the application of CCE to the binary black hole
problem. Beginning with the work in~\cite{reis1}, CCE has become an important tool for
gravitational wave data analysis (see Section~\ref{sec:ccebh}). This application of CCE
to this problem was developed as a major part of the PhD thesis of
Christian Reisswig~\cite{reisdipl}. 

In previous reviews, I tried to include material on the treatment of boundaries
in the computational mathematics and fluid dynamics literature because of its
relevance to the CCM problem. The fertile growth of this subject has warranted a
separate \emph{Living Review} on boundary conditions, which
is presently under construction and will appear soon~\cite{sarbtig}.
In anticipation of this, I will
not attempt to keep this subject up to date except for material of direct
relevance to CCM. See~\cite{oS07,oscolv} for independent reviews of boundary
conditions that have been used in numerical relativity.

The well-posedness of the associated initial-boundary value problem, i.e.
that there exists a unique solution which depends continuously on the data,
is a necessary condition for a successful numerical treatment. In addition
to the forthcoming \emph{Living Review}~\cite{sarbtig}, this subject is covered
in the review~\cite{friedrend} and the book~\cite{greenbook}.


If well-posedness can be established using energy
estimates obtained by integration by parts with respect to the coordinates defining
the numerical grid then the analogous finite difference estimates obtained
by \emph{summation by parts}~\cite{kreissch} provide guidance for a stable finite-difference
evolution algorithm. See the forthcoming \emph{Living Review}~\cite{sarbtig}
for a discussion of the application of summation
by parts to numerical relativity.

The problem of computing the evolution of a neutron star, in close orbit about a
black hole is another problem of clear importance to the
new gravitational wave detectors. The
interaction with the black hole could be strong enough to produce a drastic
change in the emitted waves, say by tidally disrupting the star, so that a
perturbative calculation would be inadequate. The understanding of such nonlinear
phenomena requires well behaved numerical simulations of hydrodynamic systems
satisfying Einstein's equations. Several numerical relativity codes  for treating
the problem of a neutron star near a black hole have been developed, as described
in the \emph{Living Review} on ``Numerical Hydrodynamics in General
Relativity'' by Font~\cite{tonirev}. Although most of these efforts concentrate
on Cauchy evolution, the characteristic approach has shown remarkable robustness
in dealing with a single black hole or relativistic star. In this vein,
axisymmetric studies of the oscillation and gravitational
collapse of relativistic stars have been achieved (see Section~\ref{sec:ahydro})
and progress has been made in the 3D simulation of a body in close orbit about a
Schwarzschild black hole (see Sections~\ref{sec:3dekg} and \ref{sec:3dhydro}).

\newpage


\section{The Characteristic Initial Value Problem}
\label{sec:civp}

Characteristics have traditionally played an important role in the
analysis of hyperbolic partial differential equations. However, the use
of characteristic hypersurfaces to supply the foliation underlying an
evolution scheme has been mainly restricted to relativity. This is
perhaps natural because in curved spacetime there is no longer a
preferred Cauchy foliation provided by the Euclidean 3-spaces allowed
in Galilean or special relativity. The method of shooting along
characteristics is a standard technique in many areas of computational
physics, but evolution based upon characteristic hypersurfaces is quite
uniquely limited to relativity.

Bondi's initial use of null coordinates to describe radiation
fields~\cite{1bondi} was followed by a rapid development of other null
formalisms. These were distinguished either as metric based approaches, as
developed for axisymmetry by Bondi, Metzner and van der Burg~\cite{bondi} and
generalized to 3 dimensions by Sachs~\cite{sachs}, or as null tetrad approaches
in which the Bianchi identities appear as part of the system of equations, as
developed by Newman and Penrose~\cite{NP}.

At the outset, null formalisms were applied to construct asymptotic solutions
at null infinity by means of $1/r$ expansions. Soon afterward, Penrose~\cite{Penrose}
devised
the conformal compactification of null infinity $\mathcal{I}$ (``scri''), thereby
reducing to geometry the asymptotic quantities describing the physical
properties of the radiation zone, most notably the Bondi mass and news
function. The characteristic initial value problem rapidly
became an important tool for the clarification of fundamental conceptual issues
regarding gravitational radiation and its energy content. It laid bare and
geometrised the gravitational far field.

The initial focus on asymptotic solutions clarified the kinematic properties of
radiation fields but could not supply the dynamical properties relating the
waveform to a specific source. It was soon realized that instead of carrying out
a $1/r$ expansion, one could reformulate the approach in terms of the
integration of ordinary differential equations along the characteristics (null
rays)~\cite{tam}. The integration constants supplied on some inner boundary then
played the role of sources in determining the specific waveforms obtained at
infinity. In the double-null initial value problem of Sachs~\cite{sachsdn}, the
integration constants are supplied at the intersection of outgoing and ingoing
null hypersurfaces. In the worldtube-nullcone formalism, the sources were
represented by integration constants on a timelike worldtube~\cite{tam}. These
early formalisms have gone through much subsequent revamping. Some have been
reformulated to fit the changing styles of modern differential geometry. Some
have been reformulated in preparation for implementation as computational
algorithms. The articles in~\cite{southam} give a representative sample of
formalisms. Rather than including a review of the extensive literature on
characteristic formalisms in general relativity, I concentrate here on those
approaches which have been implemented as computational evolution schemes.

All characteristic evolution schemes share the same skeletal form. The
fundamental ingredient is a foliation by null hypersurfaces
$ u = \mathrm{const.} $
which are generated by a two-dimensional set of null rays, labeled by
coordinates $x^A$, with a coordinate $\lambda$ varying along the rays.
In $(u,\lambda,x^A)$ null coordinates, the main set of Einstein
equations take the schematic form
\begin{equation}
  F_{,\lambda} = H_F [F,G]
\end{equation}
and
\begin{equation}
  G_{,u\lambda} = H_G [F,G,G_{,u}].
\end{equation}
Here $F$ represents a set of hypersurface variables, $G$ a set of evolution
variables, and $H_F$ and $H_G$ are nonlinear hypersurface operators, i.e.\ they
operate locally on the values of $F$, $G$ and $G_{,u}$ intrinsic to a single null
hypersurface. In the Bondi formalism, these hypersurface equations have a
hierarchical structure in which the members of the set $F$ can be integrated in
turn in terms of the characteristic data for the evolution variables and the
computed values of prior members of the hierarchy. In addition to these main
Einstein equations, there is a subset of four subsidiary Einstein equations which
are satisfied by virtue of the Bianchi identities, provided that they are
satisfied on a hypersurface transverse to the characteristics. These equations
have the physical interpretation as conservation laws. Mathematically they are
analogous to the constraint equations of the canonical formalism. But they are
not elliptic since they may be intrinsic to null or timelike hypersurfaces,
rather than spacelike Cauchy hypersurfaces.

Computational implementation of characteristic evolution may be based upon
different versions of the formalism (i.e.\ metric or tetrad) and different
versions of the initial value problem (i.e.\ double null or worldtube-nullcone).
The performance and computational requirements of the resulting evolution codes
can vary drastically. However, most characteristic evolution codes share certain
common advantages:
\begin{itemize}
\item The characteristic initial data is free, i.e.\ there are no elliptic constraints on
  the data. This eliminates the need for time consuming iterative
  constraint solvers with their accompanying artificial boundary conditions. This
  flexibility and control in prescribing initial data has the
  trade-off of limited experience with prescribing physically
  realistic characteristic initial data.
\item The coordinates are very``rigid'', i.e.\ there is very little
  remaining gauge freedom.
\item The constraints satisfy ordinary differential equations along
  the characteristics which force any constraint violation to fall off
  asymptotically as $1/r^2$.
\item No second time derivatives appear so that the number of basic
  variables is at most half the number for the corresponding version
  of the Cauchy problem.
\item The main Einstein equations form a system of coupled ordinary
  differential equations with respect to the parameter $\lambda$
  varying along the characteristics. This allows construction of an
  evolution algorithm in terms of a simple march along the
  characteristics.
\item In problems with isolated sources, the radiation zone can be
  compactified into a finite grid boundary with the metric rescaled by
  $1/r^2$ as an implementation of Penrose's conformal boundary at future
  null infinity $\mathcal{I}^+$.
  Because $\mathcal{I}^+$ is a null hypersurface, no extraneous outgoing
  radiation condition or other artificial boundary condition is
  required. The analogous treatment in the Cauchy problem requires the
  use of hyperboloidal spacelike hypersurfaces asymptoting to null
  infinity~\cite{Fhprbloid}. For reviews of the hyperboloidal approach
  and its status in treating the associated three-dimensional
  computational problem, see~\cite{Hhprbloid,joerg}.
\item The grid domain is exactly the region in which waves propagate,
  which is ideally efficient for radiation studies. Since each null
  hypersurface of the foliation extends to infinity, the radiation is
  calculated immediately (in retarded time).
\item In black hole spacetimes, a large redshift at null infinity
  relative to internal sources is an indication of the formation of an
  event horizon and can be used to limit the evolution to the exterior
  region of spacetime. While this can be disadvantageous for late time
  accuracy, it allows the possibility of identifying the event horizon
  ``on the fly'', as opposed to Cauchy evolution where the event
  horizon can only be located after the evolution has been completed.
\end{itemize}
Perhaps most important from a practical view, characteristic evolution
codes have shown remarkably robust stability and were the first to
carry out long term evolutions of moving black holes~\cite{wobb}.

Characteristic schemes also share as a common disadvantage the necessity either
to deal with caustics or to avoid them altogether. The scheme to tackle the
caustics head on by including their development and structure as part of the
evolution~\cite{friedst1,friedst2} is perhaps a great idea
still ahead of its time but one that should not
be forgotten. There are only a handful of structurally stable caustics, and they
have well known algebraic properties. This makes it possible to model their
singular structure in terms of Pad\'e approximants. The structural stability of
the singularities should in principle make this possible, and algorithms to
evolve the elementary caustics have been proposed~\cite{cstew,padstew}. In the
axisymmetric case, cusps and folds are the only structurally stable caustics, and
they have already been identified in the horizon formation occurring in
simulations of head-on collisions of black holes and in the temporarily toroidal
horizons occurring in collapse of rotating matter~\cite{sci,toroid}. In a generic
binary black hole horizon, where axisymmetry is broken, there is a closed curve
of cusps which bounds the two-dimensional region on the event horizon where the
black holes initially form and merge~\cite{ndata,asym}.

\subsection{The worldtube-nullcone problem}
\label{sec:worldtube-nullcone}

A version of the characteristic initial value problem for Einstein's equations
which avoids caustics is the worldtube-nullcone problem, where
boundary data is given on a timelike worldtube and on an initial outgoing null
hypersurface~\cite{tam}. The underlying physical picture
is that the worldtube data represent the outgoing gravitational
radiation emanating from interior matter sources, while ingoing radiation
incident on the system is represented by the initial null data.

The well-posedness of the worldtube-nullcone problem for Einstein's
equations has not yet been established.
Rendall~\cite{rend} established the well-posedness
of the double null version of the problem where data is
given on a pair of intersecting characteristic hypersurfaces. He did not
treat the characteristic problem head-on but reduced it to a standard Cauchy
problem with data on a spacelike hypersurface passing through the
intersection of the characteristic hypersurfaces. Unfortunately, this
approach cannot be applied to the null-timelike problem and it
does not provide guidance for the development of
a stable finite-difference approximation based upon characteristic
coordinates.

Another limiting case of the nullcone-worldtube problem is the Cauchy problem on
a characteristic cone, corresponding to the limit in which the timelike worldtube
shrinks to a nonsingular worldline. Choquet-Bruhat, Chru\'{s}ciel and Mart\'{\i}n-Garc\'{\i}a
established the existence of solutions to this problem
using harmonic coordinates adapted
to the null cones, thus avoiding the singular nature of
characteristic coordinates at the vertex~\cite{cbcmg1}. Again, this
does not shed light on numerical implementation in characteristic coordinates.

A necessary condition for the well-posedness of the gravitational problem is
that  the corresponding problem for the quasilinear wave equation be
well-posed. This brings our attention to the Minkowski space wave equation,
which provides the simplest version of the worldtube-nullcone problem.
The treatment of this simplified problem
traces back to Duff~\cite{Duff}, who showed existence and uniqueness for
the case of analytic data. Later, Friedlander extended existence and uniqueness
to the $C^\infty$ case for the
linear wave equation on an asymptotically flat curved space background~\cite{friedl,friedl2}.

The well-posedness of a variable coefficient or quasilinear problem
requires energy estimates for the derivatives of the linearized solutions.
Partial estimates for  characteristic boundary value
problems were first obtained by M{\"{u}}ller zum Hagen and Seifert~\cite{hagenseifert77}.
Later, Balean carried out a comprehensive study of the differentiability of
solutions of the worldtube-nullcone problem for the flat space wave
equation~\cite{bal1,bal2}.  He was able to establish the required estimates for the
derivatives tangential to the outgoing null cones but weaker estimates for the
time derivatives transverse to the cones had to be obtained from a direct
integration of the wave equation. Balean concentrated on
the differentiability of the solution rather than well-posedness.
Frittelli~\cite{frit} made the first explicit investigation of well-posedness, using the approach
of Duff, in which the characteristic formulation of the wave equation is
reduced to a canonical first order differential form, in close analogue to the
symmetric hyperbolic formulation of the Cauchy problem. The energy associated
with this first order reduction gave estimates for the derivatives of the
field tangential to the null hypersurfaces but, as in Balean's work,
weaker estimates for the time derivatives had to be obtained indirectly.
As a result,  well-posedness could not be established for variable
coefficient of quasilinear wave equations.

The basic difficulty underlying this problem can be illustrated in terms of the
1(spatial)-dimensional wave equation
\begin{equation}
             ( \partial_{\tilde t}^2 -\partial_{\tilde x}^2)\Phi=0,
             \label{eq:tilde1d}
\end{equation}
where $(\tilde t,\tilde x)$ are standard space-time coordinates.
The conserved energy
\begin{equation}
       \tilde  E(\tilde t)= \frac{1}{2} \int d\tilde x  \bigg( (
        \partial_{\tilde t}\Phi)^2  +(\partial_{\tilde x}\Phi)^2 \bigg )
       \label{eq:tildeE}
\end{equation}
leads to the well-posedness of the Cauchy problem. In characteristic
coordinates $(t=\tilde t -\tilde x, \, x=\tilde t +\tilde x)$,
the wave equation transforms into
\begin{equation}
   \partial_t \partial_x \Phi =0.
   \label{eq:1dphi}
\end{equation}
The conserved energy on the characteristics $t=\mathrm{const.}$,
\begin{equation}
       \tilde   E(t) = \int dx (\partial_ x \Phi)^2,
\end{equation}
no longer controls the time derivative $\partial_t \Phi$. 

As a result, the standard
technique for establishing well-posedness of the Cauchy problem fails.
For (\ref{eq:tilde1d}), the solutions to the Cauchy problem
with compact initial data on $\tilde t=0$ are square integrable and
well-posedness can be established using the $L_2$ norm  (\ref{eq:tildeE}).
However, In characteristic coordinates the 1-dimensional wave equation
(\ref{eq:1dphi}) admits signals traveling in the $+x$-direction with infinite
coordinate velocity. In particular, initial data of compact support
$\Phi(0,x)=f(x)$ on the characteristic $t=0$ admits the solution $\Phi = g(t)
+f(x)$, provided that $g(0)=0$.  Here $g(t)$ represents the profile of a wave
which travels from past null infinity ($x\rightarrow -\infty$) to future null
infinity  ($x\rightarrow +\infty$). Thus, without a boundary condition at past
null infinity, there is no unique solution and the Cauchy problem is ill
posed. Even with the boundary condition $\Phi(t,-\infty)=0$,  a source of
compact support $S(t,x)$ added to (\ref{eq:1dphi}), i.e.
\begin{equation}
   \partial_t \partial_x \Phi =S,
   \label{eq:1dphis}
\end{equation}
produces waves propagating to $x=+\infty$ so that although the solution is 
unique it is still not square integrable.

On the other hand, consider the modified problem obtained by
setting $\Phi=e^{ax}\Psi$,
\begin{equation} 
 \partial_t (\partial_x+a) \Psi=F \, ,  \quad \Psi(0,x)
    = e^{-ax}f(x) \,  ,\quad a>0
       \label{eq:1dpsi}
\end{equation} 
where $F=e^{-ax}  S$. With the boundary condition $\Psi(t,-\infty)=0$, the
solutions to (\ref{eq:1dpsi}) vanish at $x=+\infty$ and are square
integrable. As a result, the Cauchy problem (\ref{eq:1dpsi}) is well posed
with respect to an $L_2$ norm. For the simple example where $F=0$,
multiplication of (\ref{eq:1dpsi}) by
$(2a \Psi+\partial_x \Psi+\frac{1}{2}\partial_t \Psi)$ and integration by
parts gives
\begin{equation}
       \frac{1}{2}\partial_t  \int dx \bigg( 
             (\partial_ x \Psi)^2+2a^2 \Psi^2 \bigg)
            =\frac{a}{2} \int dx \bigg(2(\partial_ t \Psi)\partial_ x \Psi 
           -  (\partial_ t \Psi)^2 \bigg)   
         \le \frac{a}{2} \int dx  (\partial_ x \Psi)^2 .
\end{equation} 
The resulting inequality
\begin{equation}
       \partial_t  E \le \mathrm{const.} E 
\end{equation} 
for the energy
\begin{equation}
      E=\frac{1}{2}  \int dx \bigg( (\partial_ x \Psi)^2+2a^2 \Psi^2 \bigg)
      \label{eq:1denergy}
\end{equation} 
provides the estimates for $\partial_x \Psi$ and $\Psi$ which are necessary
for well-posedness. Estimates for $\partial_t \Psi$, and other higher
derivatives, follow from applying this approach to the derivatives of 
(\ref{eq:1dpsi}). The approach can be extended to include the source term $F$
and other generic lower differential order terms. This allows well-posedness
to be extended to the case of variable coefficients and, locally in time, to
the quasilinear case.

The modification in going from  
(\ref{eq:1dphis}) to (\ref{eq:1dpsi}) leads to an effective modification of
the standard energy for the problem. Rewritten in terms of the original
variable $\Phi=e^{ax}\Psi$, (\ref{eq:1denergy}) corresponds to the
energy 
\begin{equation}
     E=\frac{1}{2}  \int dx e^{-2ax}  \bigg( (\partial_ x \Phi)^2+a^2 \Phi^2 \bigg ).
     \label{eq:enorm}
\end{equation}
Thus while the Cauchy problem for (\ref{eq:1dpsi}) is ill posed with respect
to the standard $L_2$ norm it is well posed with respect to the exponentially weighted
norm (\ref{eq:enorm}). 

This technique was introduced in~\cite{krwin}
to treat the worldtube-nullcone problem for the 3-dimensional
 quasilinear wave
equation for a scalar field $\Phi$ in an asymptotically flat curved space
background with source $S$,
\begin{equation}
        g^{ab}\nabla_a \nabla_b \Phi = S (\Phi,\partial_c \Phi, x^c),
        \label{eq:qw}
\end{equation}
where the metric $g^{ab}$ and its associated covariant  derivative $\nabla_a$
are explicitly prescribed functions of $(\Phi,x^c)$. In terms of retarded spherical
null coordinates $x^a=(u=t-r,r,\theta,\phi)$, the initial-boundary value problem
consists of determining $\Phi$ in the region $(r>R,u>0)$ given data
$\Phi(u,R,\theta,\phi)$ on the timelike worldtube $r=R$
and  $\Phi(0,r,\theta,\phi)$ on the initial null hypersurface $u=0$.
It was shown that this quasilinear wave problem is well posed
on a domain extending to future null infinity subject to smoothness and asymptotic falloff
conditions on the data. The treatment was based upon energy
estimates obtained by integration by parts with respect to the characteristic
coordinates. As a result, the analogous finite difference estimates obtained
by \emph{summation by parts}~\cite{kreissch} do provide guidance for the development of a stable
numerical evolution algorithm. The corresponding
worldtube-nullcone problem for Einstein's equations plays a major underlying
role in the CCM strategy. Its well-posedness appears to be confirmed by
numerical simulations but the analytic proof remains an important unresolved
problem.

\newpage


\section{Prototype Characteristic Evolution Codes}
\label{sec:proto}

Limited computer power, as well as the instabilities arising from non-hyperbolic
formulations of Einstein's equations, necessitated that the early code
development in general relativity be restricted to spacetimes with symmetry.
Characteristic codes were first developed for spacetimes with spherical symmetry.
The techniques for other special relativistic fields which propagate on null
characteristics are similar to the gravitational case. Such fields are included
in this section. We postpone treatment of relativistic fluids, whose
characteristics are timelike, until Section~\ref{sec:grace}.


\subsection{\{1\,+\,1\}-dimensional codes}
\label{sec:1d}

It is often said that the solution of the general ordinary differential equation
is essentially known, in light of the success of computational algorithms and
present day computing power. Perhaps this is an overstatement because
investigating singular behavior is still an art. But, in this spirit, it is fair
to say that the general system of hyperbolic partial differential equations in
one spatial dimension seems to be a solved problem in general relativity. Codes
have been successful in revealing important new phenomena underlying singularity
formation in cosmology~\cite{berger} and in dealing with unstable spacetimes to
discover critical phenomena~\cite{gundlach}. As described below, characteristic
evolution has contributed to a rich variety of such results.

One of the earliest characteristic evolution codes, constructed by Corkill
and Stewart~\cite{cstew,bonn}, treated spacetimes with two Killing vectors
using a grid  based upon double null coordinates, with the null
hypersurfaces intersecting in the surfaces spanned by the Killing vectors.
They simulated colliding plane waves and evolved the
Khan--Penrose~\cite{khan} collision of impulsive ($\delta$-function
curvature) plane waves to within a few numerical zones from the final
singularity, with extremely close agreement with the analytic results.
Their simulations of collisions with more general waveforms, for which
exact solutions are not known, provided input to the understanding of
singularity formation which was unforeseen in the analytic treatments of
this problem.

Many \{1\,+\,1\}-dimensional characteristic codes have been developed for
spherically symmetric systems. Here matter must be included in order to make the
system non-Schwarzschild. Initially the characteristic evolution of matter was
restricted to simple cases, such as massless Klein--Gordon fields, which allowed
simulation of gravitational collapse and radiation effects in the simple context
of spherical symmetry. Now, characteristic evolution of matter is progressing to
more complicated systems. Its application to hydrodynamics has made significant
contributions to general relativistic astrophysics, as reviewed in
Section~\ref{sec:grace}.

The synergy between analytic and computational approaches has led to dramatic
results in the massless Klein--Gordon case. On the analytic side, working in a
characteristic initial value formulation based upon outgoing null cones,
Christodoulou made a penetrating study of the spherically symmetric
problem~\cite{X21987,X1991,X1993,X1994,X1999,X21999}. In a suitable function
space, he showed the existence of an open ball about Minkowski space data whose
evolution is a complete regular spacetime; he showed that an evolution with a
nonzero final Bondi mass forms a black hole; he proved a version of cosmic
censorship for generic data; and he established the existence of naked
singularities for non-generic data. What this analytic tour-de-force did not
reveal was the remarkable critical behavior in the transition to the black hole
regime, which was discovered by Choptuik~\cite{chsouth,choptprl} in simulations
using Cauchy evolution. This phenomenon has now been understood in terms of the
methods of renormalization group theory and intermediate asymptotics, and has
spawned a new subfield in general relativity, which is covered in the \emph{Living
Review} on ``Critical Phenomena in Gravitational Collapse'' by
Gundlach~\cite{gundlach}.

The characteristic evolution algorithm for the spherically symmetric
Einstein--Klein--Gordon problem provides a simple illustration of the
techniques used in the general case. It centers about the evolution scheme
for the scalar field, which constitutes the only dynamical field. Given the
scalar field, all gravitational quantities can be determined by integration
along the characteristics of the null foliation. This is a coupled problem,
since the scalar wave equation involves the curved space metric. It
illustrates how null algorithms lead to a hierarchy of equations which can
be integrated along the characteristics to effectively decouple the
hypersurface and dynamical variables.

In a Bondi coordinate system based upon outgoing null hypersurfaces
$u = \mathrm{const.}$ and a surface area coordinate $r$, the metric is
\begin{equation}
  ds^2 = -e^{2\beta} du \left( \frac{V}{r} du + 2 \, dr \right) +
  r^2 \left( d\theta^2 + \sin^2 \theta \, d\phi^2 \right).
  \label{eq:metric}
\end{equation}
Smoothness at $r=0$ allows imposition of the coordinate conditions
\begin{equation}
  \begin{array}{rcl}
    V(u,r) &=& r + \mathcal{O} (r^3) \\ [0.5 em]
    \beta(u,r) &=& \mathcal{O} (r^2).
  \end{array}
  \label{eq:bc}
\end{equation}
The field equations consist of the curved space wave equation $\Box \Phi = 0$
for the scalar field and two hypersurface equations for the metric
functions:
\begin{eqnarray}
  \beta_{,r} &=& 2 \pi r (\Phi_{,r})^2,
  \label{eq:sbeta}
  \\
  V_{,r} &=& e^{2 \beta}.
  \label{eq:sv}
\end{eqnarray}%
The wave equation can be expressed in the form
\begin{equation}
  \Box^{(2)} g - \left( \frac{V}{r} \right)_{\!\!,r}
  \frac{e^{-2 \beta} g}{r} = 0,
  \label{eq:hatwave}
\end{equation}
where $g=r\Phi$ and $\Box ^{(2)}$ is the D'Alembertian associated with
the two-dimensional submanifold spanned by the ingoing and outgoing
null geodesics. Initial null data for evolution consists of
$\Phi(u_0,r)$ at the initial retarded time $u_0$.

Because any two-dimensional geometry is conformally flat, the surface
integral of $\Box^{(2)} g$ over a null parallelogram $\Sigma$ gives
exactly the same result as in a flat 2-space, and leads to an integral
identity upon which a simple evolution algorithm can be
based~\cite{scasym}. Let the vertices of the null parallelogram be
labeled by $N$, $E$, $S$, and $W$ corresponding, respectively, to their
relative locations (North, East, South, and West) in the 2-space, as
shown in Figure~\ref{fig:nsew}. Upon integration of Equation~(\ref{eq:hatwave}),
curvature introduces an integral correction to the flat space
null parallelogram relation between the values of $g$ at the vertices:
\begin{equation}
  g_N - g_W - g_E + g_S = - \frac{1}{2} \int_\Sigma du \, dr
  \left( \frac{V}{r} \right)_{\!\!,r} \frac{g}{r}.
  \label{eq:integral}
\end{equation}

\epubtkImage{nsew.png}{%
  \begin{figure}[htbp]
    \centerline{\includegraphics[width=0.6\textwidth]{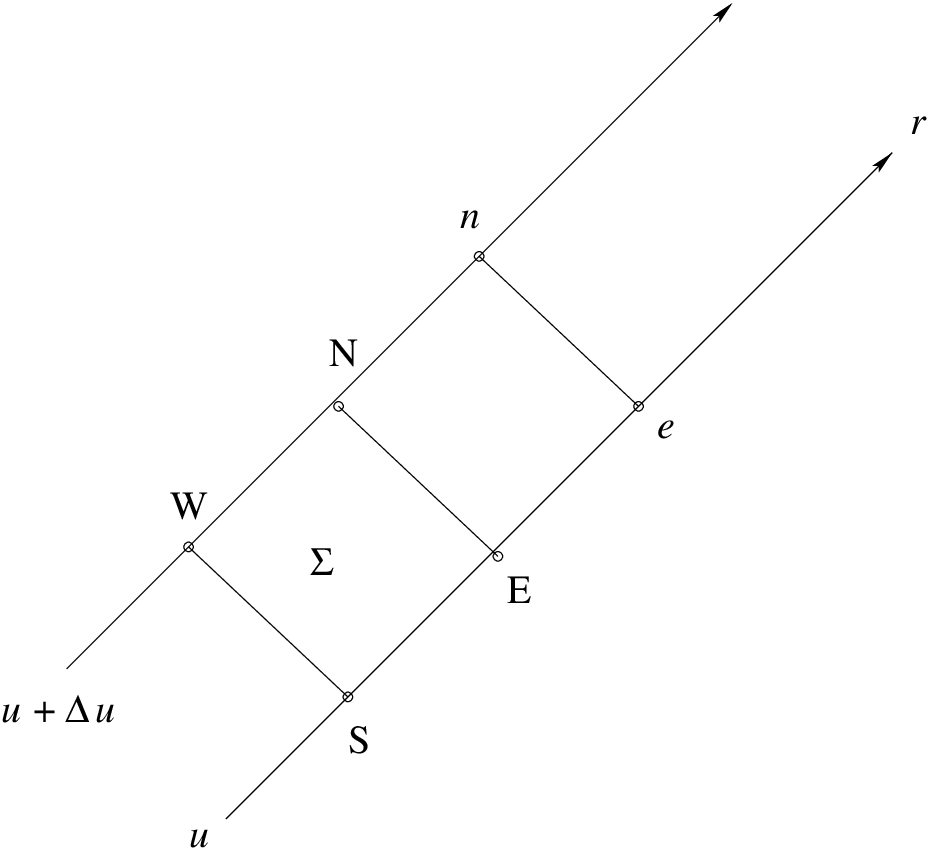}}
    \caption{The null parallelogram. After computing the field at
      point $N$, the algorithm marches the computation to $\mathcal{I}^+$
      by shifting the corners by $N\rightarrow n$, $E\rightarrow e$,
      $S\rightarrow E$, $W\rightarrow N$.}
    \label{fig:nsew}
  \end{figure}}

This identity, in one form or another, lies behind all of the null evolution
algorithms that have been applied to this system. The prime distinction between
the different algorithms is whether they are based upon double null coordinates,
or upon Bondi coordinates as in Equation~(\ref{eq:metric}). When a double null
coordinate system is adopted, the points  $N$, $E$, $S$, and $W$ can be located
in each computational cell at grid points, so that evaluation of the left hand
side of Equation~(\ref{eq:integral}) requires no interpolation. As a result, in
flat space, where the right hand side of Equation~(\ref{eq:integral}) vanishes,
it is possible to formulate an exact evolution algorithm. In curved space, of
course, there is a truncation error arising from the approximation of the
integral, e.g., by evaluating the integrand at the center of $\Sigma$.

The identity~(\ref{eq:integral}) gives rise to the following explicit marching
algorithm, indicated in Figure~\ref{fig:nsew}. Let the null parallelogram lie at
some fixed $\theta$ and $\phi$ and span adjacent retarded time levels $u_0$ and
$u_0+\Delta u$. Imagine for now that the points $N$, $E$, $S$, and $W$ lie on
the spatial grid, with $r_N-r_W=r_E-r_S=\Delta r$. If $g$ has been determined on
the entire initial cone $u_0$, which contains the points $E$ and $S$, and $g$
has been determined radially outward from the origin to the point $W$ on the
next cone $u_0+\Delta u$, then Equation~(\ref{eq:integral}) determines $g$ at
the next radial grid point $N$ in terms of an integral over $\Sigma$. The
integrand can be approximated to second order, i.e.\ to $\mathcal{O} (\Delta r
\Delta u)$, by evaluating it at the center of $\Sigma$. To this same accuracy,
the value of $g$ at the center equals its average between the points $E$ and
$W$, at which $g$ has already been determined. Similarly, the value of
$(V/r)_{,r}$ at the center of $\Sigma$ can be approximated to second order in
terms of values of $V$ at points where it can be determined by integrating the
hypersurface equations~(\ref{eq:sbeta}, \ref{eq:sv}) radially outward from
$r=0$.

After carrying out this procedure to evaluate $g$ at the point $N$, the
procedure can then be iterated to determine $g$ at the next radially outward
grid point on the $u_0+\Delta u$ level, i.e.\ point $n$ in
Figure~\ref{fig:nsew}. Upon completing this radial march to null infinity, in
terms of a compactified radial coordinate such as $x=r/(1+r)$, the field $g$ is
then evaluated on the next null cone at $u_0+2\Delta u$, beginning at the vertex
where smoothness gives the startup condition that $g(u,0)=0$.

In the compactified Bondi formalism, the vertices $N$, $E$, $S$, and $W$
of the null parallelogram $\Sigma$ cannot be chosen to lie exactly on
the grid because, even in Minkowski space, the velocity of light in
terms of a compactified radial coordinate $x$ is not constant. As a
consequence, the fields $g$, $\beta$, and $V$ at the vertices of
$\Sigma$ are approximated to second order accuracy by interpolating
between grid points. However, cancellations arise between these four
interpolations so that Equation~(\ref{eq:integral}) is satisfied to fourth
order accuracy. The net result is that the finite difference version
of Equation~(\ref{eq:integral}) steps $g$ radially outward one zone with an
error of fourth order in grid size, $\mathcal{O} ((\Delta u)^2 (\Delta x)^2)$. In
addition, the smoothness conditions~(\ref{eq:bc}) can be incorporated
into the startup for the numerical integrations for $V$ and $\beta$ to
insure no loss of accuracy in starting up the march at $r=0$. The
resulting global error in $g$, after evolving a finite retarded time,
is then $\mathcal{O} (\Delta u\Delta x)$, after compounding errors from $1/(\Delta
u\Delta x)$ number of zones.

When implemented on a grid based upon the $(u,r)$ coordinates, the stability of
this algorithm is subject to a Courant--Friedrichs--Lewy (CFL) condition
requiring that the physical domain of dependence be contained in the numerical
domain of dependence. In the spherically symmetric case, this condition
requires that the ratio of the time step to radial step be limited by
$(V/r)\Delta u \le 2\Delta r$, where $\Delta r=\Delta[x/(1-x)]$. This condition
can be built into the code using the value $V/r=e^{2H}$, corresponding to the
maximum of $V/r$ at $\mathcal{I}^+$. The strongest restriction on the time step then
arises just before the formation of a horizon, where $V/r\rightarrow \infty$ at
$\mathcal{I}^+$. This infinite redshift provides a mechanism for locating the true
event horizon ``on the fly'' and restricting the evolution to the exterior
spacetime. Points near $\mathcal{I}^+$ must be dropped in order to evolve across
the horizon due to the lack of a nonsingular compactified version of future
time infinity  $I^+$.

The situation is quite different in a double null coordinate system, in which
the vertices of the null parallelogram can be placed exactly on grid points so
that the CFL condition is automatically satisfied. A characteristic code based
upon double null coordinates was developed by Goldwirth and Piran in a study of
cosmic censorship~\cite{goldw} based upon the spherically symmetric
gravitational collapse of a massless scalar field. Their early study lacked the
sensitivity of adaptive mesh refinement (AMR) which later enabled Choptuik to
discover the critical phenomena appearing in this problem. Subsequent work by
Marsa and Choptuik~\cite{mc} combined the use of the null related ingoing
Eddington--Finklestein coordinates with Unruh's strategy of singularity
excision to construct a 1D code that ``runs forever''. Later,
Garfinkle~\cite{garf1} constructed an improved version of  the Goldwirth--Piran
double null code which was able to simulate critical phenomena without using
adaptive mesh refinement. In this treatment, as the evolution proceeds on one
outgoing null cone to the next, the grid points follow the ingoing null cones
and must be dropped as they cross the origin at $r=0$. However, after half the
grid points are lost they are  then ``recycled'' at new positions midway
between the remaining grid points. This technique is crucial for resolving the
critical phenomena associated with an $r\rightarrow  0$ size horizon. An
extension of the code~\cite{garf2} was later used to verify that scalar field
collapse in six dimensions continues to display critical phenomena.

Hamad\'e and Stewart~\cite{hamst} also applied a double null code to study
critical phenomena. In order to obtain the accuracy necessary to confirm
Choptuik's results they developed the first example of a characteristic grid
with AMR. They did this with both the standard
Berger and Oliger algorithm and their own simplified version, with both
versions giving indistinguishable results. Their simulations of critical
collapse of a massless scalar field agreed with Choptuik's values for the
universal parameters governing mass scaling and displayed the echoing
associated with discrete self-similarity. Hamad\'e, Horne, and
Stewart~\cite{hamhst} extended this study to the spherical collapse of an
axion/dilaton system and found in this case that self-similarity was a
continuous symmetry of the critical solution.

Brady, Chambers, and Gon{\c{c}}alves~\cite{BrChGo} used Garfinkle's~\cite{garf1}
double null algorithm to investigate  the effect of a \emph{massive} scalar
field on critical phenomena. The introduction of a mass term in the scalar wave
equation introduces a scale to the problem, which suggests that the critical
point behavior might differ from the massless case. They found that there are
two different regimes depending on the ratio of the Compton wavelength $1/m$ of
the scalar mass to the radial size $\lambda$ of the scalar pulse used to induce
collapse. When $\lambda m << 1$, the critical solution is the one found by
Choptuik in the $m=0$ case, corresponding to a type~II phase transition.
However, when $\lambda m >> 1$, the critical solution is an unstable soliton
star (see~\cite{SeidelSuen}), corresponding to a type~I phase transition where
black hole formation turns on at a finite mass.

A code based upon Bondi coordinates, developed by Husa and his
collaborators~\cite{huscrit}, has been successfully applied to spherically
symmetric critical collapse of a nonlinear $\sigma$-model coupled to gravity.
Critical phenomena cannot be resolved on a static grid based upon the Bondi
$r$-coordinate. Instead, the numerical techniques of Garfinkle were adopted by
using a dynamic grid following the ingoing null rays and by recycling radial
grid points. They studied how coupling to gravity affects the critical behavior
previously observed by Bizo\'n~\cite{bizon} and others in the Minkowski
space version of the model. For a wide range of the coupling constant, they
observe discrete self-similarity and typical mass scaling near the critical
solution. The code is shown to be second order accurate and to give second
order convergence for the value of the critical parameter.

The first characteristic code in Bondi coordinates for the self-gravitating
scalar wave problem was constructed by G\'omez and Winicour~\cite{scasym}. They
introduced a numerical compactification of $\mathcal{I}^+$ for the purpose of
studying  effects of self-gravity on the scalar radiation, particularly in the
high amplitude limit of the rescaling $\Phi\rightarrow a\Phi$. As $a
\rightarrow\infty$, the red shift creates an effective boundary layer at
$\mathcal{I}^+$ which causes the Bondi mass $M_\mathrm{B}$ and the scalar field monopole
moment $Q$ to be related by $M_\mathrm{B}\sim \pi |Q|/\sqrt{2}$, rather than the
quadratic relation of the weak field limit~\cite{scasym}. This could also be
established analytically so that the high amplitude limit provided a check on
the code's ability to handle strongly nonlinear fields. In the small amplitude
case, this work \emph{incorrectly} reported that the radiation tails from black
hole formation had an exponential decay characteristic of quasinormal modes
rather than the polynomial $1/t$ or $1/t^2$ falloff expected from
Price's~\cite{price} work on perturbations of Schwarzschild black
holes. In hindsight,
the error here was not having confidence to run the code sufficiently long to
see the proper late time behavior.

Gundlach, Price, and Pullin~\cite{gpp1,gpp2} subsequently reexamined the
issue of power law tails using a double null code similar to that developed by
Goldwirth and Piran. Their numerical simulations verified the existence of
power law tails in the full nonlinear case, thus establishing consistency with
analytic perturbative theory. They also found normal mode ringing at
intermediate time, which provided reassuring consistency with perturbation
theory and showed that there is a region of spacetime where the results of
linearized theory are remarkably reliable even though highly nonlinear behavior
is taking place elsewhere. These results have led to a methodology that has
application beyond the confines of spherically symmetric problems, most notably
in the ``close approximation'' for the binary black hole problem~\cite{jorge}.
Power law tails and quasinormal ringing have also been confirmed using Cauchy
evolution~\cite{mc}.

The study of the radiation tail decay of a scalar field was subsequently
extended by G\'omez, Schmidt, and Winicour~\cite{gsw} using a characteristic
code. They showed that the Newman--Penrose constant~\cite{conserved} for the
scalar field determines the exponent of the power law (and not the static
monopole moment as often stated). When this constant is non-zero, the tail
decays as $1/t$ on $\mathcal{I}^+$, as opposed to the $1/t^2$ decay for the
vanishing case. (They also found $t^{-n}\log t$ corrections, in addition to
the exponentially decaying contributions of the quasinormal modes.)  This code
was also used to study the instability of a topological kink in the
configuration of the scalar field~\cite{kink}. The kink instability provides
the simplest example of the turning point instability~\cite{ipser,sork} which
underlies gravitational collapse of static equilibria.

Brady and Smith~\cite{bradysmith} have demonstrated that characteristic
evolution is especially well adapted to explore properties of Cauchy horizons.
They examined the stability of the Reissner--Nordstr\"om Cauchy horizon using an
Einstein--Klein--Gordon code based upon advanced Bondi coordinates $(v,r)$ (where
the hypersurfaces $v=const$ are ingoing null hypersurfaces). They studied the
effect of a spherically symmetric scalar pulse on the spacetime structure as it
propagates across the event horizon. Their numerical methods were patterned
after the work of Goldwirth and Piran~\cite{goldw}, with modifications of the
radial grid structure that allow deep penetration inside the black hole. In
accord with expectations from analytic studies, they found that the pulse first
induces a weak null singularity on the Cauchy horizon, which then
leads to a crushing spacelike singularity as $r\rightarrow 0$. The null
singularity is weak in the sense that an infalling observer experiences a
finite tidal force, although the Newman--Penrose Weyl component $\Psi_2$
diverges, a phenomenon known as mass inflation~\cite{poisson}. These results
confirm the earlier result of Gnedin and Gnedin~\cite{gnedin} that a central
spacelike singularity would be created by the interaction of a charged black
hole with a scalar field, in accord with a physical argument by
Penrose~\cite{penrcoscen} that a small perturbation undergoes an infinite
redshift as it approaches the Cauchy horizon.

Burko~\cite{burko} has confirmed and extended these results, using a code based
upon double null coordinates which was developed with Ori~\cite{burkori} in a
study of tail decay. He found that in the early stages the
perturbation of the Cauchy horizon is weak and in agreement with the behavior
calculated by perturbation theory.

Brady, Chambers, Krivan, and Laguna~\cite{BrChKrLa} have found interesting
effects of a non-zero cosmological constant $\Lambda$ on tail decay by using
a characteristic Einstein--Klein--Gordon code to study the effect of a
massless scalar pulse on Schwarzschild--de Sitter and
Reissner--Nordstr\"om--de Sitter spacetimes. First, by constructing a
linearized scalar evolution code, they show that scalar test fields with
$\ell\ne 0$ have exponentially decaying tails, in contrast to the standard
power law tails for asymptotically flat spacetimes. Rather than decaying, the
monopole mode asymptotes at late time to a constant, which scales linearly
with $\Lambda$, in contrast to the standard no-hair result. This unusual
behavior for the $\ell=0$ case was then independently confirmed with a
nonlinear spherical characteristic code.

Using a combination of numerical and analytic techniques based upon null
coordinates, Hod and Piran have made an extensive series of investigations of
the spherically symmetric charged Einstein--Klein--Gordon system dealing with
the effect of charge on critical gravitational collapse~\cite{HodPir1} and the
late time tail decay of a charged scalar field on a Reissner--Nordstr\"om black
hole~\cite{HodPir2,HodPir2.5,HodPir3,HodPir4}. These studies culminated in a
full nonlinear investigation of horizon formation by the collapse of a charged
massless scalar pulse~\cite{HodPir5}. They track the formation of an apparent
horizon which is followed by a weakly singular Cauchy horizon which then
develops a strong spacelike singularity at $r=0$. This is in complete accord
with prior perturbative results and nonlinear simulations involving a
pre-existing black hole. Oren and Piran~\cite{OrenPir} increased the late time
accuracy of this study by incorporating an adaptive grid for the retarded time
coordinate $u$, with a refinement criterion to maintain $\Delta r/r =
\mathrm{const}$. The accuracy of this scheme is confirmed through convergence
tests as well as charge and constraint conservation. They were able to observe
the physical mechanism which prohibits black hole formation with charge to mass
ration $Q/M>1$. Electrostatic repulsion of the outer parts of the scalar pulse
increases relative to the gravitational attraction and causes the outer portion
of the charge to disperse to larger radii before the black hole is formed.
Inside the black hole, they confirm the formation of a weakly singular Cauchy
horizon which turns into a strong spacelike singularity, in accord with other
studies.

Hod extended this combined numerical-analytical double null approach to
investigate higher order corrections to the dominant power law tail~\cite{Hod2},
as well as corrections due to a general spherically symmetric scattering
potential~\cite{Hod1} and due to a time dependent potential~\cite{Hod3}. He
found $(\log t)/t$ modifications to the leading order tail behavior for a
Schwarzschild black hole, in accord with earlier results of G{\'{o}}mez et
al.~\cite{gsw}. These modifications fall off at a slow rate so that a very long
numerical evolution ($t\approx 3000 \, M$)is necessary to cleanly identify the
leading order power law decay.

The foregoing numerical-analytical work based upon characteristic evolution
has contributed to a very comprehensive classical treatment of spherically
symmetric gravitational collapse. Sorkin and Piran~\cite{SorkPir} have
investigated the question of quantum  corrections due to pair creation on the
gravitational collapse of a charged scalar field. For observers outside the
black hole, several analytic studies have indicated that such pair-production
can rapidly diminish the charge of the black hole. Sorkin and Piran apply the
same double-null characteristic code used in studying the classical
problem~\cite{HodPir5} to evolve across the event horizon and observe the
quantum effects on the Cauchy horizon. The quantum electrodynamic effects are
modeled in a rudimentary way by a nonlinear dielectric $\epsilon$ constant
that limits the electric field to the critical value necessary for pair
creation. The back-reaction of the pairs on the stress-energy and the
electric current are ignored. They found that quantum effects leave the
classical picture of the Cauchy horizon qualitatively intact but that they
shorten its ``lifetime'' by hastening the conversion of the weak null
singularity into a strong spacelike singularity.

The Southampton group has constructed a \{1\,+\,1\}-dimensional characteristic
code for spacetimes with cylindrical symmetry~\cite{cylinder1,cylinder2}. The
original motivation was to use it as the exterior characteristic code in a test
case of CCM (see Section~\ref{sec:cylmatch} for the application to matching).
Subsequently, Sperhake, Sj\"odin, and Vickers~\cite{vick1,vick2} modified the
code into a global characteristic version for the purpose of studying cosmic
strings, represented by massive scalar and vector fields coupled to gravity.
Using a Geroch decomposition~\cite{gerdec} with respect to the translational
Killing vector, they reduced the global problem to a \{2\,+\,1\}-dimensional
asymptotically flat spacetime, so that $\mathcal{I}^+$ could be compactified and
included in the numerical grid. Rather than the explicit scheme used in CCM, the
new version employs an implicit, second order in space and time,
Crank--Nicholson evolution scheme. The code showed long term stability and
second order convergence in vacuum tests based upon exact Weber--Wheeler
waves~\cite{wweb} and Xanthopoulos' rotating solution~\cite{xanth}, and in tests
of wave scattering by a string. The results show damped ringing of the string
after an incoming Weber--Wheeler pulse has excited it and then scattered to
$\mathcal{I}^+$. The ringing frequencies are independent of the details of the
pulse but are inversely proportional to the masses of the scalar and vector
fields.

Frittelli and G{\'{o}}mez~\cite{gfsymhyp} have cast the spherically symmetric
Einstein--Klein--Gordon problem in symmetric hyperbolic form, where in a
Bondi-Sachs gauge the fundamental variables are the scalar field, lapse and
shift. The Bondi-Sachs gauge conditions relate the usual ADM variables (the
3-metric and extrinsic curvature) to the lapse and shift, which obey simpler
evolution equations. The resulting Cauchy problem is well-posed and the outer
boundary condition is constraint preserving (although whether the resulting IBVP
is  well-posed is not addressed, i.e.\ whether the boundary condition is
dissipative). A numerical evolution algorithm based upon the system produces a
stable simulation of a scalar pulse $\Phi$ scattering off a black hole. The
initial data for the pulse satisfies $\partial_t \Phi=0$ so, as expected, it
contains an ingoing part, which crosses the horizon, and an outgoing part, which
leaves the grid at the outer boundary with a small amount of back reflection.


\subsubsection{Cosmology on the past null cone}
\label{sec:cosmology}

The standard approach to cosmology begins with a spacetime metric
incorporating assumptions of approximate homogeneity and
isotropy. An alternative approach, based upon observational data
on the past light cone of earth based telescopes,
was proposed in a seminal paper by Kristian and Sachs~\cite{kristian}. 
In that work, construction of the metric was based upon observational data but the use of a 
series expansion restricted the approach to nearby regions. Their ideas provided
the starting point for further developments. In particular the observational
cosmology program of  Ellis et al.~\cite{ellis} exploited
an earlier work of Temple~\cite{temple} to extend the approach to larger redshift by
using the natural observational coordinates based upon null geodesics propagating
to the telescope.

In this approach, solving the Einstein equations in the context of observational cosmology
poses two problems. First, astronomical observations are used to determine the metric
on the past null cone of the observer. Second, these are used as the \emph{final data}
for a characteristic evolution into the past, which determines the cosmological history.
A program to carry out this second step via numerical evolution has been initiated by
Bishop and his collaborators~\cite{haines,walt}. As a first step, they implemented a spherically
symmetric null code for the Einstein equations coupled with a pressure free fluid
(dust). The code was tested against solutions of the spherically symmetric but inhomogeneous
Lema{\^{\i}}tre--Tolman--Bondi model. The code is then used to compare
the Lema{\^{\i}}tre-Tolman-Bondi model with the now standard
$\Lambda$-cold-dark-matter model. Using the presently observed characteristic data, it is
shown that the past histories of these two models are distinctly different. The density of the
Lema{\^{\i}}tre--Tolman--Bondi model rises more quickly
into the past, indicating a universe which might be too young. 

\subsubsection{Adaptive mesh refinement}

The goal of computing waveforms from relativistic binaries, such as a neutron
star or stellar mass back hole spiraling into a supermassive black hole,
requires more than a stable convergent code. It is a delicate task to extract a
waveform in a spacetime in which there are multiple length scales: the size of
the supermassive black hole, the size of the star, the wavelength of the
radiation. It is commonly agreed that some form of mesh refinement is essential
to attack this problem. Mesh refinement was first applied in characteristic
evolution to solve specific spherically symmetric problems regarding critical
phenomena and singularity structure~\cite{garf1,hamst,burko}.

Pretorius and Lehner~\cite{pretlehn} have presented a general approach for AMR
to a generic characteristic code. Although the method is designed to treat 3D
simulations, the implementation has so far been restricted to the
Einstein--Klein--Gordon system in spherical symmetry. The 3D approach is modeled
after the Berger and Oliger AMR algorithm for hyperbolic Cauchy problems, which
is reformulated in terms of null coordinates. The resulting characteristic AMR
algorithm can be applied to any unigrid characteristic code and is amenable to
parallelization. They applied it to the problem of a spherically symmetric
massive Klein--Gordon field propagating outward from a black hole. The non-zero
rest mass restricts the Klein--Gordon field from propagating to infinity.
Instead it diffuses into higher frequency components which Pretorius and Lehner
show can be resolved using AMR but not with a comparison unigrid code.


\subsection{\{2\,+\,1\}-dimensional codes}

One-dimensional characteristic codes enjoy a very special simplicity
due to the two preferred sets (ingoing and outgoing) of characteristic
null hypersurfaces. This eliminates a source of gauge freedom that
otherwise exists in either two- or three-dimensional characteristic
codes. However, the manner in which the characteristics of a hyperbolic
system determine domains of dependence and lead to propagation
equations for shock waves is the same as in the one-dimensional
case. This makes it desirable for the purpose of numerical evolution to
enforce propagation along characteristics as extensively as possible.
In basing a Cauchy algorithm upon shooting along characteristics, the
infinity of characteristic rays (technically, \emph{bicharacteristics})
at each point leads to an arbitrariness which, for a practical
numerical scheme, makes it necessary either to average the propagation
equations over the sphere of characteristic directions or to select out
some preferred subset of propagation equations. The latter
approach was successfully applied by Butler~\cite{butler} to the
Cauchy evolution of two-dimensional fluid flow, but there seems to have
been very little follow-up along these lines. The closest resemblance is the
use of Riemann solvers for high resolution shock capturing in hydrodynamic
codes (see Section~\ref{sec:shydro}).

The formal ideas behind the construction of two- or three-dimensional
characteristic codes are similar, although there are various technical options
for treating the angular coordinates which label the null rays. Historically,
most characteristic work graduated first from 1D to 2D because of the available
computing power.


\subsection{The Bondi problem}
\label{sec:bondiprob}

The first characteristic code based upon the original Bondi equations for a
twist-free axisymmetric spacetime was constructed by J.~Welling in his PhD
thesis at Pittsburgh~\cite{isaac} . The spacetime was foliated by a family of
null cones, complete with point vertices at which regularity conditions were
imposed. The code accurately integrated the hypersurface and evolution
equations out to compactified null infinity. This allowed studies of the Bondi
mass and radiation flux on the initial null cone, but it could not be used as a
practical evolution code because of instabilities.

These instabilities came as a rude shock and led to a retreat to the
simpler problem of axisymmetric scalar waves propagating in Minkowski
space, with the metric
\begin{equation}
  ds^2= - du^2 - 2 \, du \, dr +
  r^2 \left( d\theta^2 + \sin^2 \theta \, d\phi^2 \right)
  \label{eq:mink}
\end{equation}
in outgoing null cone coordinates. A null cone code for this problem was
constructed using an algorithm based upon Equation~(\ref{eq:integral}), with the angular
part of the flat space Laplacian replacing the curvature terms in the integrand
on the right hand side. This simple setting allowed one source of instability
to be traced to a subtle violation of the CFL condition near the vertices of
the cones. In terms of the grid spacing $\Delta x^{\alpha}$, the CFL condition
in this coordinate system takes the explicit form
\begin{equation}
  \frac{\Delta u}{\Delta r} < - 1 +
  \left[ K^2 + (K - 1)^2 - 2 K (K - 1) \cos \Delta \theta \right]^{1/2},
  \label{eq:cfl}
\end{equation}
where the coefficient $K$, which is of order 1, depends on the
particular startup procedure adopted for the outward integration. Far
from the vertex, the condition~(\ref{eq:cfl}) on the time step $\Delta
u$ is quantitatively similar to the CFL condition for a standard
Cauchy evolution algorithm in spherical coordinates. But
condition~(\ref{eq:cfl}) is strongest near the vertex of the cone
where (at the equator $\theta =\pi/2$) it implies that
\begin{equation}
  \Delta u < K \, \Delta r \, (\Delta \theta)^2.
  \label{eq:cfl0}
\end{equation}
This is in contrast to the analogous requirement
\begin{equation}
  \Delta u < K \, \Delta r \, \Delta \theta
\end{equation}
for stable Cauchy evolution near the origin of a spherical coordinate
system. The extra power of $\Delta \theta$ is the price that must be
paid near the vertex for the simplicity of a characteristic code.
Nevertheless, the enforcement of this condition allowed efficient
global simulation of axisymmetric scalar waves. Global studies of
backscattering, radiative tail decay, and solitons were carried out
for nonlinear axisymmetric waves~\cite{isaac}, but three-dimensional
simulations extending to the vertices of the cones were impractical
at the time on existing machines.

Aware now of the subtleties of the CFL condition near the vertices, the
Pittsburgh group returned to the Bondi problem, i.e.\ to evolve the Bondi
metric~\cite{bondi}
\begin{equation}
  ds^2 = \left( \frac{V}{r} e^{2 \beta} - U^2 r^2 e^{2 \gamma} \right) du^2 +
  2 e^{2 \beta} du \, dr + 2 U r^2 e^{2 \gamma} du \, d\theta -
  r^2 \left (e^{2 \gamma} d\theta^2 + e^{-2\gamma} \sin^2 \theta \, d\phi^2 \right),
  \label{eq:bmetric}
\end{equation}
by means of the three hypersurface equations
\begin{eqnarray}
  \beta_{,r} &=& \frac{1}{2} r (\gamma_{,r})^2,
  \label{eq:beta}
  \\
  \left[ r^4 e^{2 (\gamma - \beta)} U_{,r} \right]_{,r} &=&
  2 r^2 \left[ r^2  \left( \frac{\beta}{r^2} \right)_{\!\!,r\theta} -
  \frac{(\sin^2 \theta \, \gamma)_{,r\theta}}{\sin^2 \theta} +
  2 \gamma_{,r} \gamma_{,\theta} \right],
  \label{eq:U}
  \\
  V_{,r} &=& - \frac{1}{4} r^4 e^{2 (\gamma - \beta)}(U_{,r})^2 +
  \frac{(r^4 \sin \theta \, U)_{,r\theta}}{2 r^2 \sin \theta}
  \nonumber \\
  && + e^{2 (\beta - \gamma)} \left[ 1 -
  \frac{(\sin \theta \, \beta_{,\theta})_{,\theta}}{\sin \theta} +
  \gamma_{,\theta\theta} + 3 \cot \theta \, \gamma_{,\theta} -
  (\beta_{,\theta})^2 - 2 \gamma_{,\theta}
  (\gamma_{,\theta} - \beta_{,\theta}) \right], \qquad
  \label{eq:V}
\end{eqnarray}%
and the evolution equation
\begin{eqnarray}
  4 r ( r \gamma)_{,ur} & = & \left\{ 2 r \gamma_{,r} V -
  r^2 \left[ 2 \gamma_{,\theta} U + \sin \theta
  \left( \frac{U}{\sin \theta} \right)_{\!\!,\theta} \right]
  \right\}_{,r} \!\!\!\! - 2 r^{2} \frac{(\gamma_{,r} U
  \sin \theta)_{,\theta}}{\sin \theta}
  \nonumber \\
  && + \frac{1}{2} r^{4} e^{2 (\gamma - \beta)}
  (U_{,r})^2 + 2 e^{2 (\beta - \gamma)} \left[ (\beta_{,\theta})^2 +
  \sin \theta \left( \frac{\beta_{,\theta}}{\sin \theta}
  \right)_{\!\!,\theta} \right].
  \label{eq:gammaev}
\end{eqnarray}%

The beauty of the Bondi equations is that they form a clean hierarchy. Given
$\gamma$ on an initial null hypersurface, the equations can be integrated
radially to determine $\beta$, $U$, $V$, and $\gamma_{,u}$ on the hypersurface
(in that order) in terms of integration constants determined by boundary
conditions, or smoothness conditions if extended to the vertex of a null cone.
The initial data $\gamma$ is unconstrained except for smoothness conditions.
Because $\gamma$ represents an axisymmetric spin-2 field, it must be
$\mathcal{O} (\sin^2 \theta)$ near the poles of the spherical coordinates and
must consist of $l\ge 2$ spin-2 multipoles.

In the computational implementation of this system by the Pittsburgh
group~\cite{papa}, the null hypersurfaces were chosen to be complete null cones
with nonsingular vertices, which (for simplicity) trace out a geodesic worldline
$r=0$. The smoothness conditions at the vertices were formulated in local
Minkowski coordinates.

The vertices of the cones were not the chief source of
difficulty. A null parallelogram marching algorithm, similar to that
used in the scalar case, gave rise to another instability that sprang up
throughout the grid. In order to reveal the source of this instability,
physical considerations suggested looking at the linearized version of
the Bondi equations, where they can be related to the wave equation.
If this relationship were sufficiently simple, then the scalar wave
algorithm could be used as a guide in stabilizing the evolution of
$\gamma$. A scheme for relating $\gamma$ to solutions $\Phi$ of the
wave equation had been formulated in the original paper by Bondi,
Metzner, and van der Burgh~\cite{bondi}. However, in that scheme, the
relationship of the scalar wave to $\gamma$ was nonlocal in the
angular directions and was not useful for the stability analysis.

A local relationship between $\gamma$ and solutions of the wave equation was
found~\cite{papa}. This provided a test bed for the null evolution algorithm
similar to the Cauchy test bed provided by Teukolsky waves~\cite{teuk}. More
critically, it allowed a simple von~Neumann linear stability analysis of the
finite difference equations, which revealed that the evolution would be unstable
if the metric quantity $U$ was evaluated on the grid. For a stable algorithm, the
grid points for $U$ must be staggered between the grid points for $\gamma$,
$\beta$, and $V$. This unexpected feature emphasizes the value of linear
stability analysis in formulating stable finite difference approximations.

It led to an axisymmetric code~\cite{papath,papa} for the global Bondi problem
which ran stably, subject to a CFL condition, throughout the regime in which
caustics and horizons did not form. Stability in this regime was verified
experimentally by running arbitrary initial data until it radiated away to
$\mathcal{I}^+$. Also, new exact solutions as well as the linearized  null
solutions were used to perform extensive convergence tests that established
second order accuracy. The code generated a large complement of highly accurate
numerical solutions for the class of asymptotically flat, axisymmetric vacuum
spacetimes, a class for which no analytic solutions are known. All results of
numerical evolutions in this regime were consistent with the theorem of
Christodoulou and Klainerman~\cite{XKlain} that weak initial data evolve
asymptotically to Minkowski space at late time.

An additional global check on accuracy was performed  using Bondi's
formula relating mass loss to the time integral of the square of the
news function. The Bondi mass loss formula is not one of the equations
used in the evolution algorithm but follows from those equations as a
consequence of a global integration of the Bianchi identities. Thus it
not only furnishes a valuable tool for physical interpretation but it
also provides a very important calibration of numerical accuracy and
consistency.

An interesting feature of the evolution arises in regard to
compactification. By construction, the $u$-direction is timelike at
the origin where it coincides with the worldline traced out by the
vertices of the outgoing null cones. But even for weak fields, the
$u$-direction generically becomes spacelike at large distances along an
outgoing ray. Geometrically, this reflects the property that $\mathcal{I}^+$
is itself a null hypersurface so that all internal directions are
spacelike, except for the null generator. For a flat space time, the
$u$-direction picked out at the origin leads to a null evolution
direction at $\mathcal{I}^+$, but this direction becomes spacelike under a
slight deviation from spherical symmetry. Thus the evolution
generically becomes ``superluminal'' near $\mathcal{I}^+$. Remarkably,
this leads to no adverse numerical effects. This remarkable property
apparently arises from the natural way that causality is built into the
marching algorithm so that no additional resort to numerical
techniques, such as ``causal differencing''~\cite{Alliance97b}, is
necessary.


\subsubsection{The conformal-null tetrad approach}

Stewart has implemented a characteristic evolution code which handles the
Bondi problem by a null tetrad, as opposed to metric, formalism~\cite{stewbm}.
The geometrical algorithm underlying the evolution scheme, as outlined
in~\cite{friedst1,friedst2}, is Friedrich's~\cite{fried} conformal-null
description of a compactified spacetime in terms of a first order system of
partial differential equations. The variables include the metric, the
connection, and the curvature, as in a Newman--Penrose formalism, but in
addition the conformal factor (necessary for compactification of $\mathcal{I}$)
and its gradient. Without assuming any symmetry, there are more than 7 times as
many variables as in a metric based null scheme, and the corresponding
equations do not decompose into as clean a hierarchy. This disadvantage,
compared to the metric approach, is balanced by several advantages:
\begin{itemize}
\item The equations form a symmetric hyperbolic system so that
  standard theorems can be used to establish that the system is
  well-posed.
\item Standard evolution algorithms can be invoked to ensure numerical
  stability.
\item The extra variables associated with the curvature tensor are not
  completely excess baggage, since they supply essential physical
  information.
\item The regularization necessary to treat $\mathcal{I}^+$ is built in
  as part of the formalism so that no special numerical regularization
  techniques are necessary as in the metric case. (This last advantage
  is somewhat offset by the necessity of having to locate
  $\mathcal{I}$ by tracking the zeroes of the conformal factor.)
\end{itemize}

The code was intended to study gravitational waves from an axisymmetric star.
Since only the vacuum equations are evolved, the outgoing radiation from the
star is represented by data ($\Psi_4$ in Newman--Penrose notation) on an ingoing
null cone forming the inner boundary of the evolved domain. This inner boundary
data is supplemented by Schwarzschild data on the initial outgoing null cone,
which models an initially quiescent state of the star. This provides the
necessary data for a double-null initial value problem. The evolution would
normally break down where the ingoing null hypersurface develops caustics. But
by choosing a scenario in which a black hole is formed, it is possible to
evolve the entire region exterior to the horizon. An obvious test bed is the
Schwarzschild spacetime for which a numerically satisfactory evolution was
achieved (although convergence tests were not reported).

Physically interesting results were obtained by choosing data corresponding to
an outgoing quadrupole pulse of radiation. By increasing the initial amplitude
of the data $\Psi_4$, it was possible to evolve into a regime where the energy
loss due to radiation was large enough to drive the total Bondi mass negative.
Although such data is too grossly exaggerated to be consistent with an
astrophysically realistic source, the formation of a negative mass was an
impressive test of the robustness of the code.


\subsubsection{Axisymmetric mode coupling}
\label{sec:papamode}

Papadopoulos~\cite{papamode} has carried out an illuminating study of mode
mixing by computing the evolution of a pulse emanating outward from an
initially Schwarzschild white hole of mass $M$. The evolution proceeds along a
family of ingoing null hypersurfaces with outer boundary at $r=60\,M$. The
evolution is stopped before the pulse hits the outer boundary in order to avoid
spurious effects from reflection and the radiation is inferred from data at
$r=20\,M$. Although gauge ambiguities arise in reading off the waveform at a
finite  radius, the work reveals interesting nonlinear effects: (i)
modification of the light cone structure governing the principal part of the
equations and hence the propagation of signals; (ii) modulation of the
Schwarzschild potential by the introduction of an angular dependent ``mass
aspect''; and (iii) quadratic and higher order terms in the evolution equations
which couple the spherical harmonic modes. A compactified version of this
study~\cite{zlochmode} was later carried out with the 3D PITT code, which
confirms these effects as well as new effects which are not present in the
axisymmetric case (see Section~\ref{sec:mode} for details).

\subsubsection{Spectral approach to the Bondi problem}

Oliveira and Rodrigues~\cite{oliveira} have taken the first
step in developing a code based upon
the Galerkin spectral method to evolve the axisymmetric Bondi problem. The
strength of spectral methods is to provide high accuracy relative to
computational effort. The spectral decomposition reduces  the partial
differential evolution system to a system of ordinary differential equations for
the spectral coefficients. Several numerical tests were performed to verify
stability and convergence, including linearized gravitational waves  and the
global energy momentum conservation law relating the Bondi mass to the 
radiated energy flux. The main feature of the Galerkin method is that each basis
function is chosen to automatically satisfy the boundary conditions, in this
case the regularity conditions on the Bondi variables on the axes of symmetry
and at the vertices of the outgoing null cones and the asymptotic flatness
condition at infinity.  Although  $\mathcal{I}^+$ is not explicitly compactified,
the choice of radial basis functions allows verification of the asymptotic
relations governing the coefficients of the leading  gauge dependent terms of
the metric quantities.

It will be interesting to see if the approach
can be applied to highly nonlinear problems and generalized to the full 3D case.
There has been little other effort in applying spectral methods to characteristic
evolution, although the approach offers a distinct advantage in handling the
vertices of the null cones.


\subsubsection{Twisting axisymmetry}
\label{sec:axiev}

The Southampton group, as part of its goal of combining Cauchy and
characteristic evolution, has developed a code~\cite{south1,south2,pollney}
which extends the Bondi problem to full axisymmetry, as described by the
general characteristic formalism of Sachs~\cite{sachs}. By dropping the
requirement that the rotational Killing vector be twist-free, they were able to
include rotational effects, including radiation in the ``cross'' polarization
mode (only the ``plus'' mode is allowed by twist-free axisymmetry). The null
equations and variables were recast into a suitably regularized form to allow
compactification of null infinity. Regularization at the vertices or caustics
of the null hypersurfaces was not necessary, since they anticipated matching to
an interior Cauchy evolution across a finite worldtube.

The code was designed to insure standard Bondi coordinate conditions at
infinity, so that the metric has the asymptotically Minkowskian form
corresponding to null-spherical coordinates. In order to achieve this, the
hypersurface equation for the Bondi metric variable $\beta$ must be integrated
radially inward from infinity, where the integration constant is specified. The
evolution of the dynamical variables proceeds radially outward as dictated by
causality~\cite{pollney}. This differs from the Pittsburgh code in which all
the equations are integrated radially outward, so that the coordinate
conditions are determined at the inner boundary and the metric is
asymptotically flat but not asymptotically Minkowskian. The Southampton scheme
simplifies the formulae for the Bondi news function and mass in terms of the
metric. It is anticipated that the inward integration of $\beta$ causes no
numerical problems because this is a gauge choice which does not propagate
physical information. However, the code has not yet been subject to convergence
and long term stability tests so that these issues cannot be properly assessed
at the present time.

The matching of the Southampton axisymmetric code
to a Cauchy interior is discussed in Section~\ref{sec:aximatch}.


\subsection{The Bondi mass}

Numerical calculations of asymptotic quantities such as the Bondi mass  must
pick off non-leading terms in an asymptotic expansion about infinity. This is
similar to the experimental task of determining the mass of an object by
measuring its far field. For example, in an asymptotically inertial
Bondi frame at $\mathcal{I}^+$ (in which the metric takes an asymptotically
Minkowski form in null spherical coordinates)), the mass aspect ${\cal
M}(u,\theta,\phi)$ is picked off from the asymptotic expansion of Bondi's
metric quantity $V$ (see Equation~(\ref{eq:V})) of the form $V = r- 2\mathcal{M}
+\mathcal{O} (1/r)$. In gauges which incorporate some of the properties of an
asymptotically inertial frame, such as the null quasi-spherical
gauge~\cite{bartnumeth} in which the angular metric is conformal to the unit
sphere metric, this can be a straightforward computational problem. However,
the job can be more difficult if the gauge does not correspond to a standard
Bondi frame at $\mathcal{I}^+$. One must then deal with an arbitrary
coordinatization of $\mathcal{I}^+$ which is determined by the details of the
interior geometry. As a result, $V$ has a more complicated asymptotic behavior,
given in the axisymmetric case by
\begin{eqnarray}
  V - r &=& \frac{r^2 (L \sin \theta)_{,\theta}}{\sin \theta} +
  r e^{2 (H - K)} \times
  \nonumber \\
  && \left[ \left( 1 - e^{- 2 (H - K)} \right) +
  \frac{2 (H_{,\theta} \sin \theta)_{,\theta}}{\sin \theta} +
  K_{,\theta \theta} + 3 K_{,\theta} \cot \theta + 4 (H_{,\theta})^2 -
  4 H_{,\theta} K_{,\theta} - 2 (K_{,\theta})^2 \right]
  \nonumber \\
  && - 2 e^{2 H} \mathcal{M} + \mathcal{O} (r^{-1}),
  \label{eq:wasym}
\end{eqnarray}%
where $L$, $H$, and $K$ are gauge dependent functions of $(u,\theta)$
which would vanish in an inertial Bondi frame~\cite{tam,isaac}. The calculation
of the Bondi mass requires regularization of this expression by
numerical techniques so that the coefficient $\mathcal{M}$ can be picked
off. The task is now similar to the experimental determination of the
mass of an object by using non-inertial instruments in a far zone which
contains $\mathcal{O} (1/r)$ radiation fields. But it has been done!

It was accomplished in Stewart's code by re-expressing the formula for
the Bondi mass in terms of the well-behaved fields of the conformal
formalism~\cite{stewbm}. In the Pittsburgh code, it was accomplished by
re-expressing the Bondi mass in terms of renormalized metric variables
which regularize all calculations at $\mathcal{I}^+$ and made them second
order accurate in grid size~\cite{mbondi}. The calculation of the Bondi news
function (which provides the waveforms of both polarization modes) is
an easier numerical task than the Bondi mass. It has also been
implemented in both of these codes, thus allowing the important check
of the Bondi mass loss formula.

An alternative approach to computing the Bondi mass is to adopt a gauge which
corresponds more closely to an inertial Bondi frame at $\mathcal{I}^+$ and
simplifies the asymptotic limit. Such a choice is the null quasi-spherical gauge
in which the angular part of the metric is proportional to the unit sphere
metric, and as a result the gauge term $K$ vanishes in
Equation~(\ref{eq:wasym}). This gauge was adopted by Bartnik and Norton at
Canberra in their development of a 3D characteristic evolution
code~\cite{bartnumeth} (see Section~\ref{sec:3d} for further discussion). It
allowed accurate computation of the Bondi mass as a limit as $r
\rightarrow\infty$ of the Hawking mass~\cite{bartint}.

Mainstream astrophysics is couched in Newtonian concepts, some of which
have no well defined extension to general relativity. In order to
provide a sound basis for relativistic astrophysics, it is crucial to
develop general relativistic concepts which have well defined and
useful Newtonian limits. Mass and radiation flux are
fundamental in this regard. The results of characteristic codes show
that the energy of a radiating system can be evaluated rigorously and
accurately according to the rules for asymptotically flat spacetimes,
while avoiding the deficiencies that plagued the ``pre-numerical'' era
of relativity: (i) the use of coordinate dependent concepts such as
gravitational energy-momentum pseudotensors; (ii) a rather loose notion
of asymptotic flatness, particularly for radiative spacetimes; (iii)
the appearance of divergent integrals; and (iv) the use of
approximation formalisms, such as weak field or slow motion
expansions, whose errors have not been rigorously estimated.

Characteristic codes have extended the role of the Bondi mass from that of a
geometrical construct in the theory of isolated systems to that
of a highly accurate computational tool. The Bondi mass loss formula
provides an important global check on the preservation of the Bianchi
identities. The mass loss rates themselves have important astrophysical
significance. The numerical results demonstrate that computational
approaches, rigorously based upon the geometrical definition of mass in
general relativity, can be used to calculate radiation losses in highly
nonlinear processes where perturbation calculations would not be
meaningful.

Numerical calculation of the Bondi mass has been used to explore both
the Newtonian and the strong field limits of general
relativity~\cite{mbondi}. For a quasi-Newtonian system of radiating
dust, the numerical calculation joins smoothly on to a post-Newtonian
expansion of the energy in powers of $1/c$, beginning with the
Newtonian mass and mechanical energy as the leading terms. This
comparison with perturbation theory has been carried out to $\mathcal{O} (1/c^7)$,
at which stage the computed Bondi mass peels away from the
post-Newtonian expansion. It remains strictly positive, in contrast to
the truncated post-Newtonian behavior which leads to negative values.

A subtle feature of the Bondi mass stems from its role as one component of the
total energy-momentum 4-vector, whose calculation requires identification of
the translation subgroup of the Bondi--Metzner--Sachs group~\cite{bms}. This
introduces boost freedom into the problem. Identifying the translation subgroup
is tantamount to knowing the conformal transformation to an inertial Bondi
frame~\cite{tam} in which the time slices of $\mathcal{I}^+$ have unit sphere
geometry. Both Stewart's code and the Pittsburgh code adapt the coordinates to
simplify the description of the interior sources. This results in a non-standard
foliation of $\mathcal{I}^+$. The determination of the conformal factor which
relates the 2-metric $h_{AB}$ of a slice of $\mathcal{I}^+$ to the unit sphere
metric is an elliptic problem equivalent to solving the second order partial
differential equation for the conformal transformation of Gaussian curvature.
In the axisymmetric case, the PDE reduces to an ODE with respect to the angle
$\theta$, which is straightforward to solve~\cite{mbondi}. The integration
constants determine the boost freedom along the axis of symmetry.

The non-axisymmetric case is more complicated. Stewart~\cite{stewbm}
proposes an approach based upon the dyad decomposition
\begin{equation}
  h_{AB} \, dx^A \, dx^B = m_A \, dx^A \, {\bar m}_B \, dx^B.
\end{equation}
The desired conformal transformation is obtained by first relating
$h_{AB}$ conformally to the flat metric of the complex plane. Denoting
the complex coordinate of the plane by $\zeta$, this relationship can
be expressed as $d\zeta = e^f m_A \, dx^A$. The conformal factor $f$ can
then be determined from the integrability condition
\begin{equation}
  m_{[A} \partial_{B]} f = \partial_{\,[A}m_{B]}.
\end{equation}
This is equivalent to the classic Beltrami equation for finding
isothermal coordinates. It would appear to be a more effective scheme
than tackling the second order PDE directly, but numerical
implementation has not yet been carried out.

\newpage


\section{3D Characteristic Evolution}
\label{sec:3d}

The initial work on 3D characteristic evolution led to two
independent codes, one developed at Canberra and the other at Pittsburgh
(the PITT code),  both with the capability to study gravitational waves in
single black hole spacetimes at a level not yet mastered 
at the time by Cauchy codes. 
The Pittsburgh group established robust stability and second
order accuracy of a fully nonlinear code which
was able to calculate the waveform at null
infinity~\cite{cce,high} and to track a dynamical black hole and excise its
internal singularity from the computational grid~\cite{excise,wobb}. The
Canberra group implemented an independent nonlinear code which
accurately evolved the exterior region of a Schwarzschild black hole. Both
codes pose data on an initial null hypersurface and on a worldtube boundary,
and evolve the exterior spacetime out to a compactified version of null
infinity, where the waveform is computed. However, there are essential
differences in the underlying geometrical formalisms and numerical techniques
used in the two codes and in their success in evolving generic black hole
spacetimes. Recently two new codes have evolved from the PITT code
by introducing a new choice of spherical coordinates~\cite{leo,reisswig}.

\subsection{Coordinatization of the sphere}
\label{sec:sphercoor}

Any characteristic code extending to  $\mathcal{I}^+$ requires the ability
to handle tensor fields and their derivatives on
the sphere. Spherical coordinates and spherical harmonics are natural analytic
tools for the description of radiation, but their implementation in
computational work requires dealing with the impossibility of smoothly covering
the sphere with a single coordinate grid. Polar coordinate singularities in
axisymmetric systems can be regularized by standard tricks. In the absence of
symmetry, these techniques do not generalize and would be especially
prohibitive to develop for tensor fields. Because of the 
natural use of null-spherical coordinates in characteristic evolution, this
differs from Cauchy evolution where spherical harmonics can be properly
described in a Cartesian coordinate system. 

The development of grids smoothly covering the sphere has had a long history in
computational meteorology that has led to two distinct approaches: (i) the
stereographic approach in which the sphere is covered by two overlapping patches
obtained by stereographic projection about the North and South
poles~\cite{browning}; and (ii) the cubed-sphere approach in which the sphere is
covered by the 6 patches obtained by a projection of the faces of a
circumscribed cube~\cite{ronchi}. A discussion of the advantages of each of
these methods and a comparison of their performance in a standard fluid testbed
are given in~\cite{browning}. In numerical relativity, the stereographic method
has been reinvented in the context of the characteristic evolution
problem~\cite{eth}; and the cubed-sphere method has been reinvented in building
an apparent horizon finder~\cite{thornburgf}. The cubed sphere module, including
the interpatch transformations, has been integrated into the
Cactus toolkit~\cite{cactus} and applied to black hole excision
and numerous other problems in numerical
relativity~\cite{thornburge,cubed1,cubed2,cubed3,cubed4,cubed5,cubed6,cubed7,cubed8,cubed9}.
Perhaps the most
ingenious treatment of the sphere, based upon a toroidal map, was devised by the
Canberra group in building their characteristic code~\cite{bartnumeth}. These
methods are described below.

\subsubsection{Stereographic grids}

Motivated by problems in meteorology, G.L.~Browning,  J.J.~Hack and
P.N.~Swartztrauber~\cite{browning} developed the first finite
difference scheme based upon a composite mesh with two overlapping
stereographic coordinate patches, each having a circular boundary
centered about the North or South poles. Values for quantities
required at ghost points beyond the boundary of one of the patches
were interpolated from values in the other patch. Because a circular
boundary does not fit regularly on a stereographic grid, dissipation
was found necessary to remove the short wavelength error resulting
from the inter-patch interpolations. They used the shallow water
equations as a testbed to  compare their approach to existing spectral
approaches in terms of computer time, execution rate and
accuracy. Such comparisons of different numerical methods can be
difficult. Both the finite difference and spectral approaches gave
good results and were  competitive in terms of overall operation count
and memory requirements. For the particular initial data sets tested,
the spectral approach had an advantage but not enough to give clear
indication of the suitability of one method over another. The spectral
method with M modes requires $O(M^3)$ operations per time step
compared with $O(N^2)$ for a finite difference method on a $N\times N$
grid. However, assuming that the solution is analytic, the accuracy of
the spectral method is $O(e^{-M})$ compared to, say, $O(N^{-6})$ for a
sixth order finite difference method.
Hence, for comparable accuracy, $M=O(\ln N)$ which implies that the
operation count for the spectral and finite difference methods would be
$O[(\ln N)^3]$ and $O(N^2)$, respectively. Thus for sufficiently
high accuracy, i.e.\ large $N$, the spectral method  requires fewer
operations. The issue of spectral \emph{vs} finite difference methods
thus depends on the nature of the smoothness of the physical problem
being addressed and the accuracy desired. For smooth $C^\infty$ solutions
the spectral convergence rate is still faster than any power law.

The Pitt null code was first developed using two stereographic patches with
square boundaries, each overlapping the equator. This has recently been modified
based  upon the approach advocated in~\cite{browning}, which retains the
original stereographic coordinates but shrinks the overlap region by masking a
circular boundary near the equator. The original square boundaries aligned with
the grid and did not require numerical dissipation. However, the corners of  the
square boundary, besides being a significant waste of economy,  were a prime
source of inaccuracy. The resolution at the corners is only 1/9th that at the
poles due to the stretching of the stereographic map. Near the equator,  the
resolution is approximately 1/2 that at the poles. The use of a circular
boundary requires an angular version of numerical dissipation to control the
resulting high frequency error (see Section~\ref{sec:dissip}).

A crucial ingredient of the PITT code is the $\eth$-module~\cite{competh} which
incorporates a computational version of the Newman--Penrose
\emph{eth}-formalism~\cite{eth}. The underlying method can be applied
to any smooth coordinatization $x^A$ of the sphere based upon several
patches. The unit sphere metric $q_{AB}$, defined by these
coordinates, is decomposed in each patch in terms of a complex basis
vector $q_A$, 
\begin{equation}
     q_{AB}=q_{(A}\bar q_{B)}
\end{equation}
Vector and tensor fields on the sphere, and their covariant derivatives, are
then represented by their basis components. For example, the vector field $U^A$
is represented by the complex spin-weight 1 field $U=U^A q_A$. The covariant
derivative $D^A$ associated with $q_{AB}$ is then expressed in terms of the
$\eth$ operator according to
\begin{equation}
    q_A q_B D^A U^B =\eth U\, , \quad \bar q_A q_B D^A U^B =\bar \eth U .
\end{equation}
The eth-calculus simplifies the underlying equations, avoids spurious coordinate
singularities and allows accurate differentiation of tensor fields on the sphere
in a computationally efficient and clean way. Its main weakness is the numerical
noise introduced by interpolations between the patches. 

\subsubsection{Cubed sphere grids}

C.~Ronchi, R.~Iacono and P.S.~Paolucci~\cite{ronchi}, developed the
``cubed-sphere'' approach as a new gridding method for solving global
meteorological problems. The method decomposes the sphere into the 6 identical
regions obtained by projection of a cube circumscribed on its surface. This
gives a variation of the composite mesh method in which the 6 domains butt up
against each other along shared grid boundaries. As a result, depending
upon the implementation, either no inter-grid interpolations or only 1-dimensional
interpolations are necessary (as opposed to the 2-dimensional
interpolations necessary for a stereographic grid), which results in enhanced accuracy.
See~\cite{sarbtig} for a review of abutting grid techniques in numerical relativity.
The symmetry of the scheme, in which the six patches have the same geometric
structure and grid, also allows efficient use of parallel computer
architectures.  Their tests of the cubed sphere method based upon the simulation
of shallow water waves in spherical geometry show that the numerical solutions
are as accurate as those with spectral methods, with substantial saving in
execution time. Recently, the cubed-sphere method has also been developed for
application to characteristic evolution in numerical
relativity~\cite{reisswig,leo}. The \emph{eth}-calculus is used to treat tensor
fields on the sphere in the same way as in the stereographic method except the
interpatch  transformations now involve 6, rather than 2, sets of basis vectors.

\subsubsection{Toroidal grids}
\label{sec:toroidal}

The Canberra group treats fields on the sphere  by taking advantage of the
existence of a smooth map from the torus to the sphere~\cite{bartnumeth}. The
pullback of this map allows functions on the sphere to be expressed in terms of
toroidal coordinates. The intrinsic topology of these toroidal coordinates allow
them to take advantage of of fast-Fourier transforms to implement a highly
efficient pseudo-spectral treatment. This ingenious method has
apparently not yet been adopted in other fields.


\subsection{Geometrical formalism}

The PITT code uses a standard Bondi--Sachs null coordinate system,
\begin{equation}
  ds^2 =
  - \left( e^{2 \beta} \frac{V}{r} - r^2 h_{AB} \, U^A \, U^B \right) du^2 -
  2 e^{2 \beta} du \, dr - 2 r^2 h_{AB} \, U^B \, du \, dx^A +
  r^2 h_{AB} \, dx^A\, dx^B,
  \label{eq:umet}
\end{equation}
where
\begin{equation}
      \det(h_{AB})=\det(q_{AB})
   \label{eq:qdet}
\end{equation}
for some standard choice $q_{AB}$ of the unit
sphere metric. This generalizes Equation~(\ref{eq:bmetric}) to the
three-dimensional case. The characteristic version of Einstein's equations are  
The hypersurface equations derive from the ${G_\mu}^\nu \nabla_\nu u$ components
of the Einstein tensor. They take the explicit form~\cite{newtgnc}
\begin{eqnarray}
  \beta_{,r} &=& \frac{1}{16} r \, h^{AC} \, h^{BD} \, h_{AB,r} \, h_{CD,r},
  \label{eq:3beta} \\
  \left( r^4 e^{-2 \beta} h_{AB} U^B_{,r} \right)_{,r} &=&
  2 r^4 \left( r^{-2} \beta_{,A} \right)_{,r} -
  r^2 h^{BC} D_C (h_{AB,r})
  \label{eq:3u} \\
  2 e^{-2 \beta} V_{,r} &=& \mathcal{R} - 2 D^{A} D_{A} \beta -
  2 (D^A \beta) D_A \beta + r^{-2} e^{-2 \beta} D_A
  \left( \! \left( r^4 U^A \right)_{,r} \! \right)
  \nonumber \\
  && - \frac{1}{2} r^4 e^{- 4 \beta} h_{AB} \, U^A_{,r} \, U^B_{,r},
  \label{eq:3v}
\end{eqnarray}%
where $D_A$ is the covariant derivative and $\mathcal{R}$ the curvature scalar of
the conformal 2-metric $h_{AB}$ of the $r=\mathrm{const.}$ surfaces, and capital Latin
indices are raised and lowered with $h_{AB}$. Given the null data $h_{AB}$ on
an outgoing null hypersurface, this hierarchy of equations can be integrated
radially in order to determine $\beta$, $U^A$ and $V$ on the hypersurface in
terms of integration constants on an inner boundary. The evolution equations
for the $u$-derivative of the null data derive from the trace-free part of the
angular components of the Einstein tensor, i.e.\ the components $m^A m^B G_{AB}$
where $h^{AB}=2m^{(A}\bar m^{B)}$.
They take the explicit form
\begin{eqnarray}
  m^A m^B \biggl( \!\!\!\! && ( r h_{AB,u})_{,r} -
  \frac{1}{2r} (r V h_{AB,r})_{,r} -
  \frac{2}{r} e^\beta D_A D_B e^\beta + r h_{AC} D_B (U^C_{,r})
  \nonumber \\
  && - \frac{r^3}{2} e^{-2 \beta} h_{AC} \, h_{BD} \, U^C_{,r} \, U^D_{,r} +
  2 D_A \, U_B + \frac{r}{2} h_{AB,r} \, D_C \, U^C
  \nonumber \\
  && + r U^C D_C (h_{AB,r}) + r h_{AD,r} \, h^{CD} (D_B U_C - D_C U_B)
  \biggr) = 0.
  \label{eq:hev}
\end{eqnarray}
A compactified treatment of null infinity is achieved by introducing the radial
coordinate $x=r/(R+r)$, where $R$ is a scale parameter adjusted to the
size of the inner boundary. Thus $x=1$ at  $\mathcal{I}^+$.

The Canberra code employs a null quasi-spherical (NQS) gauge (not to be
confused with the quasi-spherical approximation in which quadratically
aspherical terms are ignored~\cite{cce}). The NQS gauge takes advantage of the
possibility of mapping the angular part of the Bondi metric conformally onto a
unit sphere metric, so that $h_{AB}\rightarrow q_{AB}$. The required
transformation $x^A \rightarrow y^A(u,r,x^A)$ is in general dependent upon $u$
and $r$ so that the NQS angular coordinates $y^A$ are not constant
along the outgoing null rays, unlike the Bondi--Sachs angular coordinates.
Instead the coordinates $y^A$ display the analogue of a shift on the null
hypersurfaces $ u = \mathrm{const} $. In addition, the NQS spheres $ (u,r) =
\mathrm{const.}$ are not the same as the Bondi spheres. The radiation content of
the metric is contained in a shear vector describing this shift. This results
in the description of the radiation in terms of a spin-weight 1 field, rather
than the spin-weight 2 field associated with $h_{AB}$ in the Bondi--Sachs
formalism. In both the Bondi--Sachs and NQS gauges, the independent
gravitational data on a null hypersurface is the conformal part of its
degenerate 3-metric. The Bondi--Sachs null data consist of $h_{AB}$, which
determines the intrinsic conformal metric of the null hypersurface. In the
NQS case, $h_{AB}=q_{AB}$ and the shear vector comprises the only
non-trivial part of the conformal 3-metric. Both the Bondi--Sachs and NQS gauges
can be arranged to coincide in the special case of shear-free Robinson--Trautman
metrics~\cite{derry,bartgauge}.

The formulation of Einstein's equations in the NQS gauge is
presented in~\cite{bartee}, and the associated gauge freedom arising from
$(u,r)$ dependent rotation and boosts of the unit sphere is discussed
in~\cite{bartgauge}. As in the PITT code, the main equations involve
integrating a hierarchy of hypersurface equations along the radial null
geodesics extending from the inner boundary to null infinity. In the
NQS gauge the source terms for these radial ODEs are rather
simple when the unknowns are chosen to be the connection coefficients.
However, as a price to pay for this simplicity, after the radial integrations
are performed on each null hypersurface a first order elliptic equation must
be solved on each $ r = \mathrm{const.} $ cross-section to reconstruct
the underlying metric.

\subsubsection{Worldtube conservation laws}
\label{sec:worldtube-conservation}

The components of Einstein's equations independent of the hypersurface and evolution
equations,
\begin{eqnarray}
        h^{AB}G_{AB}&=&0 \label{eq:triv} \\
        G_A^r&=&0 \label{eq:consu}\\
        G_u^r&=&0 \label{eq:consA},
\end{eqnarray}
were called supplementary conditions by Bondi et al.~\cite{bondi}
and Sachs~\cite{sachs}. They showed that the Bianchi identity 
$$
    \nabla_\mu G^\mu_\nu=\frac {1}{\sqrt{-g}} (\sqrt{-g}  G^\mu_\nu)_{,\mu}
          +\frac{1}{2} g^{\rho\sigma}_{,\nu} G_{\rho\sigma}=0
$$   
implies that these equations need only be satisfied on a worldtube $r=R(u,x^A)$.
When the hypersurface and evolution equations are satisfied, the Bianchi
identity for $\nu=r$ reduces to  $h^{AB}G_{AB}=0$ so that (\ref{eq:triv})
becomes trivially satisfied. The Bianchi identity for $\nu=A$ then reduces to
$$
          (r^2 G_A^r)_{,r} =0,
$$
so that $G_A^r=0$ if it is set to zero at $r=R(u,x^A)$. When that is the case, 
the Bianchi identity for $\nu=u$ then reduces to
$$
          (r^2 G_u^r)_{,r} =0,
$$
so that $G_u^r=0$ also vanishes if it vanishes for $r=R(u,x^A)$.

As a result, the supplementary conditions  can be replaced by the condition 
that the Einstein tensor satisfy
\begin{equation}
    \xi^{\mu}G_{\mu}^{\nu}N_{\nu}=0
    \label{eq:Gconr}
\end{equation}
on the worldtube, where $\xi^{\mu}$ is any vector field tangent to the worldtube 
and $N_\mu$ is the unit normal to the worldtube.
Since $\xi^\mu N_\mu=0$, we can further replace (\ref{eq:Gconr}) by the worldtube
condition on the Ricci tensor
\begin{equation}
    \xi^{\mu}R_{\mu}^{\nu}N_{\nu}=0.
    \label{eq:Rcon}
\end{equation}
The Ricci identity
\begin{equation}
    \xi^{\mu}R_{\mu}^{\nu}= \nabla_\mu \nabla^{(\nu}\xi^{\mu)}
       + \nabla_\mu \nabla^{[\nu}\xi^{\mu]} -\nabla^\nu \nabla_\mu \xi^\mu
       \label{eq:xicon}
\end{equation}
then gives rise to the strict Komar conservation law~\cite{komar}
$$
       \nabla_\mu \nabla^{[\nu}\xi^{\mu]} =0
$$
when $\xi^\mu$ is a Killing vector corresponding to an exact symmetry. More
generally, (\ref{eq:xicon}) gives rise to the flux conservation law
$$
           P_\xi(u_2)-P_\xi(u_1) = \int_{u_1}^{u_2} dS_\nu \{
     \nabla^\nu \nabla_\mu \xi^\mu - \nabla_\mu \nabla^{(\nu}\xi^{\mu)} \}
$$
where
$$
           P_\xi =\oint dS_{\mu\nu} \nabla^{[\nu}\xi^{\mu]} 
$$
and $dS_{\mu\nu}$ and $dS_\nu$ are, respectively, the appropriate surface and
3-volume elements on the worldtube. For the limiting case when $R\rightarrow
\infty$, these flux conservation laws govern the energy-momentum, angular
momentum and supermomentum corresponding to the generators of the
Bondi--Metzner--Sachs asymptotic symmetry group~\cite{tam}. For an asymptotic time
translation, they give rise to the Bondi mass loss relation.

These conservation laws (\ref{eq:Rcon}) can also be expressed in terms of 
the intrinsic metric of the worldtube.
$$
   H_{\mu\nu}=g_{\mu\nu}-N_\mu N_\nu
$$
and its extrinsic curvature 
$$
        K_{\mu\nu}=H_\mu^\rho \nabla_\rho N_\nu. \nonumber
$$
This leads to the worldtube analogue of the momentum
constraint for the Cauchy problem,
\begin{equation}
 0 ={\cal D}_\mu (
  K^\mu_\nu - \delta^\mu_\nu K^\rho_\rho)\quad(= H_\nu^\mu 
    G_{\mu\rho} N^\rho),
    \label{eq:bmom}
\end{equation}
where ${\cal D}_\mu$ is the covariant derivative associated with $H_{\mu\nu}$.
These are equivalent to the conservation conditions (\ref{eq:Rcon}) and allow
the conserved quantities to be expressed in terms of the extrinsic curvature of
the boundary. For any vector field $\xi^\mu$ tangent to the worldtube,
(\ref{eq:bmom}) implies
$$
   {\cal D}_\mu (\xi^\nu K^\mu_\nu-\xi^\mu K^\rho_\rho)=
       {\cal D}^{(\mu}  \xi^{\nu)} (K_{\mu\nu}-H_{\mu\nu} K^\rho_\rho).
$$
In particular, this shows that $\xi^\mu$ need only be a Killing vector for the
3-metric $H^{\mu\nu}$ to obtain a strict conservation law on the boundary.

The worldtube conservation laws  can also be interpreted 
as a symmetric hyperbolic system governing the evolution of certain
components of the extrinsic curvature~\cite{josh}. This leads to the

\bigskip
\noindent 
\textbf{Worldtube Theorem}:

\medskip

\noindent \emph{Given $H_{ab}$, $m^a m^b K_{ab}$ and $K$, the worldtube
constraints constitute a well-posed initial-value problem which determines the
remaining components of the extrinsic curvature $K_{ab}$}.

\bigskip

These extrinsic curvature components
are related to the integration constants for the Bondi--Sachs system, which
leads to possible applications of the worldtube theorem. 
One application is to waveform extraction. In that case, the data
$(H_{ab}, m^a m^b K_{ab},K)$ necessary to apply the worldtube
theorem are supplied by the numerical results of a 3\,+\,1 Cauchy evolution.
The remaining components of the extrinsic curvature
can then be determined by means of a well-posed initial value problem on the
boundary. The integration constants  $(\beta, V,U^A, U^A_{,r}, h_{AB})$, for the
Bondi-Sachs equations at $r=R(u,x^A)$ are then determined. This
approach can be used to enforce the constraints in the numerical computation of
waveforms at  $\mathcal{I}^+$by means of Cauchy-characteristic extraction (see
Section~\ref{sec:cce}).

Another possible application,is to the characteristic initial-boundary value problem, for
which boundary data consistent with the constraints must be prescribed
\emph{apriori}, i.e. independent of the evolution. The object is to obtain a
well-posed version of the characteristic initial-boundary value problem.
However, the complicated coupling between the
Bondi--Sachs evolution system and the boundary constraint system prevents any
definitive results.

\subsubsection{Angular dissipation}
\label{sec:dissip}

For a \{3\,+\,1\} evolution algorithm based upon a system of wave equations, or
any other symmetric hyperbolic system, numerical dissipation can be added in the
standard Kreiss--Oliger form~\cite{kreisoligd}. Dissipation cannot be added to
the \{2\,+\,1\,+\,1\} format of characteristic evolution in this standard way
for \{3\,+\,1\} Cauchy evolution. In the original version of the PITT code,
which used square stereographic patches with boundaries aligned with the grid,
numerical dissipation was only introduced in the radial
direction~\cite{luisdis}. This was sufficient to establish numerical stability.
In the new version of the code with circular stereographic patches,  whose
boundaries fit into the stereographic grid in an irregular way, angular
dissipation is necessary to suppress the resulting high frequency error.

Angular dissipation can be introduced in the following way~\cite{strateg}. In terms of
the spin-weight 2 variable 
\begin{equation}
    J=q^A q^B h_{AB},
    \label{eq:J}
\end{equation}
the evolution equation (\ref{eq:hev}) takes the form
\begin{equation}
   \partial_u \partial_r (rJ)= S,
\label{eq:phiev}
\end{equation}
where $S$ represents the right hand side terms.
We add angular dissipation to the $u$-evolution through the
modification
\begin{equation}
    \partial_u \partial_r (rJ)
    +\epsilon_{u} h^3 \eth^2  
      \bar\eth^2 \partial_r (rJ) = S,
\end{equation}
where $h$ is the discretization size and $\epsilon_{u}\ge 0$ is an adjustable
parameter independent of $h$.
This leads to 
\begin{equation} 
     \partial_u \bigg( |\partial_r (rJ)|^2\bigg) 
     + 2\epsilon_{u}  h^3 \Re\{ \partial_r (r\bar J) 
       \eth^2 \bar\eth^2 \bigg (\partial_r (rJ) \bigg ) \} 
      = 2\Re \{\bigg ( \partial_r (r\bar J)\bigg ) S \}.
\end{equation}
Integration over the unit sphere with solid angle element $d\Omega$
then gives
\begin{equation} 
   \partial_u \oint | \partial_r (rJ)|^2  d\Omega 
     +2\epsilon_{u}  h^3 \oint 
           |\bar\eth^2 \bigg (\partial_r (rJ) \bigg )|^2 d\Omega 
         =2\Re \oint \bigg (\partial_r (r \bar J)) S d\Omega.
\end{equation}
Thus the $\epsilon_{u}$-term has the effect of damping high frequency noise as
measured by the $L_2$ norm of $\partial_r (rJ)$ over the sphere. 

Similarly, dissipation can be introduced in the radial integration of
(\ref{eq:phiev}) through the substitution
\begin{equation}
    \partial_u \partial_r (rJ)  \rightarrow   \partial_u \partial_r (rJ) 
  +\epsilon_{r} h^3 \eth^2  \bar\eth^2 \partial_u (rJ),  
\end{equation}
with $\epsilon_{r}\ge 0$. Angular dissipation can also be introduced in the
hypersurface equations, e.g.\ in (\ref{eq:3v}) through the
substitution
\begin{equation} 
   \partial_r V  \rightarrow  \partial_r V 
   +\epsilon_V h^3 \bar \eth  \eth \eth \bar \eth  V .
\end{equation}

\subsubsection{First versus second differential order}
\label{sec:fvssec}

The PITT code was originally formulated in the second differential form of
Equations~(\ref{eq:3beta}, \ref{eq:3u}, \ref{eq:3v}, \ref{eq:hev}), which in the
spin-weighted version leads to an economical number of 2 real and 2 complex
variables. Subsequently, the variable
\begin{equation}
  Q_A = r^2 e^{-2 \beta} h_{AB} \, U^B_{,r}
\end{equation}
was introduced to reduce Equation~(\ref{eq:3u}) to two first order radial
equations, which simplified the startup procedure at the boundary. Although the
resulting code was verified to be stable and second order accurate, its
application to problems involving strong fields and gradients led to numerical
errors which made small scale effects of astrophysical importance difficult to
measure.

In particular, in initial attempts to simulate a white hole fission,
G{\'{o}}mez~\cite{gomezfo} encountered an oscillatory error pattern in the
angular directions near the time of fission. The origin of the problem was
tracked to numerical error of an oscillatory nature introduced by $\eth^2$ terms
in the hypersurface and evolution equations. G{\'{o}}mez's solution was to
remove the offending second angular derivatives by introducing additional
variables and reducing the system to first differential order in the angular
directions. This suppressed the oscillatory mode and subsequently improved
performance in the simulation of the white hole fission problem~\cite{fiss} (see
Section~\ref{sec:fission}).

This success opens the issue of whether a completely first differential order
code might perform even better, as has been proposed by G{\'{o}}mez and
Frittelli~\cite{gomfrit}. By the use of $\partial_u h_{AB}$ as a fundamental
variable, they cast the Bondi system into Duff's first order quasilinear 
canonical form~\cite{Duff}. At the analytic level this provides standard
uniqueness and existence theorems (extending previous work for the linearized
case~\cite{lehnfrit}) and is a starting point for establishing the estimates
required for well-posedness. 

At the numerical level, G{\'{o}}mez and Frittelli point out that this first
order formulation provides a bridge between the characteristic and Cauchy
approaches which allows application of standard methods for constructing
numerical algorithms, e.g.\ to take advantage of shock-capturing schemes.
Although true shocks do not exist for vacuum gravitational fields, when coupled
to hydro the resulting shocks couple back to form steep gradients which might
not be captured by standard finite difference approximations. In particular, the
second derivatives needed to compute gravitational radiation from stellar
oscillations have been noted to be a troublesome source of inaccuracy in the
characteristic treatment of hydrodynamics~\cite{sieb2002a}. Application of
standard versions of AMR is also facilitated by the first order form.

The benefits of this completely first order approach are not simple to decide
without code comparison. The part of the code in which the $\eth^2$ operator
introduced the  oscillatory error mode in~\cite{gomezfo} was not identified,
i.e.\ whether it originated in the inner boundary treatment or in the
interpolations between stereographic patches where second derivatives might be
troublesome. There are other possible ways to remove the oscillatory angular
modes, such as adding angular dissipation (see Section~\ref{sec:dissip}). The
finite difference algorithm in the original PITT code only introduced numerical
dissipation in the radial direction~\cite{luisdis}. The economy of variables and
other advantages of a second order scheme~\cite{krort} should not be abandoned
without further tests and investigation.


\subsubsection{Numerical methods}
\label{sec:nummeth}

The PITT code is an explicit  finite difference evolution algorithm based upon
retarded time steps on a uniform three-dimensional null coordinate grid based
upon the stereographic coordinates and a compactified radial coordinate. The
straightforward numerical implementation of the finite difference equations has
facilitated code development. The Canberra code uses an assortment of novel and
elegant numerical methods. Most of these involve smoothing or filtering and have
obvious advantage for removing short wavelength noise but would be unsuitable
for modeling shocks.

There have been two recent projects, to improve the performance of the PITT code
by  using the cubed-sphere method to coordinatize the sphere. They both include
an adaptation of the \emph{eth}-calculus to handle the transformation of
spin-weighted variables between the six patches. 

In one of these projects, G{\'{o}}mez, Barreto and Frittelli develop the
cubed-sphere approach into an efficient, highly  parallelized 3D code, the LEO
code, for the characteristic evolution of the coupled Einstein--Klein--Gordon
equations  in the Bondi--Sachs formalism~\cite{leo}. This code was demonstrated
to be convergent and its high accuracy in the linearized regime with a
Schwarzschild background was demonstrated by the simulation of the quasinormal
ringdown of the scalar field and its energy-momentum conservation.

Because the characteristic evolution scheme constitutes a radial integration
carried out for each angle on the sphere of null directions, the natural way to
parallelize the code is distribute the angular grid among processors. Thus given
$M\times M$ processors  one can distribute the $N\times N$ points in each
spherical patch (cubed-sphere or stereographic), assigning to each processor
equal square grids of extent $N/M$ in each direction. To be effective this
requires that the communication time between processors scales effectively. This
depends upon the ghost point location necessary to supply nearest neighbor data
and is facilitated in the cubed-sphere approach because the ghost points are
aligned on 1-dimensional grid lines, whose pattern is invariant under grid size.
In the stereographic approach, the ghost points are arranged in an irregular
pattern which changes in an essentially random way under rescaling and requires
a more complicated parallelization algorithm.  

Their goal is to develop the LEO code for application to black hole- neutron
star binaries in a close orbit regime where the absence of caustics make a pure
characteristic evolution possible. Their first anticipated application is the
simulation of a boson star orbiting a black hole, whose dynamics is described by
the Einstein--Klein--Gordon equations. They point out that characteristic
evolution of such systems of astrophysical interest have been limited in the
past by resolution due to the lack off the necessary computational power, 
parallel infrastructure and mesh refinement.  Most characteristic code
development has been geared toward single processor machines whereas the current
computational platforms are designed toward performing high resolution
simulations in reasonable times by parallel processing.

At the same time the LEO code was being developed, Reisswig et
al.~\cite{reisswig} also constructed a characteristic code for the Bondi--Sachs
problem based upon the cubed-sphere infrastructure of
Thornburg~\cite{thornburgf,thornburge}. They retain the original second order
differential form of the angular operators. 

The Canberra code handles fields on the sphere by means of a 3-fold
representation: (i)~as discretized functions on a spherical grid uniformly
spaced in standard $(\theta,\phi)$ coordinates, (ii)~as fast-Fourier transforms
with respect to $(\theta,\phi)$ (based upon the smooth map of the torus onto the
sphere), and (iii)~as a spectral decomposition of scalar, vector, and tensor
fields in terms of spin-weighted spherical harmonics. The grid values are used
in carrying out nonlinear algebraic operations; the Fourier representation is
used to calculate $(\theta,\phi)$-derivatives; and the spherical harmonic
representation is used to solve global problems, such as the solution of the
first order elliptic equation for the reconstruction of the metric, whose unique
solution requires pinning down the $\ell=1$ gauge freedom. The sizes of the grid
and of the Fourier and spherical harmonic representations are coordinated. In
practice, the spherical harmonic expansion is  carried out to 15th order in
$\ell$, but the resulting coefficients must then be projected into the $\ell \le
10$ subspace in order to avoid inconsistencies between the spherical harmonic,
grid, and Fourier representations.

The Canberra code solves the null hypersurface equations by combining an 8th
order Runge--Kutta integration with a convolution spline to interpolate field
values. The radial grid points are dynamically positioned to approximate
ingoing null geodesics, a technique originally due to Goldwirth and
Piran~\cite{goldw} to avoid the problems with a uniform $r$-grid near a horizon
which arise from the degeneracy of an areal coordinate on a stationary horizon.
The time evolution uses the method of lines with a fourth order Runge--Kutta
integrator, which introduces further high frequency filtering.


\subsubsection{Stability}
\label{sec:stability}

\begin{description}

\item[PITT code]~\\
  Analytic stability analysis  of the finite difference equations has been
  crucial in the development of a stable evolution algorithm, subject to the
  standard Courant--Friedrichs--Lewy (CFL) condition for an explicit  code.
  Linear stability analysis on Minkowski and Schwarzschild backgrounds showed
  that certain field variables must be represented on the
  half-grid~\cite{papa,cce}. Nonlinear stability analysis was essential in
  revealing and curing a mode coupling instability that was not present in the
  original axisymmetric version of the code~\cite{high,luisdis}. This has led to
  a code whose stability persists even in the regime that the $u$-direction,
  along which the grid flows, becomes spacelike, such as outside the velocity of
  light cone in a rotating coordinate system. Severe tests were used to verify
  stability. In the linear regime, \emph{robust stability} was established by
  imposing random initial data on the initial characteristic hypersurface and
  random constraint violating boundary data on an inner worldtube. (Robust
  stability was later adopted as one of the standardized tests for Cauchy
  codes~\cite{apples}.) The code ran stably for 10,000 grid crossing times under
  these conditions~\cite{cce}. The PITT code was the first 3D general
  relativistic code to pass this robust stability test. The use of random data
  is only possible in sufficiently weak cases where terms quadratic in the field
  gradients are not dominant. Stability in the highly nonlinear regime was
  tested in two ways. Runs for a time of $ 60,000 \, M$ were carried out for a
  moving, distorted Schwarzschild black hole (of mass $M$), with the marginally
  trapped surface at the inner boundary tracked and its interior excised from
  the computational grid~\cite{wobb,stablett}. At the time, this was by far the
  longest simulation of a dynamic black hole. Furthermore, the scattering of a
  gravitational wave off a Schwarzschild black hole was successfully carried out
  in the extreme nonlinear regime where the backscattered Bondi news was as
  large as $N=400$ (in dimensionless geometric units)~\cite{high}, showing that
  the code can cope with the enormous power output $N^2 c^5/G \approx 10^{60}
  \mathrm{\ W}$ in conventional units. This exceeds the power that would be
  produced if, in 1 second, the entire galaxy were converted to gravitational
  radiation.

\end{description}

\begin{description}

\item[Cubed-sphere codes]~\\
The characteristic codes using the cubed-sphere grid~\cite{leo, reisswig}  are
based upon the same variables and equations as the PITT code, with the same
radial integration scheme. Thus it should be expected that stability be
maintained since the main difference is that the interpatch interpolations are
simpler, i.e.\ only 1-dimensional.  This appears to be the case in the reported
tests, although robust stability was not directly confirmed. In particular, the
LEO code showed no sign of instability in long time, high resolution 
simulations of  the quasinormal ringdown of a scalar field scattering off a
Schwarzschild black hole.  Angular dissipation was not necessary.

\end{description}

\begin{description}

\item[Canberra code]~\\
  Analytic stability analysis of the underlying finite difference
  equations is impractical because of the extensive mix of spectral
  techniques, higher order methods, and splines. Although there is no
  clear-cut CFL limit on the code, stability tests show that there is
  a limit on the time step. The damping of high frequency modes due to
  the implicit filtering would be expected to suppress numerical
  instability, but the stability of the Canberra code is nevertheless
  subject to two qualifications~\cite{bartacc,bartnumsol,bartnumeth}:
  (i) At late times (less than $100 \, M$), the evolution terminates as
  it approaches an event horizon, apparently because of a breakdown of
  the NQS gauge condition, although an analysis of how and why this
  should occur has not yet been given. (ii) Numerical instabilities
  arise from dynamic inner boundary conditions and restrict the inner
  boundary to a fixed  Schwarzschild horizon. Tests in the extreme
  nonlinear regime were not reported.

\end{description}


\subsubsection{Accuracy}

\begin{description}

\item[PITT code]~\\
  The designed second order accuracy has been verified in an extensive number
  of testbeds~\cite{cce,high,wobb,Zlochower,zlochmode}, including new exact
  solutions specifically constructed in null coordinates for the purpose of
  convergence tests:
  \begin{itemize}
  \item Linearized waves on a Minkowski background in null cone
    coordinates.
  \item Boost and rotation symmetric solutions~\cite{boostro,boostrot}.
  \item Schwarzschild in rotating coordinates.
  \item Polarization symmetry of nonlinear twist-free axisymmetric
    waveforms.
  \item Robinson--Trautman waveforms from perturbed Schwarzschild black
    holes.
  \item Nonlinear Robinson--Trautman waveforms utilizing an
    independently computed solution of the Robinson--Trautman equation.
  \item Perturbations of a Schwarzschild black hole utilizing an
    independently computed solution of the Teukolsky equation.

\end{itemize}

In addition to these testbeds, a set of linearized solutions has recently been
obtained in the Bondi--Sachs gauge for either Schwarzschild or Minkowski
backgrounds~\cite{bishlin}. The solutions are generated by the introduction of a
thin shell of matter whose density varies with time and angle. This gives rise
to an exterior field containing gravitational waves. For a Minkowski background,
the solution is given in exact analytic form and, for a Schwarzschild
background, in terms of a power series. The solutions are parametrized by
frequency and spherical harmonic decomposition. They supply a new and very
useful testbed for the calibration and further development of characteristic
evolution codes for Einstein's equations, analogous to the role of the Teukolsky
waves in Cauchy evolution. The PITT code showed clean second order convergence
in both the $L_2$ and $L_\infty$ error norms in tests based upon waves in a
Minkowski background. However, in applications involving very high resolution
or  nonlinearity, there was excessive short wavelength noise which degraded
convergence. Recent improvements in the code~\cite{ccetool} have now established
clean second order convergence in the nonlinear regime.

It would be of great value to increase the accuracy of the code to higher order.
However, the marching algorithm which combines the radial integration
of the hypersurface and evolution equations does not fall into the standard
categories that have been studied in computational mathematics. In particular,
there are no energy estimates for the analytic problem which would could serve
as a guide to design a higher order stable algorithm. This is a an important area
for future investigation.
 
\end{description}

\begin{description}

\item[Cubed-sphere codes]~\\
  Convergence of the cubed-sphere code developed by Reisswig et
  al~\cite{reisswig} was also checked using linearized waves in a Minkowski
  background. The convergence rate for the $L_2$ error norm was approximately
  second order accurate but in some cases there was significant degradation.
  They conjectured that the underlying source of error arises at the corners
  of the six patches. Comparison with the $L_\infty$ error norm would
  discriminate such a localized source of error but such results were not
  reported.

  The designed convergence rate of the $\eth$ operator used in the LEO
  code was verified for second, fourth and eighth order finite difference
  approximations, using the spin-weight 2 spherical harmonic ${}_2 Y_{43}$
  as a test. Similarly, the convergence of the integral relations governing
  the orthonormality of the spin-weighted harmonics was verified. The
  code includes coupling to a Klein--Gordon scalar field. Although
  convergence of the evolution code was not explicitly checked,  high
  accuracy in the linearized regime with Schwarzschild background was
  demonstrated in the simulation of quasinormal ringdown of the scalar field
  and in the energy-momentum conservation of the scalar field. 

\end{description}

\begin{description}

\item[Canberra code]~\\
  The complexity of the algorithm and NQS gauge makes it problematic
  to establish accuracy by direct means. Exact solutions do not
  provide an effective convergence check, because the Schwarzschild
  solution is trivial in the NQS gauge and other known solutions in
  this gauge require dynamic inner boundary conditions which
  destabilize the present version of the code. Convergence to
  linearized solutions is a possible check but has not yet been
  performed. Instead indirect tests by means of geometric consistency
  and partial convergence tests are used to calibrate accuracy. The
  consistency tests were based on the constraint equations, which are
  not enforced during null evolution except at the inner boundary. The
  balance between mass loss and radiation flux through $\mathcal{I}^+$
  is a global consequence of these constraints. No appreciable growth
  of the constraints was noticeable until within $5 \, M$ of the final
  breakdown of the code. In weak field tests where angular resolution
  does not dominate the error, partial convergence tests based upon
  varying the radial grid size verify the 8th order convergence in the
  shear expected from the Runge--Kutta integration and splines. When
  the radial source of error is small, reduced error with smaller time
  step can also be discerned.
  
  In practical runs, the major source of inaccuracy is the spherical
  harmonic resolution, which was restricted to $\ell \le 15$ by
  hardware limitations. Truncation of the spherical harmonic expansion
  has the effect of modifying the equations to a system for which the
  constraints are no longer satisfied. The relative error in the
  constraints is 10\super{-3} for strong field
  simulations~\cite{bartint}.

\end{description}



\subsubsection{Nonlinear scattering off a Schwarzschild black hole}

A natural physical application of a characteristic evolution code is the
nonlinear version of the classic problem of scattering off a Schwarzschild black
hole, first solved perturbatively by Price~\cite{price}. Here the inner
worldtube for the characteristic initial value problem consists of the ingoing
branch of the $r=2m$ hypersurface (the past horizon), where Schwarzschild data
are prescribed. The nonlinear problem of a gravitational wave scattering off a
Schwarzschild black hole is then posed in terms of data on an outgoing null cone
which describe an incoming pulse with compact support. Part of the energy of
this pulse falls into the black hole and part is backscattered to 
$\mathcal{I}^+$. This problem has been investigated using both the PITT and
Canberra codes.

The Pittsburgh group studied the backscattered waveform (described by the Bondi
news function) as a function of incoming pulse amplitude. The computational
eth-module smoothly handled the complicated time dependent transformation
between the non-inertial computational frame at $\mathcal{I}^+$ and the inertial
(Bondi) frame necessary to obtain the standard ``plus'' and ``cross''
polarization modes. In the perturbative regime, the news corresponds to the
backscattering of the incoming pulse off the effective Schwarzschild potential.
When the energy of the pulse is no larger than the central Schwarzschild mass, 
the backscattered waveform  still depends roughly linearly on the amplitude of
the incoming pulse. However, for very high amplitudes the waveform behaves quite
differently. Its amplitude is greater than that predicted by linear scaling and
its shape drastically changes and exhibits extra oscillations. In this very high
amplitude case, the mass of the system is completely dominated by the incoming
pulse, which essentially backscatters off itself in a nonlinear way.

The Canberra code was used to study the change in Bondi mass due to the
radiation~\cite{bartint}. The Hawking mass $M_\mathrm{H}(u,r)$ was calculated as
a function of radius and retarded time, with the Bondi mass $M_\mathrm{B}(u)$
then obtained by taking the limit $r\rightarrow \infty$. The limit had good
numerical behavior. For a strong initial pulse with $\ell=4$ angular dependence,
in a run from $u=0$ to $u=70$ (in units where the interior Schwarzschild mass is
1), the Bondi mass dropped from 1.8 to 1.00002, showing that almost half of the
initial energy of the system was backscattered and that a surprisingly
negligible amount of energy fell into the black hole. A possible explanation is
that the truncation of the spherical harmonic expansion cuts off wavelengths
small enough to effectively penetrate the horizon. The Bondi mass decreased
monotonically in time, as necessary theoretically, but its rate of change
exhibited an interesting pulsing behavior whose time scale could not be
obviously explained in terms of quasinormal oscillations. The Bondi mass loss
formula was confirmed with relative error of less than 10\super{-3}. This is
impressive accuracy considering the potential sources of numerical error
introduced by taking the limit of the Hawking mass with limited resolution. The
code was also used to study the appearance of logarithmic terms in the
asymptotic expansion of the Weyl tensor~\cite{bartnort}. In addition, the
Canberra group studied the effect of the initial pulse amplitude on the waveform
of the backscattered radiation, but did not extend their study to the very high
amplitude regime in which qualitatively interesting nonlinear effects occur.


\subsubsection{Black hole in a box}

The PITT code has also been implemented to evolve along an advanced time
foliation by \emph{ingoing} null cones, with data given on a worldtube at their
\emph{outer} boundary and on the initial \emph{ingoing} null cone. The code was
used to evolve a black hole in the region interior to the worldtube by
implementing a horizon finder to locate the marginally trapped surface (MTS) on
the ingoing cones and excising its singular interior~\cite{excise}. The code
tracks the motion of the MTS and measures its area during the evolution. It was
used to simulate a distorted ``black hole in a box''~\cite{wobb}. Data at the
outer worldtube was induced from a Schwarzschild or Kerr spacetime but the
worldtube was allowed to move relative to the stationary trajectories; i.e.\
with respect to the grid the worldtube is fixed but the black hole moves inside
it. The initial null data consisted of a pulse of radiation which subsequently
travels outward to the worldtube where it reflects back toward the black hole.
The approach of the system to equilibrium was monitored by the area of the MTS,
which also equals its Hawking mass. When the worldtube is stationary (static or
rotating in place), the distorted black hole inside evolved to equilibrium with
the boundary. A boost or other motion of the worldtube with respect to the
black hole did not affect this result. The marginally trapped surface always
reached equilibrium with the outer boundary, confirming that the motion of the
boundary was ``pure gauge''.

This was the first code that ran ``forever'' in a dynamic black hole simulation,
even when the worldtube wobbled with respect to the
black hole to produce artificial periodic time dependence. An initially
distorted, wobbling black hole was evolved for a time of $60,000 \, M$, longer by
orders of magnitude than permitted by the stability of other existing black
hole codes at the time. This exceptional performance opens a promising new
approach to handle the inner boundary condition for Cauchy evolution of black
holes by the matching methods reviewed in Section~\ref{sec:ccm}.

Note that setting the pulse to zero is equivalent to prescribing shear free
data on the initial null cone. Combined with Schwarzschild boundary data on the
outer worldtube, this would be complete data for a Schwarzschild space time.
However, the evolution of such shear free null data combined with Kerr boundary
data would have an initial transient phase before settling down to a Kerr black
hole. This is because the twist of the shear-free Kerr null congruence implies
that Kerr data specified on a null hypersurface are not generally shear free.
The event horizon is an exception but Kerr null data on other null
hypersurfaces have not been cast in explicit analytic form. This makes the Kerr
spacetime an awkward testbed for characteristic codes. (Curiously, Kerr data on
a null hypersurface with a conical type singularity do take a simple analytic
form, although unsuitable for numerical evolution~\cite{lun}.) Using some
intermediate analytic results of Israel and Pretorius~\cite{IsrPret}, Venter
and Bishop~\cite{bishkerr} have recently constructed a numerical algorithm for
transforming the Kerr solution into Bondi coordinates and in that way provide
the necessary null data numerically.


\subsection{Characteristic treatment of binary black holes}
\label{sec:nullbbh}

An important application of characteristic evolution is the calculation of the
waveform emitted by binary black holes, which is possible during the very
interesting nonlinear domain from merger to ringdown~\cite{ndata,kyoto}. The
evolution is carried out along a family of ingoing null hypersurfaces which
intersect the horizon in topological spheres. It is restricted to the period
following the merger, for otherwise the ingoing null hypersurfaces would
intersect the horizon in disjoint pieces corresponding to the individual black
holes. The evolution proceeds \emph{backward} in time on an ingoing null
foliation to determine the exterior spacetime in the post-merger era. It is an
example of the characteristic initial value problem posed on an intersecting
pair of null hypersurfaces~\cite{sachsdn,haywdn}, for which existence theorems
apply in some neighborhood of the initial null
hypersurfaces~\cite{hagenseifert77,helmut81a,fried}. Here one of the null
hypersurfaces is the event horizon $\mathcal{H}^+$ of the binary black holes. The
other is an ingoing null hypersurface $J^+$ which intersects $\mathcal{H}^+$ in a
topologically spherical surface $\mathcal{S}^+$ approximating the equilibrium of
the final Kerr black hole, so that $J^+$ approximates future null infinity
$\mathcal{I}^+$. The required data for the analytic problem consists of the
degenerate conformal null metrics of $\mathcal{H}^+$ and $J^+$ and the metric and
extrinsic curvature of their intersection $\mathcal{S}^+$.

The conformal metric of $\mathcal{H}^+$ is provided by the conformal horizon model
for a binary black hole horizon~\cite{ndata,asym}, which treats the horizon in
stand-alone fashion as a three-dimensional manifold endowed with a degenerate
metric $\gamma_{ab}$ and affine parameter $t$ along its null rays. The metric
is obtained from the conformal mapping $\gamma_{ab}=\Omega^2 \hat \gamma_{ab}$
of the intrinsic metric $\hat \gamma_{ab}$ of a flat space null hypersurface
emanating from a convex surface $\mathcal{S}_0$ embedded at constant time in
Minkowski space. The horizon is identified with the null hypersurface formed
by the inner branch of the boundary of the past of $\mathcal{S}_0$, and its
extension into the future. The flat space null hypersurface expands forever as
its affine parameter $\hat t$ (Minkowski time) increases, but the conformal
factor is chosen to stop the expansion so that the cross-sectional area of the
black hole approaches a finite limit in the future. At the same time, the
Raychaudhuri equation (which governs the growth of surface area) forces a
nonlinear relation between the affine parameters $t$ and $\hat t$. This is what
produces the nontrivial topology of the affine $t$-slices of the black hole
horizon. The relative distortion between the affine parameters $t$ and $\hat
t$, brought about by curved space focusing, gives rise to the trousers shape of
a binary black hole horizon.

\epubtkImage{pants.png}{%
  \begin{figure}[htbp]
    \centerline{\includegraphics[height=7cm]{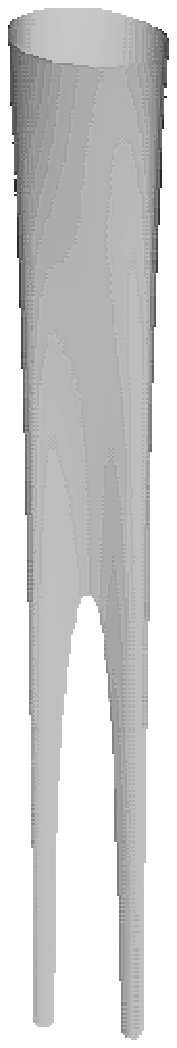}}
    \caption{Trousers shaped event horizon obtained by the
      conformal model.}
    \label{fig:pants}
  \end{figure}}

An embedding diagram of the horizon for an axisymmetric head-on collision,
obtained by choosing $\mathcal{S}_0$ to be a prolate spheroid, is shown in
Figure~\ref{fig:pants}~\cite{ndata}. The black hole event horizon associated
with a triaxial ellipsoid  reveals new features not seen in the  degenerate case
of the head-on collision~\cite{asym}, as depicted in Figure~\ref{fig:bbh2}. If
the degeneracy is slightly broken, the individual black holes form with
spherical topology but as they approach, an effective tidal distortion produces
two sharp pincers on each black hole just prior to merger. At merger, the two
pincers join to form a single temporarily toroidal black hole. The inner hole of
the torus subsequently closes up to produce first a peanut shaped black hole and
finally a spherical black hole. No violation of topological
censorship~\cite{fsw} occurs because the hole in the torus closes up
superluminally. Consequently, a causal curve passing through the torus at a
given time can be slipped below the bottom of a trouser leg to yield a causal
curve lying entirely outside the hole~\cite{toroid}. In the degenerate
axisymmetric limit, the pincers reduce to a point so that the individual holes
have teardrop shape and they merge without a toroidal phase. Animations of this
merger can be viewed at~\cite{psc}.

\epubtkImage{bbh2-all.png}{%
  \begin{figure}[htbp]
    \centerline{\includegraphics[width=0.8\textwidth]{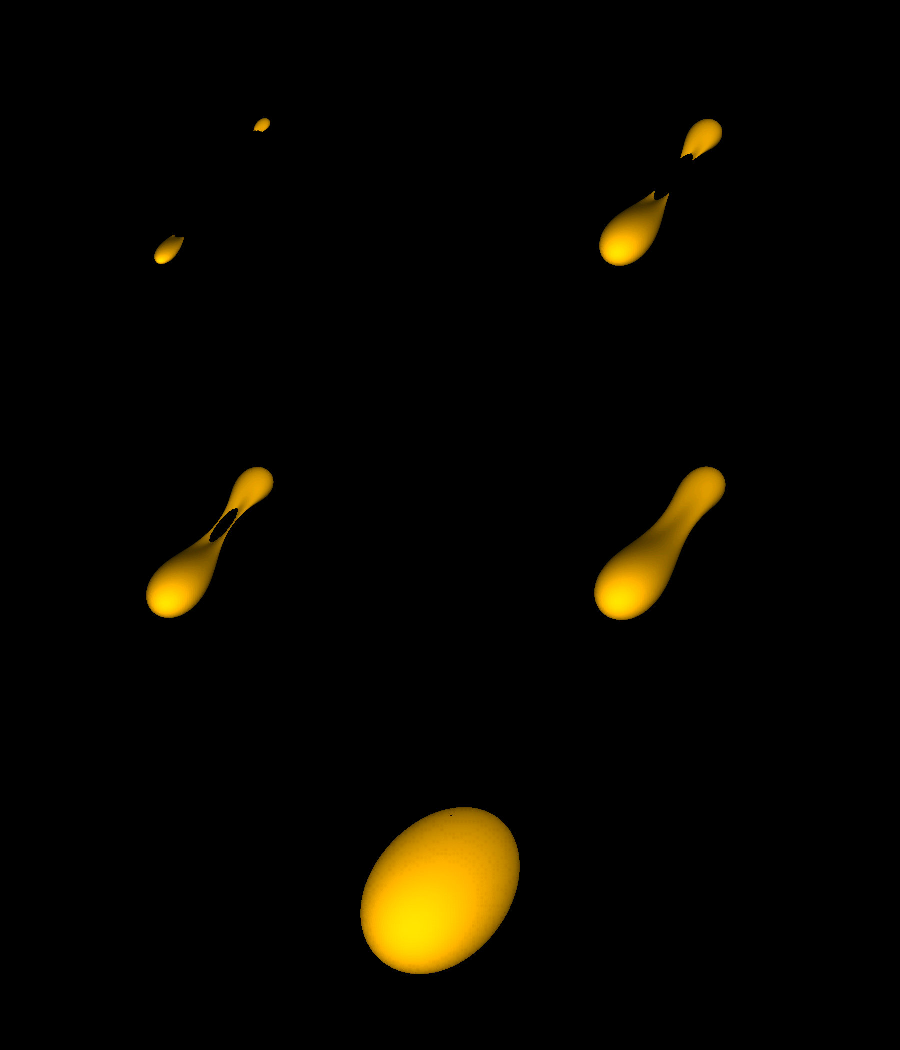}}
    \caption{Upper left: Tidal distortion of approaching black holes
      Upper right: Formation of sharp pincers just prior to
      merger. Middle left: Temporarily toroidal stage just after
      merger. Middle right: Peanut shaped black hole after the hole in
      the torus closes. Lower: Approach to final equilibrium.}
    \label{fig:bbh2}
  \end{figure}}

The conformal horizon model determines the data on $\mathcal{H}^+$ and ${\cal
S}^+$. The remaining data necessary to evolve the exterior spacetime are given
by the conformal geometry of $J^+$, which constitutes the outgoing radiation
waveform. The determination of the merger-ringdown waveform proceeds in two
stages. In the first stage, this outgoing waveform is set to zero and the
spacetime is evolved backward in time to calculate the incoming radiation
entering from $\mathcal{I}^-$. (This incoming radiation is eventually absorbed by
the black hole.) From a time reversed point of view, this evolution describes
the outgoing waveform emitted in the fission of a white hole, with the
physically correct initial condition of no ingoing radiation. Preliminary
calculations show that at late times the waveform is entirely quadrupolar
($\ell =2$) but that a strong octopole mode ($\ell =4$) exists just before
fission. In the second stage of the calculation, this waveform could be used to
generate the physically correct outgoing waveform for a black hole merger. The
passage from the first stage to the second is the nonlinear equivalent of first
determining an inhomogeneous solution to a linear problem and then adding the
appropriate homogeneous solution to satisfy the boundary conditions. In this
context, the first stage supplies an advanced solution and the second stage the
homogeneous retarded minus advanced solution. When the evolution is carried out
in the perturbative regime of a Kerr or Schwarzschild background, as in the
close approximation~\cite{jorge}, this superposition of solutions is simplified
by the time reflection symmetry~\cite{kyoto}. The second stage has been carried
out in the perturbative regime of the close approximation using a
characteristic code which solves the Teukolsky equation, as described in
Section~\ref{sec:schwpert}. More generally, beyond the perturbative regime, the
merger-ringdown waveform must be obtained by a more complicated inverse
scattering procedure, which has not yet been attempted.

There is a complication in applying the PITT code to this double null evolution
because a dynamic horizon does not lie precisely on $r$-grid points. As a
result, the $r$-derivative of the null data, i.e.\ the ingoing shear of ${\cal
H}$, must also be provided in order to initiate the radial hypersurface
integrations. The ingoing shear is part of the free data specified at ${\cal
S}^+$. Its value on $\mathcal{H}$ can be determined by integrating (backward in
time) a sequence of propagation equations involving the horizon's twist and
ingoing divergence. A horizon code which carries out these integrations has
been tested to give accurate data even beyond the merger~\cite{hdata}.

The code has revealed new global properties of the head-on collision by
studying a sequence of data for a family of colliding black holes  which
approaches a single Schwarzschild black hole. The resulting perturbed
Schwarzschild horizon provides global insight into the close
limit~\cite{jorge}, in which the individual black holes have joined in the
infinite past. A marginally anti-trapped surface divides the horizon into
interior and exterior regions, analogous to the division of the Schwarzschild
horizon by the $r=2M$ bifurcation sphere. In passing from the perturbative to
the strongly nonlinear regime there is a rapid transition in which the
individual black holes move into the exterior portion of the horizon. The data
pave the way for the PITT code to calculate whether this dramatic time
dependence of the horizon produces an equally dramatic waveform. See
Section~\ref{sec:fission} for first stage results.


\subsection{Perturbations of Schwarzschild}
\label{sec:schwpert}

The nonlinear 3D PITT code has been calibrated in the regime of small
perturbations of a Schwarz\-schild spacetime~\cite{Zlochower,zlochmode} by
measuring convergence with respect to independent solutions of the Teukolsky
equation~\cite{teuk72}. By decomposition into spherical harmonics,  the
Teukolsky equation reduces the problem of a perturbation of a stationary black
hole to a 1D problem in the $(t,r)$ subspace perturbations for a component of
the Weyl tensor. Historically, the Teukolsky equation was first solved
numerically by Cauchy evolution. Campanelli, G{\'{o}}mez, Husa, Winicour, and
Zlochower~\cite{zlochadv,zlochret} have reformulated the Teukolsky formalism as
a double-null characteristic evolution algorithm. The evolution proceeds on a
family of outgoing null hypersurfaces with an ingoing null hypersurface as inner
boundary and with the outer boundary compactified at future null infinity. It
applies to either the Weyl component $\Psi_0$ or $\Psi_4$, as classified in the
Newman--Penrose formalism. The $\Psi_0$ component comprises constraint-free
gravitational data on an outgoing null hypersurface and $\Psi_4$ comprises the
corresponding data  on an ingoing null hypersurface. In the study of
perturbations of a Schwarzschild black hole, $\Psi_0$ is prescribed on an
outgoing null hypersurface $\mathcal{J}^-$, representing an early retarded time
approximating past null infinity, and $\Psi_4$ is prescribed on the inner white
hole horizon $\mathcal{H}^-$.

The physical setup is described in
Figure~\ref{fig:setup}. The outgoing null hypersurfaces extend to future
null infinity $\mathcal{I}^+$ on a compactified numerical grid.
Consequently, there is no need for either an artificial outer boundary
condition or an interior extraction worldtube. The outgoing radiation
is computed in the coordinates of an observer in an inertial frame at
infinity, thus avoiding any gauge ambiguity in the waveform.

\epubtkImage{fig1.png}{%
  \begin{figure}[htbp]
    \centerline{\includegraphics[width=0.45\textwidth]{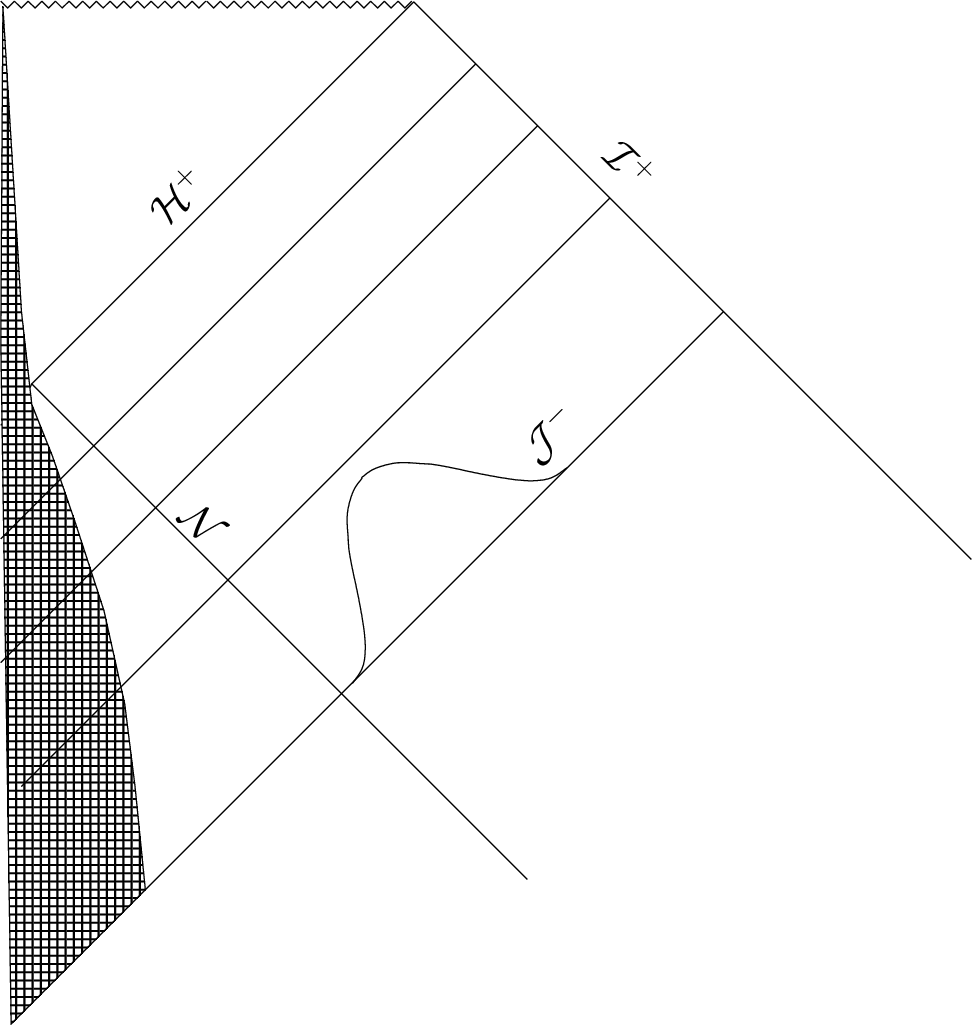}}
    \caption{The physical setup for the scattering problem. A star
      of mass $M$ has undergone spherically symmetric collapse to form
      a black hole. The ingoing null worldtube $\mathcal{N}$ lies outside
      the collapsing matter. Inside $\mathcal{N}$ (but outside the
      matter) there is a vacuum Schwarzschild metric. Outside of
      $\mathcal{N}$, data for an ingoing pulse is specified on the
      initial outgoing null hypersurface $\mathcal{J}^-$. As the pulse
      propagates to the black hole event horizon $\mathcal{H}^+$, part of
      its energy is scattered to $\mathcal{I}^+$.}
    \label{fig:setup}
  \end{figure}}

The first calculations were carried out with nonzero data for $\Psi_4$  on
$\mathcal{H}^-$ and zero data on $\mathcal{J}^-$~\cite{zlochadv} (so that no
ingoing radiation entered the system). The resulting simulations were highly
accurate and tracked the quasi-normal ringdown of a perturbation consisting of a
compact pulse through 10 orders of magnitude and tracked the final power law
decay through an additional 6 orders of magnitude. The measured exponent of the
power law decay varied from $\approx 5.8$, at the beginning of the tail, to
$\approx 5.9$ near the end, in good agreement with the predicted value of
$2\ell+2 =6$ for a quadrupole wave~\cite{price}.

The accuracy of the perturbative solutions provide a virtual exact solution
for carrying out convergence tests of the nonlinear PITT null code. In this
way, the error in the Bondi news function computed by the PITT
code was calibrated for perturbative data consisting of either an outgoing
pulse on $\mathcal{H}^-$ or an ingoing pulse on $\mathcal{J}^-$.
For the outgoing pulse, clean second order convergence
was confirmed until late times in the evolution, when small deviations
from second order arise from accumulation of roundoff and truncation error.
For the Bondi news produced by the scattering of an ingoing
pulse, clean second order convergence was again confirmed until late times
when the pulse approached the $r=2M$ black hole horizon. The late time error
arises from loss of resolution of the pulse (in the radial direction) resulting
from the properties of the compactified radial coordinate used in the code.
This type of error could be eliminated by using characteristic the AMR
techniques under development~\cite{pretlehn}.


\subsubsection{Close approximation white hole and black hole waveforms}
\label{sec:close}

The characteristic Teukolsky code has been used to study radiation from
axisymmetric white holes and black holes in the close approximation. The
radiation from an axisymmetric fissioning white hole~\cite{zlochadv} was
computed using the Weyl data on $\mathcal{H}^-$ supplied by the conformal
horizon model described in Section~\ref{sec:nullbbh}, with the fission occurring
along the axis of symmetry. The close approximation implies that the fission
takes place far in the future, i.e.\ in the region of $\mathcal{H}^-$ above the
black hole horizon $\mathcal{H}^+$. The data have a free parameter $\eta$ which
controls the energy yielded by the white hole fission. The radiation waveform
reveals an interesting dependence on the parameter $\eta$. In the large $\eta$
limit, the  waveform consists of a single pulse, followed by ringdown and tail
decay. The amplitude of the pulse scales quadratically with $\eta$ and the width
decreases with $\eta$. As $\eta$ is reduced, the initial pulse broadens and
develops more structure. In the small $\eta$ limit, the amplitude scales
linearly with $\eta$ and the shape is independent of $\eta$.

Since there was no incoming radiation, the above model gave the physically
appropriate boundary conditions for a white hole fission (in the close
approximation). From a time reversed view point, the system corresponds to a
black hole merger with no outgoing radiation at future null infinity, i.e.\ the
analog of an advanced solution with only ingoing but no outgoing radiation. In
the axisymmetric case studied, the merger corresponds to a head-on collision
between two black holes. The physically appropriate boundary conditions for a
black hole merger correspond to no ingoing radiation on $\mathcal{J}^-$ and
binary black hole data on $\mathcal{H}^+$. Because  $\mathcal{J}^-$ and
$\mathcal{H}^+$ are disjoint, the corresponding data cannot be used directly to
formulate a double null characteristic initial value problem. However, the
ingoing radiation at $\mathcal{J}^-$ supplied by the advanced solution for the
black hole merger could be used as Stage~I of a two stage approach to determine
the corresponding retarded solution. In Stage~II, this ingoing radiation is used
to generate the analogue of an \emph{advanced minus retarded} solution. A pure
retarded solution (with no ingoing radiation but outgoing radiation at
$\mathcal{I}^+$ can then be constructed by superposition. The time reflection
symmetry of the Schwarzschild background is key to carrying out this
construction.

This two stage strategy has been carried out by Husa, Zlochower, G{\'{o}}mez,
and Winicour~\cite{zlochret}. The superposition of the Stage~I and~II solutions
removes the ingoing radiation from $\mathcal{J}^-$ while modifying the close
approximation perturbation of $\mathcal{H}^+$, essentially making it ring. The
amplitude  of the radiation waveform at $\mathcal{I}^+$ has a linear dependence
on the parameter $\eta$, which in this black hole scenario governs the energy
lost in the inelastic merger process. Unlike the fission waveforms, there is
very little $\eta$-dependence in their shape and the amplitude continues to
scale linearly even for large $\eta$. It is not surprising that the retarded
waveforms from a black hole merger differs markedly from the retarded waveforms
from a white hole merger. The white hole process is directly visible at 
$\mathcal{I}^+$ whereas the merger waveform results indirectly from the black
holes through the preceding collapse of matter or gravitational energy that
formed them. This explains why the fission waveform is more sensitive to the
parameter $\eta$ which controls the shape and timescale of the horizon data.
However, the weakness of the dependence of the merger waveform on $\eta$ is
surprising and has potential importance for enabling the design of an efficient
template for extracting a gravitational wave signal from noise.


\subsubsection{Fissioning white hole}
\label{sec:fission}

In the purely vacuum approach to the binary black hole problem, the stars which
collapse to form the black holes are replaced either by imploding
gravitational waves or some past singularity as in the Kruskal picture.
This avoids hydrodynamic difficulties at the expense of a globally complicated
initial value problem. The imploding waves either emanate from a past
singularity, in which case the time-reversed application of cosmic censorship
implies the existence of an anti-trapped surface; or they emanate from ${\cal
I}^-$, which complicates the issue of gravitational radiation content in
the initial data and its effect on the outgoing waveform. These complications
are avoided in the two stage approach adopted in the close approximation
studies described in Section~\ref{sec:close}, where advanced and retarded
solutions in a Schwarzschild background can be rigorously identified and
superimposed. Computational experiments have been carried out to study the
applicability of this approach in the nonlinear regime~\cite{fiss}.

From a time reversed viewpoint, the first stage is equivalent to the
determination of the outgoing radiation from a fission of a white hole in the
absence of ingoing radiation, i.e.\ the physically appropriate ``retarded''
waveform from a white hole fission. This fission problem can be formulated in
terms of data on the white hole horizon $\mathcal{H}^-$ and data representing
the absence of ingoing radiation on a null hypersurface $J^-$ which emanates
from $\mathcal{H}^-$ at an early time. The data on $\mathcal{H}^-$ is provided
by the conformal horizon model for a fissioning white hole. This allows study of
a range of models extending from the perturbative close approximation regime, in
which the fission occurs inside a black hole event horizon, to the nonlinear
regime of a ``bare'' fission visible from $\mathcal{I}^+$. The study
concentrates on the axisymmetric spinless fission (corresponding in the time
reversed view to the head-on collision of non-spinning black holes). In the
perturbative regime, the news function agrees with the close approximation
waveforms. In the highly nonlinear regime, a bare fission was found to produce a
dramatically sharp radiation pulse, which then undergoes a damped oscillation.
Because the black hole fission is visible from $\mathcal{I}^+$, it is a more
efficient source of gravitational waves than a black hole merger and can produce
a higher fractional mass loss!


\subsection{Nonlinear mode coupling}
\label{sec:mode}

The PITT code has been used to  model the nonlinear generation of waveforms by
scattering off a Schwarzschild black hole~\cite{Zlochower,zlochmode}. The
physical setup is similar to the perturbative study in
Section~\ref{sec:schwpert}. A radially compact pulse is prescribed on an early
time outgoing null hypersurface $\mathcal{J}^-$ and Schwarzschild null data is
given on the interior white hole horizon $\mathcal{H}^-$, which is causally
unaffected by the pulse. The input pulse is standardized to ($\ell=2$, $m=0$)
and ($\ell=2$, $m=2$) quadrupole modes with amplitude $A$. The outgoing null
hypersurfaces extend to future null infinity $\mathcal{I}^+$ on a compactified
numerical grid. Consequently, there is no need for an artificial outer boundary.
The evolution code then provides the news function at $\mathcal{I}^+$, in the
coordinates of an observer in an inertial frame at infinity, thus avoiding any
gauge ambiguity in the waveform. This provides a simple setting for how the
nonlinearities generated by high amplitudes affect the waveform.

The study reveals several features of qualitative importance:
\begin{enumerate}
\item The mode coupling amplitudes consistently scale as powers $A^n$
  of the input amplitude $A$ corresponding to the nonlinear order of
  the terms in the evolution equations which produce the mode. This
  allows much economy in producing a waveform catalog: Given the order
  $n$ associated with a given mode generation, the response to any
  input amplitude $A$ can be obtained from the response to a single
  reference amplitude.
  \label{feature_1}
\item The frequency response has similar behavior but in a less
  consistent way. The dominant frequencies produced by mode coupling
  are in the approximate range of the quasinormal frequency of the
  input mode and the expected sums and difference frequencies
  generated by the order of nonlinearity.
  \label{feature_2}
\item Large phase shifts, ranging up 15\% in a half cycle relative to
  the linearized waveform, are exhibited in the news function obtained
  by the superposition of all output modes, i.e.\ in the  waveform of
  observational significance. These phase shifts, which are important
  for design of signal extraction templates, arise in an erratic way
  from superposing modes with different oscillation frequencies. This
  furnishes a strong argument for going beyond the linearized
  approximation in designing a waveform catalog for signal extraction.
  \label{feature_3}
\item Besides the nonlinear generation of harmonic modes absent in the
  initial data, there is also a stronger than linear generation of
  gravitational wave output. This provides a potential mechanism for
  enhancing the strength of the gravitational radiation produced
  during, say, the merger phase of a binary inspiral above the
  strength predicted in linearized theory.
  \label{feature_4}
\item In the non-axisymmetric $m=2$ case, there is also considerable
  generation of radiation in polarization states not present in the
  linearized approximation. In the simulations, input amplitudes in
  the range  $A=0.1$ to $A=0.36$ lead to nonlinear generation of the
  $\oplus$ polarization mode which is of the same order of magnitude
  as the $\otimes$ mode (which would be the sole polarization in the
  linearized regime). As a result, significant nonlinear amplification
  and phase shifting of the waveform would be observed by a
  gravitational wave detector, depending on its orientation.
  \label{feature_5}
\end{enumerate}

These effects arise from the nonlinear modification of the Schwarzschild
geometry identified by Papadopoulos in his prior work on axisymmetric mode
coupling~\cite{papamode}, reported in Section~\ref{sec:papamode}. Although
Papadopoulos studied nonlinear mode generation produced by an outgoing pulse, as
opposed to the case of an ingoing pulse studied in~\cite{Zlochower,zlochmode},
the same nonlinear factors were in play and gave rise to several common
features. In both cases, the major effects arise in the region near $r=3M$.
Analogs of Features~\ref{feature_1}, \ref{feature_2}, \ref{feature_3},
and~\ref{feature_4} above are all apparent in Papadopoulos's work. At the finite
difference level, both codes respect the reflection symmetry inherent in
Einstein's equations and exhibit the corresponding selection rules arising from
parity considerations. In the axisymmetric case considered by Papadopoulos, this
forbids the nonlinear generation of a $\oplus$ mode  from a $\otimes$ mode, as
described in Feature~\ref{feature_5} above.

The evolution along ingoing null hypersurfaces in the axisymmetric work of
Papadopoulos has complementary numerical features with the evolution along
outgoing null hypersurfaces in the 3D work. The grid based upon \emph{ingoing}
null hypersurfaces avoids the difficulty in resolving effects close to $r=2M$
encountered with the grid based upon \emph{outgoing} null hypersurfaces. The
outgoing code would require AMR in order to resolve the quasinormal ringdown for
as many cycles as achieved by Papadopoulos. However, the outgoing code avoids
the late time caustic formation noted in Papadopoulos' work, as well as the
complications of gauge ambiguity and backscattering introduced by a finite outer
boundary. One attractive option would be to combine the best features of these
approaches by matching an interior evolution based upon ingoing null
hypersurfaces to an exterior evolution based upon outgoing null hypersurfaces,
as implemented in~\cite{luis2m} for spherically symmetric
Einstein--Klein--Gordon waves.

The waveform of relevance to gravitational wave astronomy is the
superposition of modes with different frequency compositions and
angular dependence. Although this waveform results from a complicated
nonlinear processing of the input signal, which varies with choice of
observation angle, the response of the individual
modes to an input signal of arbitrary amplitude can be obtained by
scaling the response to an input of standard reference amplitude. This
offers an economical approach to preparing a waveform catalog.
Nonlinear mode coupling has also been extensively studied
by gauge invariant second order perturbative methods
(see~\cite{modec1,modec2,modec3} and references therein).


\subsection{3D Einstein--Klein--Gordon system}
\label{sec:3dekg}

The Einstein--Klein--Gordon (EKG) system can be used to simulate many
interesting physical phenomena. In 1D, characteristic EKG codes have been used
to simulate critical phenomena and the perturbation of black holes (see
Section~\ref{sec:1d}), and a Cauchy EKG code has been used to study boson star
dynamics~\cite{SeidelSuen}. Extending these codes to 3D would open up a new
range of possibilities, e.g., the possibility to study radiation from a boson
star orbiting a black hole. A first step in that direction has been achieved
with the incorporation a massless
scalar field into the PITT code~\cite{barretoekg}. Since the scalar and
gravitational evolution equations have the same basic form, the same evolution
algorithm could be utilized. The code was tested to be second order convergent
and stable. It was applied to the fully nonlinear simulation of an asymmetric
pulse of ingoing scalar radiation propagating toward a Schwarzschild black hole.
The resulting scalar radiation and gravitational news backscattered to
$\mathcal{I}^+$ was computed. The amplitudes of the scalar and gravitational
radiation modes exhibited the expected power law scaling with respect to the
initial pulse amplitude. In addition, the computed ringdown frequencies agreed
with the results from perturbative quasinormal mode calculations.

The LEO code~\cite{leo} developed by G{\'{o}}mez et al.\ has been applied to the
characteristic evolution of the coupled Einstein--Klein--Gordon fields, using the
cubed-sphere coordinates. The long term plan is to simulate a boson star
orbiting a black hole. In simulations of a scalar pulse incident on a
Schwarzschild black hole, they find the interesting result that scalar energy
flow into the black hole reaches a maximum at spherical harmonic index $\ell=2$,
and then decreases for larger $\ell$ due to the centrifugal barrier preventing
the harmonics from effective penetration. The efficient parallelization allows
them to perform large simulations with resolution never achieved before. 
Characteristic evolution of such systems of astrophysical interest have been
limited in the past by resolution. They note that at the finest resolution
considered in~\cite{matter2}, it would take 1.5 months on the fastest current
(single) processor to track a star in close orbit around a black hole. This is
so even though the grid in question is only $81\times 123$ points, which
is moderate by today's standards.

\newpage


\section{Cauchy-Characteristic Matching}
\label{sec:ccm}

Characteristic evolution has many advantages over Cauchy evolution. Its main
disadvantage is the existence of either a caustic, where neighboring
characteristics focus, or a milder version consisting of a crossover between two
distinct characteristics. The vertex of a light cone is a highly symmetric
caustic which already strongly limits the time step for characteristic evolution
because of the CFL condition~(\ref{eq:cfl0}). It does not appear possible for a
single characteristic coordinate system to cover the entire exterior region of a
binary black hole spacetime without developing very complicated caustics and
crossovers. This limits the waveform determined by a purely characteristic
evolution to the post merger period.

CCM is a way to avoid this limitation by combining the strong points of
characteristic and Cauchy evolution into a global evolution~\cite{Bis2}. One of
the prime goals of computational relativity is the simulation of the inspiral
and merger of binary black holes. Given the appropriate data on a worldtube
surrounding a binary system, characteristic evolution can supply the exterior
spacetime and the radiated waveform. But determination of the worldtube data for
a binary requires an interior Cauchy evolution. CCM is designed to solve such
global problems. The potential advantages of CCM over traditional boundary
conditions are
\begin{itemize}
\item accurate waveform and polarization state at infinity,
\item computational efficiency for radiation problems in terms of both
  the grid domain and the computational algorithm,
\item elimination of an artificial outer boundary condition on the
  Cauchy problem, which eliminates contamination from back-reflection
  and clarifies the global initial value problem, and
\item a global picture of the spacetime exterior to the event horizon.
\end{itemize}

These advantages have been realized in model tests
(see Sections~\ref{sec:1dccm}\,--\,\ref{sec:linccm}), but CCM has not yet been
achieved in fully nonlinear 3-dimensional general relativity. The early
attempts to implement CCM in general relativity employed the
Arnowitt--Deser--Misner (ADM)~\cite{adm} formulation,
with explicit lapse and shift, for the Cauchy evolution.
A major problem in this application has since been identified
with the weakly hyperbolic nature of this system. Even at the analytic
level of the Cauchy problem there are secularly growing modes
with arbitrarily fast rates, i.e. the Cauchy problem is ill-posed.
Such power law instabilities of the Cauchy problem
can be converted to exponentially growing instabilities by the introduction of
lower order or nonlinear terms.
See~\cite{weak} for discussions relevant to the stability of the ADM formulation.

Such behavior can also be made worse by the imposition of boundary conditions.
Linearized studies~\cite{belath,cauchboun,apples2} of ADM
evolution-boundary algorithms with prescribed values of lapse and
shift have shown the following:
\begin{itemize}
\item On analytic grounds, those ADM boundary algorithms which supply values
  for all components of the metric (or extrinsic curvature) are
  inconsistent.
\item A consistent boundary algorithm allows free specification
  of the transverse-traceless components of the metric with respect to
  the boundary.
\item Using such a boundary algorithm, linearized ADM evolution can be
  carried out in a bounded domain for thousands of crossing times without sign
  of exponential growth, even though there are the secularly growing modes
  whose rates increase with resolution.
\end{itemize}

Such results contributed to the early belief that long term evolutions
might be possible by means of ADM evolution.
The linearized tests satisfied the original criterion for robust stability,
i.e. that there be no exponential growth when the initial Cauchy data and free
boundary data are prescribed as random numbers (in the linearized
regime)~\cite{cauchboun}. However, it was subsequently shown that
the weakly hyperbolic nature of ADM led to uncontrolled
power law instabilities. In the nonlinear regime, it is symptomatic of
weakly hyperbolic systems that such instabilities become exponential.
This has led to  refined criteria for robust stability as a standardized
test~\cite{apples2}.

CCM cannot work unless the Cauchy and characteristic codes have robustly stable
boundaries. This is necessarily so because interpolations continually introduce
short wavelength noise into the neighborhood of the boundary. It has been
demonstrated that the PITT characteristic code has a robustly stable boundary
(see Section~\ref{sec:stability}), but robustness of the Cauchy boundary has
only recently been studied.


\subsection{Computational boundaries}

Boundary conditions are both the most important and the most difficult part of a
theoretical treatment of most physical systems. Usually, that's where all the
physics is. And, in computational approaches, that's usually where all the agony
is. Computational boundaries for hyperbolic systems pose special difficulties.
Even with an analytic form of the correct physical boundary condition in hand,
there are seemingly infinitely more unstable numerical implementations than
stable ones. In general, a stable problem places more boundary requirements on
the numerical algorithm than on the corresponding partial differential
equations. Furthermore, the methods of linear stability analysis are often more
unwieldy to apply to the boundary than to the interior evolution algorithm.
Only if there is an energy estimate for the analytic problem is there a straightforward
way to proceed. In that case the integration by parts underlying the energy
conservation law can be converted into a
\emph{summation by parts}~\cite{kreissch} construction
of a stable finite difference algorithm. (See the forthcoming Living Review~\cite{sarbtig},)
For problems with constraints, such as
general relativity, there is the additional complication that the boundary condition
must enforce the constraints. 

The von~Neumann stability analysis of the interior algorithm linearizes the
equations, while assuming a uniform grid with periodic boundary conditions, and
checks that the discrete Fourier modes do not grow exponentially. There is an
additional stability condition that a boundary introduces into this analysis.
Consider the one-dimensional case. The mode $e^{kx}$, with $k$ real, is not
included in the von~Neumann analysis for periodic boundary conditions. However,
for the half plane problem in the domain $x\le 0$, one can legitimately
prescribe such a mode  as initial data as long as $k>0$ so that it has finite
energy. Thus the stability of such boundary modes must  be checked. In the case
of an additional boundary, e.g.\ for a problem in the domain $-1\le x \le 1$, the
Godunov--Ryaben'kii theory gives as a necessary condition for stability the
combined von~Neumann stability of the interior and the stability of the allowed
boundary modes~\cite{sod}. The Kreiss condition~\cite{kreiss2,gks} strengthens this
result by providing a sufficient condition for stability.

The correct physical formulation of any Cauchy problem for an isolated system
also involves asymptotic conditions at infinity. These conditions must ensure
not only that the total energy and energy loss by radiation are both finite,
but they must also ensure the proper $1/r$ asymptotic falloff of the radiation
fields. However, when treating radiative systems computationally, an outer
boundary is often established artificially at some large but finite distance in
the wave zone, i.e.\ many wavelengths from the source. Imposing an appropriate
radiation boundary condition at a finite distance is a difficult task even in
the case of a simple radiative system evolving on a fixed geometric background.
Gustaffson and Kreiss have shown in general that the construction of a
nonreflecting boundary condition for an isolated system requires knowledge of
the solution in a neighborhood of infinity\cite{guskreis}.  

When the system is nonlinear and not amenable to an exact solution, a finite
outer boundary condition must necessarily introduce spurious physical effects
into a Cauchy evolution. The domain of dependence of the initial Cauchy data in
the region spanned by the computational grid would shrink in time along ingoing
characteristics unless data on a worldtube traced out by the outer grid boundary
is included as part of the problem. In order to maintain a causally sensible
evolution, this worldtube data must correctly substitute for the missing Cauchy
data which would have been supplied if the Cauchy hypersurface had extended to
infinity. In a scattering problem, this missing exterior Cauchy data might, for
instance, correspond to an incoming pulse initially outside the outer boundary.
In a scalar wave problem with field $\Phi$ where the initial radiation is
confined to a compact region inside the boundary, the missing  Cauchy data
outside the boundary would be $\Phi=\Phi_{,t}=0$ at the initial time $t_0$.
However, the determination of Cauchy data for general relativity is a global
elliptic constraint problem so that there is no well defined scheme to confine
it to a compact region. Furthermore, even in the scalar field case where 
$\Phi=\Phi_{,t}=0$ is appropriate Cauchy data outside the boundary at $t_0$, it
would still be a non-trivial evolution problem to correctly assign the
associated boundary data for $t\ge t_0$.

It is common practice in computational physics to impose an artificial boundary
condition (ABC), such as an outgoing radiation condition, in an attempt to
approximate the proper data for the exterior region. This ABC may cause partial
reflection of an outgoing wave back into the system~\cite{Lind1,Orsz,Hig86,Ren},
which contaminates the accuracy of the interior evolution and the calculation of
the radiated waveform. Furthermore, nonlinear waves intrinsically backscatter,
which makes it incorrect to try to entirely eliminate incoming radiation from
the outer region. The resulting error is of an analytic origin, essentially
independent of computational discretization. In general, a systematic reduction
of this error can only be achieved by moving the computational boundary to
larger and larger radii. There has been recent progress in designing
such {\em absorbing}
boundary conditions for the gravitational field~\cite{sarbuch}. See the review~\cite{sarbtig}
for details on this subject. 

A traditional ABC for the wave equation is the Sommerfeld condition. For a
scalar field $\Phi$ satisfying the Minkowski space wave equation
\begin{equation}
    \eta^{\alpha\beta}\partial_\alpha \partial_\beta \Phi=S,
\end{equation}
with a smooth source $S$ of compact support emitting outgoing radiation,  the
exterior retarded field has the form
\begin{equation}
    \Phi=\frac{f(t-r,\theta,\phi)}{r} +\frac{g(t-r,\theta,\phi)}{r^2}
    +\frac{h(t,r,\theta,\phi)}{r^3} ,
    \label{eq:outgoing}
\end{equation}
where $f$, $g$ and $h$ and their derivatives are smooth bounded functions. The
simplest case is the monopole radiation
\begin{equation}
    \Phi=\frac{f(t-r)}{r}
\end{equation}
which satisfies $(\partial_t+\partial_r) (r\Phi)=0$. This motivates
the use of the Sommerfeld condition
\begin{equation}
  \frac{1}{r}(\partial_t+\partial_r)(r\Phi)|_R= q(t,R,\theta,\phi)
\end{equation}
on a finite boundary $r=R$. 

A homogeneous Sommerfeld condition,  i.e.$q=0$, is exact only in  the spherically
symmetric case. The Sommerfeld boundary data $q$ in the general case
(\ref{eq:outgoing}) falls off as $1/R^3$, so that a homogeneous Sommerfeld
condition introduces an error, which is small only for large $R$. As an example,
for the dipole solution
\begin{equation}
    \Phi_{\mathrm{Dipole}}=\partial_z{\frac{f(t-r)}{r}}
                 =-\left( \frac{f'(t-r)}{r} + \frac{f(t-r)}{r^2} \right)
                  \cos\theta
\end{equation}
we have
\begin{equation}
    q= \frac{f(t-r)\cos\theta}{R^3}.
\end{equation}
A homogeneous Sommerfeld condition at $r=R$ would lead to a solution
$\tilde \Phi_{Dipole}$ containing a reflected ingoing wave. For large
$R$,
\begin{equation}
  \tilde \Phi_{\mathrm{Dipole}} \sim \Phi_{\mathrm{Dipole}}
  + \kappa \frac{F(t+r-2R)\cos\theta}{r},
\end{equation}
where $\partial_t f(t)=F(t)$ and the reflection coefficient has asymptotic
behavior $\kappa=O(1/R^2)$. More precisely, the Fourier mode
\begin{equation}
  \tilde \Phi_{\mathrm{Dipole}}(\omega) = \partial_z\bigg (\frac{e^{i\omega (t-r)}}{r}
         + \kappa_\omega \frac{e^{i\omega (t+r-2R)}}{r} \bigg ),
\end{equation}
satisfies the homogeneous boundary condition
$(\partial_t+\partial_r)(r\tilde\Phi_{Dipole}(\omega)|_R=0$ with reflection
coefficient
\begin{equation}
          \kappa_\omega =\frac{1}{2\omega^2 R^2 +2i\omega R-1}
                 \sim \frac{1}{2\omega^2 R^2} .
\label{eq:skappa}
\end{equation}

Much work has been done on formulating boundary conditions, both exact and
approximate, for linear problems in situations that are not spherically
symmetric. These
boundary conditions are given various names in the literature, e.g., absorbing
or non-reflecting. A variety of ABCs have been reported for linear problems.
See the articles~\cite{giv,Ren,tsy,ryab,jcp97} for general discussions.

Local ABCs have been extensively applied to linear problems with varying
success~\cite{Lind1,Eng77,Bay80,Tre86,Hig86,Bla88,Jia90}. Some of these
conditions are local approximations to exact integral representations of the
solution in the exterior of the computational domain~\cite{Eng77}, while others
are based on approximating the dispersion relation of the so-called one-way
wave equations~\cite{Lind1,Tre86}. Higdon~\cite{Hig86} showed that this last
approach is essentially equivalent to specifying a finite number of angles of
incidence for which the ABCs yield perfect transmission. Local ABCs have
also been derived for the linear wave equation by considering the asymptotic
behavior of outgoing solutions~\cite{Bay80}, thus generalizing the Sommerfeld
outgoing radiation condition. Although this type of ABC is relatively simple to
implement and has a low computational cost, the final accuracy is often limited
because the assumptions made about the behavior of the waves are rarely met in
practice~\cite{giv,tsy}.

The disadvantages of local ABCs have led some workers to consider exact
nonlocal boundary conditions based on integral representations of the infinite
domain problem~\cite{Tin86,giv,tsy}. Even for problems where the Green's
function is known and easily computed, such approaches were initially dismissed
as impractical~\cite{Eng77}; however, the rapid increase in computer power has
made it possible to implement exact nonlocal ABCs for the linear wave equation
and Maxwell's equations in 3D~\cite{deM,kell}. If properly implemented, this
method can yield numerical solutions to a linear problem which converge to the
exact infinite domain problem in the continuum limit, while keeping the
artificial boundary at a fixed distance. However, due to nonlocality, the
computational cost per time step usually grows at a higher power with grid size
($\mathcal{O} (N^4)$ per time step in three dimensions) than in a local
approach~\cite{giv,deM,tsy}.

The extension of ABCs to \emph{nonlinear} problems is much more difficult.
The problem is normally treated by linearizing the region between the
outer boundary and infinity, using either local or nonlocal linear
ABCs~\cite{tsy,ryab}. The neglect of the nonlinear terms in this
region introduces an unavoidable error at the analytic level. But even
larger errors are typically introduced in prescribing the outer
boundary data. This is a subtle global problem because the correct
boundary data must correspond to the continuity of fields and their
normal derivatives when extended across the boundary into the
linearized exterior. This is a clear requirement for any consistent
boundary algorithm, since discontinuities in the field or its
derivatives would otherwise act as a spurious sheet source on the
boundary, which contaminates both the interior and the exterior
evolutions. But the fields and their normal derivatives constitute an
overdetermined set of data for the boundary problem. So it
is necessary to solve a global linearized problem, not just an exterior
one, in order to find the proper data. The designation ``exact ABC'' is
given to an ABC for a nonlinear system whose only error is due to
linearization of the exterior. An exact ABC requires the use of global
techniques, such as the difference potentials method, to eliminate back
reflection at the boundary~\cite{tsy}.

There have been only a few applications of ABCs to strongly nonlinear
problems~\cite{giv}. Thompson~\cite{Tho87} generalized a previous nonlinear ABC
of Hedstrom~\cite{Hed79} to treat 1D and 2D problems in gas dynamics. These
boundary conditions performed poorly in some situations because of their
difficulty in adequately modeling the field outside the computational
domain~\cite{Tho87,giv}. Hagstrom and Hariharan~\cite{Hag88} have overcome these
difficulties in 1D gas dynamics by a clever use of Riemann invariants. They
proposed a heuristic generalization of their local ABC to 3D, but this approach
has not yet been validated.

In order to reduce the level of approximation at the analytic level, an
artificial boundary for a nonlinear problem must be placed sufficiently far from
the strong-field region. This sharply increases the computational cost in
multi-dimensional simulations~\cite{Eng77}. There is no numerical method which
converges (as the discretization is refined) to the infinite domain exact
solution of a strongly nonlinear wave problem in multi-dimensions, while keeping
the artificial boundary fixed. Attempts to use compactified Cauchy hypersurfaces
which extend the domain to spatial infinity have failed because the phase of
short wavelength radiation varies rapidly in spatial directions~\cite{Orsz}.
Characteristic evolution avoids this problem by approaching infinity along the
phase fronts.

CCM is a strategy that eliminates this nonlinear source of error. In the
simplest version of CCM, Cauchy and characteristic evolution algorithms are
pasted together in the neighborhood of a worldtube to form a global evolution
algorithm. The characteristic algorithm provides an \emph{outer} boundary
condition for the interior Cauchy evolution, while the Cauchy algorithm supplies
an \emph{inner} boundary condition for the characteristic evolution. The
matching worldtube provides the geometric framework necessary to relate the two
evolutions. The Cauchy foliation slices the worldtube into spherical
cross-sections. The characteristic evolution is based upon the outgoing null
hypersurfaces emanating from these slices, with the evolution proceeding from
one hypersurface to the next by the outward radial march described in
Section~\ref{sec:1d}. There is no need to truncate spacetime at a finite
distance from the source, since compactification of the radial null coordinate
used in the characteristic evolution makes it possible to cover the infinite
space with a finite computational grid. In this way, the true waveform may be
computed up to discretization error by the finite difference algorithm.


\subsection{The computational matching strategy}
\label{sec:3dccm}

CCM evolves a mixed spacelike-null initial value problem in which Cauchy data is
given in a spacelike hypersurface bounded by a spherical boundary $\mathcal{S}$
and characteristic data is given on a null hypersurface emanating from
$\mathcal{S}$. The general idea is not entirely new. An early mathematical
investigation combining spacelike and characteristic hypersurfaces appears in
the work of Duff~\cite{Duff}. The three chief ingredients for computational
implementation are: (i) a Cauchy evolution module, (ii) a characteristic
evolution module and, (iii) a module for matching the Cauchy and characteristic
regions across their interface. In the simplest scenario, the interface is the
timelike worldtube which is traced out by the flow of $\mathcal{S}$ along the
worldlines of the Cauchy evolution, as determined by the choice of lapse and
shift. Matching provides the exchange of data across the worldtube to allow
evolution without any further boundary conditions, as would be necessary in
either a purely Cauchy or purely characteristic evolution. Other versions of CCM
involve a finite overlap between the characteristic and Cauchy regions.

The most important application of CCM is anticipated to be the waveform and
momentum recoil in the binary black hole inspiral and merger. The 3D Cauchy
codes being applied to simulate this problem employ a single Cartesian
coordinate patch.  In principle, the application of CCM to this problem might
seem routine, tantamount to translating into finite difference form the textbook
construction of an atlas consisting of overlapping coordinate patches. In
practice, it is a complicated project. The computational strategy has been
outlined in~\cite{vishu}. The underlying algorithm consists of the following
main submodules:
\begin{itemize}
\item The \emph{boundary module} which sets the grid structures. This
  defines masks identifying which points in the Cauchy grid are to be
  evolved by the Cauchy module and which points are to be interpolated
  from the characteristic grid, and vice versa. The reference
  structures for constructing the mask is the inner characteristic
  boundary, which in the Cartesian Cauchy coordinates is the
  `` spherical'' extraction worldtube $x^2+y^2+z^2=R_E^2$, and the outer
  Cauchy boundary $x^2+y^2+z^2=R_I^2$,
  where the Cauchy boundary data is injected. The choice of lapse and
  shift for the Cauchy evolution
  governs the dynamical and geometrical properties of these worldtubes.
\item The \emph{extraction module} whose input is Cauchy grid data in
  the neighborhood of the extraction worldtube at $R_E$
  and whose output is the inner
  boundary data for the exterior characteristic evolution. This module
  numerically implements the transformation from Cartesian \{3\,+\,1\}
  coordinates to spherical null coordinates. The algorithm makes no
  perturbative assumptions and is based upon interpolations of the
  Cauchy data to a set of prescribed grid points near $R_E$. The
  metric information is then used to solve for the null geodesics
  normal to the slices of the extraction worldtube. This provides the Jacobian
  for the transformation to null coordinates in the neighborhood of
  the worldtube. The characteristic evolution module is then used to
  propagate the data from the worldtube to null infinity, where the
  waveform is calculated.
\item The \emph{injection module} which completes the interface by
  using the exterior characteristic evolution to inject the outer
  boundary data for the Cauchy evolution at $R_I$. This is the inverse of the
  extraction procedure but must be implemented with $R_I  > R_E$
  to allow for overlap between the Cauchy and characteristic domains. The
  overlap region can be constructed either to have a fixed physical
  size or to shrink to zero in the continuum limit. In the latter
  case, the inverse Jacobian describing the transformation from null
  to Cauchy coordinates can be obtained to prescribed accuracy in
  terms of an affine parameter expansion along the null geodesics
  emanating from the worldtube. The numerical stability of this element
 of the  scheme is not guaranteed.
\end{itemize}
The above strategy provides a model of how Cauchy and characteristic codes
can be pieced together as modules to form a global evolution code.

The full advantage of CCM lies in the numerical treatment of
nonlinear systems where its error converges to zero in the continuum
limit for any size outer boundary and extraction
radius~\cite{Bis,Bis2,Clarke}. For high
accuracy, CCM is also very efficient. For small
target error $\varepsilon$, it has been shown on the assumption
of unigrid codes that the relative
amount of computation required for CCM ($A_\mathrm{CCM}$) compared to that
required for a pure Cauchy calculation ($A_\mathrm{C}$) goes to zero,
$A_\mathrm{CCM}/A_\mathrm{C} \rightarrow O$ as
$\varepsilon \rightarrow O$~\cite{cce,vishu}. An important factor here
is the use of a compactified characteristic evolution, so that the
whole spacetime is represented on a finite grid. From a numerical
point of view this means that the only error made in a calculation of
the radiation waveform at infinity is the controlled error due to the
finite discretization. 

The accuracy of a Cauchy algorithm which uses an
ABC requires a large grid domain in order to avoid error from
nonlinear effects in its exterior. Improved numerical techniques,
such as the design of Cauchy grids whose resolution decreases with radius,
has improved the efficiency of this approach. Nevertheless,
the computational demands of
CCM are small since the interface problem involves one less
dimension than the evolution problem and characteristic evolution
algorithms are more efficient than Cauchy algorithms. CCM also
offers the possibility of using a small matching radius,
consistent with the requirement that it lie in the region
exterior to any caustics. This is advantageous in simulations of stellar
collapse, in which the star extends over the entire computational
grid, although it is then necessary to include the matter in the
characteristic treatment.

At present, the computational strategy of CCM is mainly the tool of numerical
relativists, who are used to dealing with dynamical coordinate systems. The
first discussion of its potential was given in~\cite{Bis} and its feasibility
has been more fully explored in~\cite{Clarke,cylinder1,cylinder2,Ccprl,harm}.
Recent work has been stimulated by the requirements of the binary black hole
problem, where CCM is one of the strategies to provide boundary conditions and
determine the radiation waveform. However, it also has inherent advantages in
dealing with other hyperbolic systems in computational physics, particularly
nonlinear 3-dimensional problems. A detailed study of the stability and accuracy
of CCM for linear and nonlinear wave equations has been presented
in~\cite{jcp97}, illustrating its potential for a wide range of problems.


\subsection{The outer Cauchy boundary in numerical relativity}
\label{sec:outercb}

A special issue arising in general relativity is whether the boundary
conditions on an artificial outer worldtube preserve the constraints.
It is typical of hyperbolic reductions of
the Einstein equations that the Hamiltonian and momentum constraints propagate
in a domain of dependence dictated by the characteristics. Unless the boundary
conditions enforce these constraints, they will be violated outside the domain
of dependence of the initial Cauchy hypersurface. This issue of a
constraint-preserving initial boundary value problem has only recently been
addressed~\cite{stewartbc}. The first fully nonlinear treatment of a well-posed
constraint preserving formulation of the Einstein initial-boundary value problem
(IBVP) has subsequently been given by Friedrich and Nagy~\cite{friednag}. Their
treatment is based upon a frame formulation in which the evolution variables are
the tetrad, connection coefficients and Weyl curvature. Although this system has
not yet been implemented computationally, it has spurred the investigation of
simpler treatments of Einstein equations which give rise to a constraint
preserving IBVP under various
restrictions~\cite{lehnercp,harm,reulacp,frittellicp,gundlachcp,ruiz,hKjW06}.
See~\cite{oS07,oscolv} for reviews.

The successful implementation of CCM for Einstein's equations requires a well-posed
initial-boundary value problem for the artificial outer boundary of the Cauchy
evolution. This is particularly cogent for dealing with waveform extraction in the simulation
of black holes by BSSN formulations. There is no well-posed outer boundary theory
for the BSSN formulation and the strategy is to place the boundary out far enough so that it
does no harm. The harmonic formulation has a simpler mathematical structure as
a system of coupled quasilinear wave equations which is more amenable to an
analytic treatment. 

Standard harmonic coordinates satisfy the covariant wave equation 
\begin{equation}
    \Gamma^\alpha=-\Box x^\alpha
   = - \frac{1}{ \sqrt{-g}} \partial_\beta \gamma^{\alpha\beta}=0 \, ,
      \quad \gamma^{\alpha\beta}=\sqrt{-g}g^{\alpha\beta} .
       \label{eq:hc}
\end{equation}
This can easily be generalized to include gauge forcing~\cite{friedred},
whereby $\Gamma^\alpha =f^\alpha (x^\beta,g^{\beta\gamma})$. For simplicity of
discussion, I will set $\Gamma^\alpha=0$, although gauge forcing is an essential
tool in simulating black holes~\cite{fP05}. 

When $\Gamma^\alpha=0$, Einstein's equations reduce to the
10 quasilinear wave equations
\begin{equation}
     g^{\mu\nu} \partial_\mu \partial_\nu
      \gamma^{\alpha\beta}  =S^{\alpha\beta} ,
		\label{eq:redhar}
\end{equation}
where $S^{\alpha\beta}$ does not enter the principle part and vanishes in the
linearized approximation. Straightforward techniques can be applied to formulate
a well-posed IBVP for the system (\ref{eq:redhar}). The catch is that Einstein's
equations are not necessarily satisfied unless the constraints are also
satisfied.

In the harmonic formalism, the constraints can be reduced to the harmonic
coordinate conditions (\ref{eq:hc}). For the resulting IBVP to be constraint
preserving, these harmonic conditions must be built into the boundary condition.
Numerous early attempts to accomplish this failed because (\ref{eq:hc}) contains
derivatives tangent to the boundary, which do not fit into the standard methods
for obtaining the necessary energy estimates. The use of pseudo-differential
techniques developed for similar problems in elasticity theory has led to the
first well-posed formulation of the IBVP for the harmonic Einstein
equations~\cite{hKjW06}. Subsequently, well-posedness
was also obtained using energy estimates by means of a novel, non-conventional choice of the
energy for the harmonic system~\cite{krreulsarw1}. 
A Cauchy evolution code, the Abigel code, based upon a discretized version
of these energy estimates was found to be stable, convergent and constraint preserving
in nonlinear boundary tests~\cite{cpsomm}.  These results were confirmed using an
independent harmonic code developed at the Albert Einstein Institute~\cite{seiler}. A
linearized version of the Abigel code has been used to successfully carry out
CCM (see Section~\ref{sec:linccm}). 

Given a well-posed IBVP, there is the additional complication of the correct specification
of boundary data. Ideally, this data would be supplied by matching to a solution
extending to infinity, e.g. by CCM. In the formulations
of~\cite{hKjW06} and~\cite{krreulsarw1}, the
boundary conditions are of the  Sommerfeld type for which homogeneous
boundary data , i.e. zero boundary values, is a good approximation
in the sense that the reflection coefficients for gravitational waves
fall off as $O(1/R^3)$ as
the  boundary radius $R\rightarrow \infty$~\cite{krreulsarw2}. A second differential
order boundary condition based upon requiring the Newman--Penrose~\cite{NP}
Weyl tensor component  $\psi_0=0$ has also been shown
to be well-posed by means of pseudo-differential techniques~\cite{ruiz}. For this
$\psi_0$ condition, the reflection coefficients fall off at an addition power of $1/R$.
In the present state of the art of black hole simulations, the $\psi_0$ condition comes
closest to a satisfactory treatment of the outer boundary~\cite{psiocit}.


\subsection{Perturbative matching schemes}

In numerous analytic studies outside of general relativity,
matching techniques have successfully cured pathologies in perturbative
expansions~\cite{aliney}. Matching is a strategy for obtaining a global
solution by patching together solutions obtained using different
coordinate systems for different regions. By adopting each coordinate
system to a length scale appropriate to its domain, a globally
convergent perturbation expansion is sometimes possible in cases where
a single coordinate system would fail.

In general relativity, Burke showed that matching could be used to eliminate
some of the divergences arising in perturbative calculations of gravitational
radiation~\cite{burke}. Kates and Kegles further showed that use of an exterior
null coordinate system in the matching scheme could eliminate problems in the
perturbative treatment of a scalar radiation field on a Schwarzschild
background~\cite{kk}. The Schwarzschild light cones have drastically different
asymptotic behavior from the artificial Minkowski light cones used in
perturbative expansions based upon a flat space Green function. Use of the
Minkowski light cones leads to \emph{nonuniformities} in the expansion of the
radiation fields which are eliminated by the use of true null coordinates in
the exterior. Kates, Anderson, Kegles, and Madonna extended this work to the
fully general relativistic case and reached the same conclusion~\cite{kakm}.
Anderson later applied this approach to the slow motion approximation of a
binary system and obtained a derivation of the radiation reaction effect on the
orbital period which avoided some objections to earlier
approaches~\cite{and87}. The use of the true light cones was also essential in
formulating as a mathematical theorem that the Bondi news function satisfies
the Einstein quadrupole formula to leading order in a Newtonian
limit~\cite{quad}. Although questions of mathematical consistency still remain
in the perturbative treatment of gravitational radiation, it is clear that the
use of characteristic methods pushes these problems to a higher perturbative
order.

One of the first computational applications of characteristic matching was a
hybrid numerical-analytical treatment by Anderson and Hobill of the test problem
of nonlinear 1D scalar waves~\cite{Hobill,Hobill2,Hobill3}. They matched an
inner numerical solution to a  far field solution which was obtained by a
perturbation expansion. A key ingredient is that the far field is solved in
retarded null coordinates $(u,r)$. Because the transformation from null
coordinates $(u,r)$ to Cauchy coordinates $(t,r)$ is known analytically for this
problem, the matching between the null and Cauchy solutions is quite simple.
Causality was enforced by requiring that the system be stationary prior to some
fixed time. This eliminates extraneous incoming radiation in a physically
correct way in a system which is stationary prior to a fixed time but it is
nontrivial to generalize, say, to the problem of radiation from an orbiting
binary.

Later, a global, characteristic, numerical study of the self-gravitating version
of this problem confirmed that the use of the true null cones is essential in
getting the correct radiated waveform~\cite{finw}. For quasi-periodic radiation,
the phase of the waveform is particular sensitive to the truncation of the outer
region at a finite boundary. Although a perturbative estimate would indicate an
$\mathcal{O} (M/R)$ error, this error accumulates over many cycles to produce an
error of order $\pi$ in the phase.

Anderson and Hobill proposed that their method be extended to general
relativity by matching a numerical solution to an analytic $1/r$ expansion in
null coordinates. Most perturbative-numerical matching schemes that have been
implemented in general relativity have been based upon perturbations of a
Schwarzschild background using the standard Schwarzschild time
slicing~\cite{ab1,ab2,ab3,all1,rupright,rezzmatz,nagar}. It would be
interesting to compare results with an analytic-numeric matching scheme based
upon the true null cones. Although the full proposal by Anderson and Hobill
has not been carried out, characteristic techniques have been
used~\cite{lous,zlochadv,zlochret} to study the radiation content of numerical
solutions by treating the far field as a perturbation of a Schwarzschild
spacetime. 

Most metric based treatments of gravitational radiation are based upon
perturbations of the Schwarzschild metric and solve the underlying
Regge--Wheeler~\cite{regge} and 
~\cite{zerilli} equations using
traditional spacelike Cauchy hypersurfaces. At one level, these approaches
\emph{extract} the radiation from a numerical solution in a region with outer
boundary $\mathcal{B}$ by using data on an inner worldtube $\mathcal{W}$ to
construct the perturbative solution. Ambiguities are avoided by use of
Moncrief's gauge invariant perturbation quantities~\cite{moncrief}. For this to
work, $\mathcal{W}$  must not only be located in the far field, i.e.\ many
wavelengths from the source, but, because of the lack of proper outer boundary
data, it is necessary that the boundary $\mathcal{B}$  be sufficiently far
outside $\mathcal{W}$ so that the extracted radiation is not contaminated by
back-reflection for some significant window of time.  This poses extreme
computational requirements in a 3D problem. This extraction strategy has also
been carried out using characteristic evolution in the exterior of $\mathcal{W}$
instead of a perturbative solution, i.e.\ Cauchy-characteristic
extraction~\cite{cce} (see Section~\ref{sec:cce}).

A study by Babiuc, Szil{\'{a}}gyi, Hawke, and Zlochower carried out  in the
perturbative regime~\cite{babiuc} shows that CCE compares favorably with Zerilli
extraction and has advantages at small extraction radii.  When the extraction
worldtube is sufficiently large, e.g.\ $r=200\lambda$, where $\lambda$ is the
characteristic wavelength of the radiation, the Zerilli and CCE methods both
give excellent results. However, the accuracy of CCE remains unchanged at small
extraction radii, e.g.\ $r=10\lambda$, whereas the Zerilli approach shows error
associated with near zone effects. This flexibility to apply CCE to small extraction
radii has proved advantageous in the simulations of stellar
collapse~\cite{reis3,collapsar} discussed in Sec.~\ref{sec:stelcoll}.

The contamination of the extracted radiation by back-reflection can only be
completely eliminated by matching to an exterior solution which \emph{injects} the
physically appropriate boundary data on $\mathcal{W}$. Cauchy-perturbative
matching~\cite{rupright,rezzmatz} has been implemented using the same modular
structure described for CCM in Section~\ref{sec:3dccm}. Nagar and
Rezzolla~\cite{nagar} have given a  review of this approach. At present,
perturbative matching and CCM share the common problem of long term stability
of the outer Cauchy boundary in 3D applications.


\subsection{Cauchy-characteristic matching for 1D gravitational systems}
\label{sec:1dccm}

The first numerical implementations of CCM were 1D feasibility studies.
These model problems provided a controlled environment for the
development of CCM, in which either exact solutions or independent numerical
solutions were known. In the following studies CCM worked like
a charm in a variety of 1D applications, i.e.\ the matched evolutions were
essentially transparent to the presence of the interface.


\subsubsection{Cylindrical matching}
\label{sec:cylmatch}

The Southampton group chose cylindrically symmetric systems as their model
problem for developing matching techniques. In preliminary work, they showed
how CCM could be consistently carried out for a scalar wave evolving in
Minkowski spacetime but expressed in a nontrivial cylindrical coordinate
system~\cite{Clarke}.

They then tackled the gravitational problem. First they set up the analytic
machinery necessary for investigating cylindrically symmetric vacuum
spacetimes~\cite{cylinder1}. Although the problem involves only one spatial
dimension, there are two independent modes of polarization. The Cauchy metric
was treated in the Jordan--Ehlers--Kompaneets canonical form, using coordinates
$(t,r,\phi,z)$ adapted to the $(\phi,z)$ cylindrical symmetry. The advantage
here is that $u=t-r$ is then a null coordinate which can be used for the
characteristic evolution. They successfully recast the equations in a suitably
regularized form for the compactification of $\mathcal{I}^+$ in terms of the
coordinate $y=\sqrt{1/r}$. The simple analytic relationship between Cauchy
coordinates $(t,r)$ and characteristic coordinates $(u,y)$ facilitated the
translation between Cauchy and characteristic variables on the matching
worldtube, given by $r = \mathrm{const}$.

Next they implemented the scheme as a numerical code. The interior Cauchy
evolution was carried out using an unconstrained leapfrog scheme. It is notable
that they report no problems with instability, which have arisen in other
attempts at unconstrained leapfrog evolution in general relativity. The
characteristic evolution also used a leapfrog scheme for the evolution between
retarded time levels $u$, while numerically integrating the hypersurface
equations outward along the characteristics.

The matching interface was located at points common to both the Cauchy and
characteristic grids. In order to update these points by Cauchy evolution, it
was necessary to obtain field values at the Cauchy ``ghost'' points which lie
outside the worldtube in the characteristic region. These values were obtained
by interpolation from characteristic grid points (lying on three levels of null
hypersurfaces in order to ensure second order accuracy). Similarly, the
boundary data for starting up the characteristic integration was obtained by
interpolation from Cauchy grid values inside the worldtube.

The matching code was first tested~\cite{cylinder2} using exact Weber--Wheeler
cylindrical waves~\cite{wweb}, which come in from $\mathcal{I}^-$, pass through
the symmetry axis and expand out to $\mathcal{I}^+$. The numerical errors were
oscillatory with low growth rate, and second order convergence was confirmed.
Of special importance, little numerical noise was introduced by the interface.
Comparisons of CCM  were made with Cauchy evolutions using a standard outgoing
radiation boundary condition~\cite{piran}. At high amplitudes the standard
condition developed a large error very quickly and was competitive only for
weak waves with a large outer boundary. In contrast, the matching code
performed well even with a small matching radius. Some interesting simulations
were presented in which an outgoing wave in one polarization mode collided with
an incoming wave in the other mode, a problem studied earlier by pure Cauchy
evolution~\cite{stark}. The simulations of the collision were qualitatively
similar in these two studies.

The Weber--Wheeler waves contain only one gravitational degree of freedom. The
code was next tested~\cite{south3} using exact cylindrically symmetric
solutions, due to Piran, Safier, and Katz~\cite{katz}, which contain both
degrees of freedom. These solutions are singular at $\mathcal{I}^+$ so that the
code had to be suitably modified. Relative errors of the various metric
quantities were in the range 10\super{-4} to 10\super{-2}. The convergence rate of
the numerical solution starts off as second order but diminishes to first order
after long time evolution. This performance could perhaps be improved by
incorporating subsequent improvements in the characteristic code made by
Sperhake, Sj\"odin, and Vickers (see Section~\ref{sec:1d}).


\subsubsection{Spherical matching}

A joint collaboration between groups at Pennsylvania State University and the
University of Pittsburgh applied CCM to the EKG
system with spherical symmetry~\cite{ekgmat}. This model problem
allowed simulation of black hole formation as well as wave
propagation.

The geometrical setup is analogous to the cylindrically symmetric problem.
Initial data were specified on the union of a spacelike hypersurface and a null
hypersurface. The evolution used a 3-level Cauchy scheme in the interior and a
2-level characteristic evolution in the compactified exterior. A constrained
Cauchy evolution was adopted because of its earlier success in accurately
simulating scalar wave collapse~\cite{choptprl}. Characteristic evolution was
based upon the null parallelogram algorithm~(\ref{eq:integral}). The
matching between the Cauchy and characteristic foliations was achieved by
imposing continuity conditions on the metric, extrinsic curvature and scalar
field variables, ensuring smoothness of fields and their derivatives across the
matching interface. The extensive analytical and numerical studies of this
system in recent years aided the development of CCM in this non-trivial
geometrical setting by providing basic knowledge of the
expected physical and geometrical behavior, in the absence of exact solutions.

The CCM code accurately handled wave propagation and
black hole formation for all values of $M/R$ at the matching radius,
with no symptoms of instability or back-reflection. Second order
accuracy was established by checking energy conservation.


\subsubsection{Excising 1D black holes}
\label{sec:1Dexcis}

In further developmental work on the EKG model, the Pittsburgh group used CCM
to formulate a new treatment of the inner Cauchy boundary for a black hole
spacetime~\cite{excise}. In the excision strategy, the inner boundary of
the Cauchy evolution is located at an apparent horizon, which must lie inside
(or on) the event horizon~\cite{wald1984}. The physical rationale behind this apparent
horizon boundary condition is that the truncated region of
spacetime cannot causually affect the gravitational
waves radiated to infinity.  However, it should be noted that many Cauchy
formalisms contain superluminal gauge or constraint violating
modes so that this strategy is not always fully justified.

In the CCM excision strategy, illustrated in Figure~\ref{fig:1dexci}, the interior
black hole region is evolved using an \emph{ingoing} null algorithm whose inner
boundary is a marginally trapped surface, and whose outer boundary lies outside
the black hole and forms the inner boundary of a region evolved by the Cauchy
algorithm. In turn, the outer boundary of the Cauchy region is handled by
matching to an outgoing null evolution extending to $\mathcal{I}^+$. Data are
passed between the inner characteristic and central Cauchy regions using a CCM
procedure similar to that already described for an outer Cauchy boundary. The
main difference is that, whereas the outer Cauchy boundary data is induced
from the Bondi metric on an outgoing null hypersurface, the inner Cauchy
boundary is now obtained from an ingoing null hypersurface which enters the
event horizon and terminates at a marginally trapped surface.

\epubtkImage{1dexci.png}{%
  \begin{figure}[htbp]
    \centerline{\includegraphics[width=0.7\textwidth]{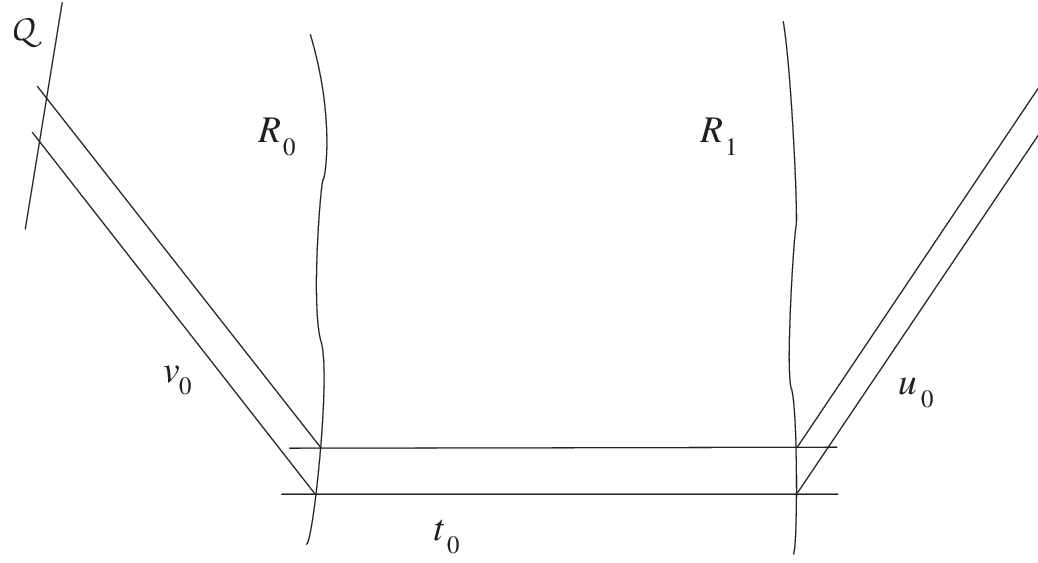}}
    \caption{Black hole excision by matching. A Cauchy evolution,
      with data at $t_0$ is matched across worldtubes $R_0$ and $R_1$
      to an ingoing null evolution, with data at $v_0$, and an
      outgoing null evolution, with data at $u_0$. The ingoing null
      evolution extends to an inner trapped boundary $Q$, and the
      outgoing null evolution extends to $\mathcal{I}^+$.}
    \label{fig:1dexci}
  \end{figure}}

The translation from an outgoing to an incoming null evolution algorithm can be
easily carried out. The substitution $\beta\rightarrow \beta+i\pi/2$ in the 3D
version of the Bondi metric~(\ref{eq:metric}) provides a simple formal recipe
for switching from an outgoing to an ingoing null formalism~\cite{excise}.

In order to ensure that trapped surfaces exist on the ingoing
null hypersurfaces, initial data were chosen which guarantee black hole
formation. Such data can be obtained from initial Cauchy data for a
black hole. However, rather than extending the Cauchy hypersurface
inward to an apparent horizon, it was truncated sufficiently far outside
the apparent horizon to avoid computational problems with the Cauchy
evolution. The initial Cauchy data were then extended into the black hole
interior as initial null data until a marginally trapped surface was
reached. Two ingredients were essential in order to arrange this.
First, the inner matching surface must be chosen to be convex, in the
sense that its outward null normals uniformly diverge and its inner
null normals uniformly converge. (This is trivial to satisfy in the
spherically symmetric case.) Given any physically reasonable matter
source, the focusing theorem guarantees that the null rays
emanating inward from the matching sphere continue to converge until
reaching a caustic. Second, the initial null data must lead to a
trapped surface before such a caustic is encountered. This is a
relatively easy requirement to satisfy because the initial null data
can be posed freely, without any elliptic or algebraic constraints
other than continuity with the Cauchy data.

A code was developed which implemented CCM at both the inner and outer
boundaries~\cite{excise}. Its performance showed that CCM provides as good a
solution to the black hole excision problem in spherical symmetry as any
previous treatment~\cite{scheel1995a,scheel1995b,mc,anninos1995}. CCM is
computationally more efficient than these pure Cauchy approaches (fewer
variables) and much easier to implement. Depending upon the Cauchy formalism
adopted, achieving stability with a pure Cauchy scheme in the region of an
apparent horizon can be quite tricky, involving much trial and error in
choosing finite difference schemes. There were no complications with stability
of the null evolution at the marginally trapped surface.

The Cauchy evolution was carried out in ingoing Eddington--Finklestein
(IEF) coordinates. The initial Cauchy data consisted of a
Schwarzschild black hole with an ingoing Gaussian pulse of scalar
radiation. Since IEF coordinates are based on ingoing null cones, it
is possible to construct a simple transformation between the IEF Cauchy
metric and the ingoing null metric. Initially there was no scalar field
present on either the ingoing or outgoing null patches. The initial
values for the Bondi variables $\beta$ and $V$ were determined by
matching to the Cauchy data at the matching surfaces and integrating
the hypersurface equations~(\ref{eq:sbeta}, \ref{eq:sv}).

As the evolution proceeds, the scalar field passes into the black hole, and the
marginally trapped surface (MTS) grows outward. The MTS is easily located
in the spherically symmetric case by an algebraic equation. In order to excise
the singular region, the grid points inside the marginally trapped surface were
identified and masked out of the evolution. The backscattered radiation
propagated cleanly across the outer matching surface to $\mathcal{I}^+$. The
strategy worked smoothly, and second order accuracy of the approach was
established by comparing it to an independent numerical solution obtained using
a second order accurate, purely Cauchy code~\cite{mc}. As discussed in
Section~\ref{sec:bbhib}, this inside-outside application of CCM has potential
application to the binary black hole problem.

In a variant of this double CCM matching scheme, Lehner~\cite{luis2m} has
eliminated the middle Cauchy region between $R_0$ and $R_1$
in Figure~\ref{fig:1dexci}. He constructed a 1D code matching the
ingoing and outgoing characteristic evolutions directly across a single
timelike worldtube. In this way, he was able to simulate the global problem of a
scalar wave falling into a black hole by purely characteristic methods.


\subsection{Axisymmetric Cauchy-characteristic matching}
\label{sec:aximatch}

The Southampton CCM project is being carried out for spacetimes with (twisting)
axial symmetry. The formal basis for the matching scheme was developed by
d'Inverno and Vickers~\cite{south1,south2}. Similar to the Pittsburgh 3D
strategy (see Section~\ref{sec:3dccm}), matching is based upon an extraction
module, which supplies boundary data for the exterior characteristic evolution,
and an injection module, which supplies boundary data for the interior Cauchy
evolution. However, their use of spherical coordinates for the Cauchy evolution
(as opposed to Cartesian coordinates in the 3D strategy) allows use of a
matching worldtube $r=R_\mathrm{m}$ which lies simultaneously on Cauchy and
characteristic gridpoints. This tremendously simplifies the necessary
interpolations between the Cauchy and characteristic evolutions, at the
expense of dealing with the $r=0$ coordinate singularity in the Cauchy
evolution. The characteristic code (see Section~\ref{sec:axiev}) is based upon a
compactified Bondi--Sachs formalism. The use of a ``radial'' Cauchy gauge, in
which the Cauchy coordinate $r$ measures the surface area of spheres,
simplifies the relation to the Bondi--Sachs coordinates. In the numerical
scheme, the metric and its derivatives are passed between the Cauchy and
characteristic evolutions exactly at $r=R_\mathrm{m}$, thus eliminating the need of a
matching interface encompassing a few grid zones, as in the 3D Pittsburgh
scheme. This avoids a great deal of interpolation error and computational
complexity.

Preliminary results in the development of the Southampton CCM code are
described by Pollney in his thesis~\cite{pollney}. The Cauchy code was based
upon the axisymmetric ADM code of Stark and Piran~\cite{starkpir} and
reproduces their vacuum results for a short time period, after which an
instability at the origin becomes manifest. The characteristic code has been
tested to reproduce accurately the Schwarzschild and boost-rotation symmetric
solutions~\cite{boostrot}, with more thorough tests of stability and accuracy
still to be carried out.


\subsection{Cauchy-characteristic matching for 3D scalar waves}
\label{sec3dsccm}

CCM has been successfully implemented in the fully 3D problem of
nonlinear scalar waves evolving in a flat spacetime~\cite{jcp97,Ccprl}.
This study demonstrated the feasibility of
matching between Cartesian Cauchy coordinates and spherical null
coordinates, the setup required to apply CCM to the binary
black hole problem. Unlike spherically or cylindrically symmetric
examples of matching, the
Cauchy and characteristic patches do not share a common coordinate
which can be used to define the matching interface. This introduces a
major complication into the matching procedure, resulting in extensive
use of intergrid interpolation. The accompanying short wavelength
numerical noise presents a challenge in obtaining a stable
algorithm.

The nonlinear waves were modeled by the equation
\begin{equation}
  c^{-2} \partial_{t}^2 \Phi = \nabla^{2} \Phi + F (\Phi) + S (t,x,y,z),
  \label{eq:swe}
\end{equation}
with self-coupling $F(\Phi)$ and external source $S$. The initial
Cauchy data $\Phi(t_0,x,y,z)$ and $\partial_t\Phi(t_0,x,y,z)$ are
assigned in a spatial region bounded by a spherical matching surface of
radius $R_\mathrm{m}$.

The characteristic initial value problem~(\ref{eq:swe}) is expressed in
standard spherical coordinates $(r,\theta,\varphi)$ and retarded time
$u=t-r+R_\mathrm{m}$:
\begin{equation}
  2 \partial_{u} \partial_{r}g =
  \partial_{r}^2 g - \frac{L^2 g}{r^2} + r (F + S),
  \label{eq:SWE}
\end{equation}
where $g = r\Phi$ and $L^2$ is the angular momentum operator
\begin{equation}
  L^2 g = - \frac{\partial_\theta (\sin \theta \, \partial_{\theta} g)}
  {\sin \theta} - \frac{\partial_\varphi^2 g}{\sin^2 \theta}.
\end{equation}
The initial null data consist of $g(r,\theta,\varphi,u_0)$ on the outgoing
characteristic cone $u_0 =t_0$ emanating at the initial Cauchy time from the
matching worldtube at $r= R_\mathrm{m}$.

CCM was implemented so that, in the continuum limit, $\Phi$ and its normal
derivatives would be continuous across the matching interface. The use of a
Cartesian discretization in the interior and a spherical discretization in the
exterior complicated the treatment of the interface. In particular, the
stability of the matching algorithm required careful attention to the details
of the intergrid matching. Nevertheless, there was a reasonably broad range
of discretization parameters for which CCM was stable.

Two different ways of handling the spherical coordinates were used. One was
based upon two overlapping stereographic grid patches and the other upon a
multiquadric approximation using a quasi-regular triangulation of the sphere.
Both methods gave similar accuracy. The multiquadric method showed a slightly
larger range of stability. Also, two separate tactics were used to implement
matching, one based upon straightforward interpolations and the other upon
maintaining continuity of derivatives in the outward null direction (a
generalization of the Sommerfeld condition). Both methods were stable for a
reasonable range of grid parameters. The solutions were second order accurate
and the Richardson extrapolation technique could be used to accelerate
convergence.

The performance of CCM was compared to traditional ABCs. As expected, the
nonlocal ABCs yielded convergent results only in linear problems, and
convergence was not observed for local ABCs, whose restrictive assumptions
were violated in all of the numerical experiments. The computational cost of
CCM was much lower than that of current nonlocal boundary conditions. In
strongly nonlinear problems, CCM appears to be the only available method which
is able to produce numerical solutions which converge to the exact solution
with a fixed boundary.


\subsection{Stable  3D linearized Cauchy-characteristic matching}
\label{sec:linccm}

Although the individual pieces of the CCM module have been calibrated to give
an accurate interface between Cauchy and characteristic evolution
modules in 3D general relativity, its stability has not yet been
established~\cite{vishu}. However, a stable version of CCM for linearized
gravitational theory has recently been demonstrated~\cite{harm}. The Cauchy
evolution is carried out using a harmonic formulation for which the reduced
equations have a well-posed initial-boundary problem. Previous attempts at CCM
were plagued by boundary induced instabilities of the Cauchy code. Although
stable behavior of the Cauchy boundary is only a necessary and not a
sufficient condition for CCM, the tests with the linearized harmonic code
matched to a linearized characteristic code were successful.

The harmonic conditions consist of wave equations for the coordinates which can
be used to propagate the gauge as four scalar waves using characteristic
evolution. This allows the extraction worldtube to be placed at a finite
distance from the injection worldtube without introducing a gauge ambiguity.
Furthermore, the harmonic gauge conditions are the only constraints on the
Cauchy formalism so that gauge propagation also insures constraint propagation.
This allows the Cauchy data to be supplied in numerically benign Sommerfeld
form,  without introducing constraint violation. Using random initial data,
robust stability of the CCM algorithm was confirmed for 2000 crossing times on a
45\super{3} Cauchy grid. Figure~\ref{fig:2D.ccm} shows a sequence of profiles  of the
metric component $\gamma^{xy}=\sqrt{-g}g^{xy}$ as a linearized wave propagates
cleanly through the spherical injection boundary and passes to the
characteristic grid, where it is propagated to $\mathcal{I}^+$.

\epubtkImage{hxyall.png}{%
  \begin{figure}[htbp]
    \centerline{\includegraphics[width=0.5\textwidth]{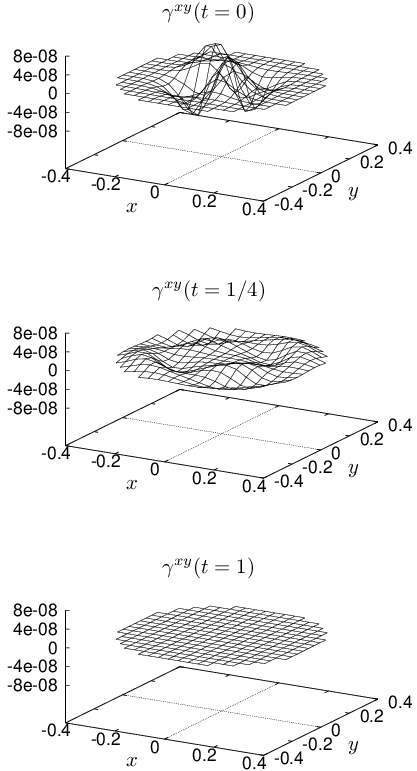}}
    \caption{Sequence of slices of the metric component
      $\gamma^{xy}$, evolved with the linear matched
      Cauchy-characteristic code. In the last snapshot, the wave has
      propagated cleanly onto the characteristic grid with negligible
      remnant noise.}
    \label{fig:2D.ccm}
  \end{figure}}


\subsection{The binary black hole inner boundary}
\label{sec:bbhib}

CCM also offers a new approach to singularity  excision in the binary black hole
problem in the manner described in Section~\ref{sec:1Dexcis} for a single
spherically symmetric black hole. In a binary system, there are computational
advantages in posing the Cauchy evolution in a frame which is co-rotating with
the orbiting black holes.  In this co-orbiting description, the Cauchy evolution
requires an inner boundary condition inside the black holes and also an outer
boundary condition on a worldtube outside of which the grid rotation is likely
to be superluminal. An outgoing characteristic code can routinely handle such
superluminal gauge flows in the exterior~\cite{high}. Thus, successful
implementation of CCM could solve the exterior boundary problem for this
co-orbiting description.

CCM also has the potential to handle the two black holes inside the Cauchy
region. As described earlier with respect to Figure~\ref{fig:1dexci}, an ingoing
characteristic code can evolve a moving black hole with long term
stability~\cite{excise,wobb}. This means that CCM might also be able to provide
the inner boundary condition for Cauchy evolution once stable matching has been
accomplished. In this approach, the interior boundary of the Cauchy evolution is
located \emph{outside} the apparent horizon and matched to a characteristic
evolution based upon ingoing null cones. The inner boundary for the
characteristic evolution is a trapped or marginally trapped surface, whose
interior is excised from the evolution.

In addition to restricting the Cauchy evolution to the region outside the black
holes, this strategy offers several other advantages. Although finding a
marginally trapped surface on the ingoing null hypersurfaces remains an
elliptic problem, there is a natural radial coordinate system $(r,\theta,\phi)$
to facilitate its solution. Motion of the black hole through the grid reduces
to a one-dimensional radial problem, leaving the angular grid intact and thus
reducing the computational complexity of excising the inner singular region.
(The angular coordinates can even rotate relative to the Cauchy coordinates in
order to accommodate spinning black holes.) The chief danger in this approach
is that a caustic might be encountered on the ingoing null hypersurface before
entering the trapped region. This is a gauge problem whose solution lies in
choosing the right location and geometry of the surface across which the Cauchy
and characteristic evolutions are matched. There is a great deal of flexibility
here because the characteristic initial data can be posed without constraints.

This global strategy is tailor-made to treat two black holes in the co-orbiting
gauge, as illustrated in Figure~\ref{fig:canbbh}. Two disjoint characteristic
evolutions based upon ingoing null cones are matched across worldtubes to
a central Cauchy region. The interior boundaries of each of these interior
characteristic regions border a trapped surface. At the outer boundary of
the Cauchy region, a matched characteristic evolution based upon outgoing null
hypersurfaces propagates the radiation to infinity.

\epubtkImage{canbbh.png}{%
  \begin{figure}[htbp]
    \centerline{\includegraphics[width=0.5\textwidth]{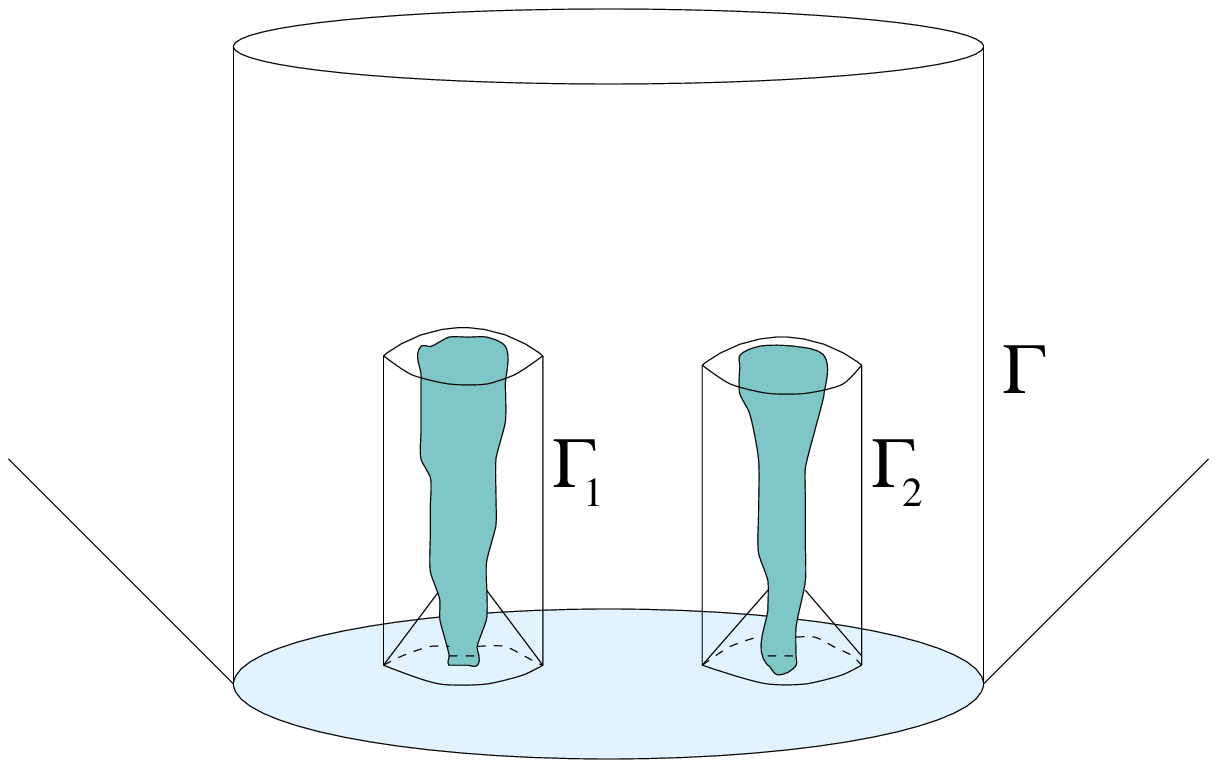}}
    \caption{CCM for binary black holes, portrayed in a
      co-rotating frame. The Cauchy evolution is matched across two
      inner worldtubes $\Gamma_1$ and $\Gamma_2$ to two ingoing null
      evolutions whose inner boundaries excise the individual black
      holes. The outer Cauchy boundary is matched across the worldtube
      $\Gamma$ to an outgoing null evolution extending to
      $\mathcal{I}^+$.}
    \label{fig:canbbh}
  \end{figure}}

Present characteristic and Cauchy codes can handle the individual pieces of this
problem. Their unification offers  a new approach to simulating the inspiral and
merger of two black holes. The individual pieces of the fully nonlinear CCM
module, as outlined in Section~\ref{sec:3dccm}, have been implemented and tested
for accuracy. The missing ingredient is long term stability in the nonlinear
gravitational case, which would open the way to future applications.



\section{Cauchy-Characteristic Extraction of Waveforms}
\label{sec:cce}

When an artificial finite outer boundary is introduced there are two broad
sources of error:
\begin{itemize}

\item The outer boundary condition

\item Waveform extraction at a finite inner worldtube.
\end{itemize}

CCM addresses both of these items. Cauchy-characteristic extraction (CCE), which
is one of the pieces of the CCM strategy, offers a means to avoid the second
source of error introduced by extraction at a finite worldtube.  In current
codes used to simulate black holes, the waveform is extracted at an interior
worldtube which must be sufficiently far inside  the outer boundary in order to
isolate it from errors introduced by the boundary condition. At this inner worldtube,
the waveform is extracted by a perturbative scheme based upon the introduction of a
background Schwarzschild spacetime. This has been carried out using the 
Regge--Wheeler--Zerilli~\cite{regge,zerilli} treatment of the perturbed metric,
as reviewed in~\cite{nagar}, and also by calculating the Newman--Penrose Weyl
component $\Psi_4$, as first done for the binary black hole problem
in~\cite{bakcamluo,fP05,mCcLpMyZ06,bakcenchokopvme}. In these approaches, errors
arise from the finite size of the extraction worldtube, from nonlinearities and
from gauge ambiguities involved in the arbitrary introduction of a background
metric. The gauge ambiguities might seem less severe in the case of $\Psi_4$ (vs
metric) extraction, but there are still delicate problems associated with the
choices of a preferred null tetrad and  preferred worldlines along which to
measure the waveform (see~\cite{lehmor} for an analysis).

CCE offers a means to avoid this error introduced by extraction at a finite
worldtube. In CCE, the inner worldtube data supplied by the Cauchy evolution is
used as boundary data for a characteristic evolution to future null infinity,
where the waveform can be unambiguously computed in terms of the Bondi news
function. By itself, CCE does not use the characteristic evolution to inject
outer boundary data for the Cauchy evolution, which can be a source of
instability in full CCM. A wide number of highly nonlinear tests involving black
holes~\cite{cce,high,Zlochower,zlochmode} have shown that
early versions of CCE were a stable procedure
which provided the gravitational waveform up to numerical error which is second
order convergent when the worldtube data is prescribed in analytic form.
Nevertheless, in nonlinear applications requiring numerical worldtube data
and high resolution, such as the inspiral of matter into a black
hole~\cite{particle}, the numerical error  was a troublesome factor in computing
the waveform. The CCE modules were first developed in a past period when
stability was the dominant issue and second order accuracy was considered
sufficient. Only recently have they  begun to be updated to include the more
accurate techniques now standard in Cauchy codes. There are two distinct ways,
geometric and numerical, that the accuracy of  CCE might be improved. In the
geometrical category, one option is to  compute $\Psi_4$ instead of the news
function as the primary description of the waveform. In the numerical category,
some standard methods for improving accuracy, such as  higher order finite
difference approximations, are straightforward to implement whereas others, such
as adaptive mesh refinement, have only been tackled for 1D characteristic
codes~\cite{pretlehn}. 

A major source of numerical error in characteristic evolution arises from the
intergrid interpolations arising from the multiple patches necessary to
coordinatize the spherical cross-sections of the outgoing null hypersurfaces.
More accurate methods, have now been developed to reduce this interpolation error, as
discussed in Section~\ref{sec:sphercoor}. In particular, the cubed-sphere method
and the stereographic method with circular patch boundaries have both shown
improvement over the original use of square stereographic patches. 
In a test problem involving a scalar
wave $\Phi$, the accuracies of the circular-stereographic and cubed-sphere
methods were compared~\cite{strateg}. For equivalent computational expense, the
cubed-sphere error in the scalar field ${\cal E}(\Phi)$ was $\approx
\frac{1}{3}$ the circular-stereographic error but the advantage was smaller for
the higher $\eth$-derivatives (angular derivatives) required in gravitational
waveform extraction. The cubed-sphere error ${\cal E}(\bar \eth \eth^2\Phi)$ was
$\approx \frac{4}{5}$ the stereographic error. However, the cubed-sphere method
has not yet been developed for extraction of gravitational waveforms at $\mathcal{I}^+$.

\subsection{Waveforms at null infinity}
\label{sec:conf}

In order to appreciate why waveforms are not easy to extract accurately
it is worthwhile to review the calculation of the required asymptotic
quantities. A simple approach to Penrose compactification is by
introducing an inverse surface area coordinate $\ell=1/r$, so that future null
infinity  $\mathcal{I}^+$ is given by $\ell=0$~\cite{tam}. In the resulting
$x^\mu=(u,\ell,x^A)$ Bondi coordinates, where $u$ is the retarded time
defined on the outgoing null hypersurfaces and $x^A$ are angular coordinates
along the outgoing null rays, the physical space-time metric $g_{\mu\nu}$ has
conformal compactification $\hat g_{\mu\nu}=\ell^{2} g_{\mu\nu}$
of the form
\begin{equation}
   \hat g_{\mu\nu}dx^\mu dx^\nu= 
           -\alpha du^2
        +2e^{2\beta}dud\ell -2 h_{AB}U^Bdudx^A + h_{AB}dx^Adx^B ,
   \label{eq:lmet}
\end{equation}
where $\alpha$, $\beta$, $U^A$ and $h_{AB}$ are smooth fields at 
$\mathcal{I}^+$. 

The news function and Weyl component $\Psi_4$, which describe the radiation, are
constructed from the leading coefficients in an expansion of $\hat g_{\mu\nu}$
in powers of $\ell$. The requirement of asymptotic flatness imposes relations
between these expansion coefficients. In terms of the Einstein tensor $\hat
G_{\mu\nu}$ and covariant derivative $\hat \nabla_\mu$ associated with $\hat
g_{\mu\nu}$, the vacuum Einstein equations become 
\begin{equation}
     -\ell^2 \hat G_{\mu\nu} =2\ell (\hat\nabla_\mu \hat\nabla_\nu \ell
           -\hat g_{\mu\nu} \hat \nabla^\alpha \hat\nabla_\alpha \ell ) 
          +3\hat g_{\mu\nu} (\hat\nabla^\alpha \ell) \hat\nabla_\alpha \ell .
	  \label{eq:einstein}
\end{equation}
Asymptotic flatness immediately implies that $\hat g^{\ell \ell}
=(\hat\nabla^\alpha \ell) \hat\nabla_\alpha \ell =O(\ell)$ so that
$\mathcal{I}^+$ is a null hypersurface with generators in the $\hat \nabla^\mu
\ell$ direction. From (\ref{eq:einstein}) there also follows the existence of a
smooth trace-free field $\hat\Sigma_{\mu\nu}$ defined on
$\mathcal{I}^+$ by
\begin{equation}
     \hat \Sigma_{\mu\nu} := \lim_{\ell\rightarrow 0} \frac{1}{\ell}
      (\hat \nabla_\mu \hat\nabla_\nu \ell
          -\frac{1}{4}\hat g_{\mu\nu}\hat\Theta) ,
\label{eq:Sigma}
\end{equation}
where $\hat\Theta:=\hat\nabla^\mu \hat\nabla_\mu \ell$ is the expansion of
$\mathcal{I}^+$. The expansion $\hat\Theta$ depends upon the conformal factor
used to compactify $\mathcal{I}^+$. In an \emph{inertial} conformal Bondi frame,
tailored to a standard Minkowski metric at $\mathcal{I}^+$, $\hat\Theta =0$. But
this is not the case for the computational frame used in characteristic
evolution, which is determined by conditions on the inner extraction worldtube.

The gravitational waveform depends on $\hat\Sigma_{\mu\nu}$, which in turn
depends on the leading terms in the expansion of $\hat g_{\mu\nu}$:
\begin{equation}
   h_{AB}= H_{AB}+\ell c_{AB}+O(\ell^2), \quad \beta = H +O(\ell), \quad
              U^A =L^A+O(\ell) .
\end{equation}
In an inertial conformal Bondi frame, $H^{AB}=Q^{AB}$ (the unit sphere metric),
$H=L^A=0$ and the Bondi news function reduces to the simple form
\begin{equation}
    N={1\over 4}Q^A Q^B \partial_u c_{AB},
     \label{eq:inews}
\end{equation}
where $Q^A$ is a complex polarization dyad on the unit sphere, i.e.\ $Q^{AB}=
Q^{(A} \bar Q^{B)}$. The spin rotation freedom $Q^\beta \rightarrow e^{-i\gamma}
Q^\beta$ is fixed by parallel propagation along the generators of 
$\mathcal{I}^+$, so that the real and imaginary parts of $N$  correctly describe
the $\oplus$ and $\otimes$ polarization modes of inertial observers at
$\mathcal{I}^+$. 

However, in the computational frame the news function has the more complicated
form
\begin{equation}
 N ={1\over 2}Q^{\alpha} Q^{\beta} \bigg (
  \hat \Sigma_{\alpha\beta}
   -\omega{\hat \nabla}_{\alpha} {\hat \nabla}_{\beta} {1\over \omega}
     +\frac{1}{\omega}(\partial_\ell \hat g_{\alpha\beta} )
         (\hat\nabla^\mu \ell) \hat\nabla_\mu \omega \bigg ),
     \label{eq:lnews}
\end{equation}
where $\omega$ is the conformal factor relating $H_{AB}$ to the unit sphere
metric, i.e.\ $Q_{AB}=\omega^2 H_{AB}$. The conformal factor obeys  the elliptic
equation governing the conformal transformation relating the  metric of the
cross-sections of $\mathcal{I}^+$  to the unit sphere metric,
\begin{equation}
     {\cal R}=2(\omega^2+H^{AB}D_A D_B \log \omega),
\label{eq:conf}
\end{equation}
where ${\cal R}$ is the curvature scalar and $D_A$ the covariant derivative
associated with $H_{AB}$. By first solving (\ref{eq:conf}) at the initial
retarded time, $\omega$ can then be determined at later times by evolving it
according to the asymptotic relation
\begin{equation}
     {\hat n}^{\alpha} \partial_{\alpha} \log \omega
    =-\frac{1}{2}e^{-2H}D_AL^A, \quad  {\hat n}^{\alpha} =\hat \nabla^\alpha \ell.
\label{eq:omegadot}
\end{equation}
All of these procedures introduce numerical error which presents a
a challenge for computational accuracy, especially because of the
appearance of second angular derivatives of $\omega$ in the news function
(\ref{eq:lnews}).
 
Similar complications appear in $\Psi_4$ extraction. Asymptotic flatness implies
that the Weyl tensor vanishes at $\mathcal{I}^+$, i.e.\ $\hat
C_{\mu\nu\rho\sigma}=O(\ell)$. This is the conformal space statement of the
peeling property~\cite{Penrose}. Let $(\hat n^\mu, \hat \ell^\mu, \hat m^\mu)$
be an orthonormal null tetrad such that $\hat n^\mu=\hat \nabla^\mu \ell$ and 
$\hat \ell^\mu \partial_\mu=\partial_\ell$ at  $\mathcal{I}^+$. Then the
radiation is described by the limit 
\begin{equation}
      \hat \Psi:=-\frac{1}{2} \lim_{\ell \rightarrow 0}\frac{1}{\ell}
    \hat n^\mu \hat m^\nu \hat n^\rho \hat m^\sigma \hat C_{\mu\nu\rho\sigma},
\end{equation}
which corresponds in Newman--Penrose notation to $-(1/2)\bar \psi_4^0$. 
The main calculational result in~\cite{strateg} is that
\begin{equation}
       \hat \Psi=\frac{1}{2}\hat n^\mu \hat m^\nu \hat m^\rho \bigg(
	         \hat \nabla_\mu  \hat \Sigma_{\nu\rho}
		  -\hat \nabla_\nu \hat \Sigma_{\mu\rho}\bigg),
\label{eq:psisigma}
\end{equation}
which is independent of the freedom
$\hat m^\mu \rightarrow \hat m^\mu + \lambda \hat n^\mu$
in the choice of $m^\mu$.
In inertial Bondi coordinates, this reduces to
\begin{equation}
       \hat  \Psi = \frac {1}{4} Q^A Q^B\partial_u^2  c_{AB}, 
\end{equation}
which is related to the Bondi news function by
\begin{equation}
     \hat \Psi =\partial_u N
\label{eq:PsiNu}
\end{equation}
so that
5
\begin{equation}
    N_\Psi = N|_{u=0} +\int_0^u \hat \Psi du ,
\end{equation}
with $N_\Psi=N$ up to numerical error.

As in the case of the news function, the general expression (\ref{eq:psisigma}) 
for $\hat \Psi$ must be used. This  challenges numerical accuracy due to the
large number of terms and the appearance of third angular derivatives. For
instance, in the linearized approximation, the value of $\hat \Psi$ on
$\mathcal{I}^+$ is given by the fairly complicated expression
5
\begin{equation}
   \hat \Psi=\frac{1}{2}\partial_u^2 \partial_\ell J -\frac{1}{2}\partial_u J
      -\frac{1}{2}\eth L -\frac{1}{8} \eth^2( \eth \bar L +\bar \eth L)
       + \partial_u \eth^2 H,
\label{eq:linPsi}
\end{equation}
where $J=Q^A Q^B h_{AB}$ and $L=Q^A L_A$. 
In the same approximation, the news function is given by
\begin{equation}
   N =\frac{1}{2} \partial_u \partial_\ell J
      +\frac{1}{2} \eth^2(\omega +2H).
\label{eq:linN}
\end{equation}
(The relationship (\ref{eq:PsiNu}) still holds in the linearized
approximation but in the nonlinear case, the derivative along the generators of
$\mathcal{I}^+$ is $\hat n^\mu \partial_\mu =e^{-2H}(\partial_u +L^A \partial_A)$
and (\ref{eq:PsiNu}) must be modified accordingly.)

These linearized expressions provide a starting point to compare the advantages
between computing the radiation via $N$ or $N_\Psi$. The troublesome gauge terms
involving $L$, $H$ and $\omega$ all vanish in inertial Bondi coordinates (where
$\omega=1$). One difference is that $\hat \Psi$ contains third order angular
derivatives, e.g.\ $\eth^3 \bar L$, as opposed to second angular derivatives for
$N$. This means that the smoothness of the numerical error is more crucial in
the $\hat \Psi$ approach. Balancing this, $N$ contains the $\eth^2 \omega$ term,
which is a potential source of numerical error since $\omega$ must be evolved
via (\ref{eq:omegadot}).

The accuracy of waveform extraction via the Bondi news function $N$ and its
counterpart $N_\Psi$ constructed from the Weyl curvature has been compared in a
linearized gravitational wave test problem~\cite{strateg}. The results show that
both methods are competitive, although the $\Psi_4$ approach has an edge. 

However, even though both methods were tested to be second order convergent
in test beds with analytic data,
there was still  considerable error, of the order of $5\%$ for grids of
practical size.  This error reflects the intrinsic difficulty in extracting
waveforms because of the delicate cancellation of leading order terms in the
underlying metric and connection when computing the $O(1/r)$ radiation field. 
It is somewhat analogous to the experimental task of isolating a transverse
radiation field from the longitudinal fields representing the total mass, while
in a very non-inertial laboratory. In the linearized wave test carried out
in~\cite{strateg}, the news consisted of the sum of three terms, $N=A+B+C$,
where because of cancellations $N\approx A/24$. The individual terms $A$, $B$ and
$C$ had small fractional error but the cancellations magnified the fractional
error in $N$.

The tests in~\cite{strateg} were carried out with a characteristic code using
the circular-stereographic patches. The results are in qualitative agreement
with tests of CCE using a cubed-sphere code~\cite{reisswig}, which in addition
confirmed the expectation that fourth-order finite difference approximations for
the $\eth$-operator gives improved accuracy. As demonstrated
recently~\cite{leo}, once all the necessary infrastructure for interpatch
communication is in place, an advantage of the cubed-sphere approach is that its
shared boundaries admit a highly scalable algorithm for parallel architectures.

Another alternative is to carry out a coordinate transformation in the
neighborhood of $\mathcal{I}^+$ to inertial Bondi coordinates, in which the news
calculation is then quite clean numerically. This approach was
implemented in~\cite{bishnews} and shown to be second order convergent in
Robinson--Trautman and Schwarzschild testbeds. However, it is clear that this
coordinate transformation also involves the same difficult numerical problem of
extracting a small radiation field in the presence of the large gauge effects
that are present in the primary output data.

These underlying gauge effects which complicate CCE are introduced at the inner
extraction worldtube and then propagate out to $\mathcal{I}^+$ but they are of
numerical origin and can be reduced with increased accuracy.
Perturbative waveform extraction suffers the same gauge effects but in this case they are
of analytic origin and cannot be controlled by numerical accuracy.
Lehner and Moreschi~\cite{lehmor} have shown that the delicate
gauge issues involved at $\mathcal{I}^+$ have counterparts in $\Psi_4$ extraction
of radiation on a finite worldtube. They
show how some of the analytic techniques used at $\mathcal{I}^+$ can also be used to
reduce the effect of these ambiguities on a finite worldtube, in particular the ambiguity
arising from the conformal factor $\omega$. The analogue of $\omega$ on a finite
worldtube can reduce some of the non-inertial effects that enter the
perturbative waveform. In addition, use of  normalization conventions on the null
tetrad defining  $\Psi_4$ analogous to the conventions at $\mathcal{I}^+$ can
avoid other spurious errors. This approach can also be used to reduce gauge
ambiguities in the perturbative calculation of momentum recoil in the merger of black
holes~\cite{gallolm}. 

\subsection{Application of CCE to binary black hole inspirals}
\label{sec:ccebh}

The emission of gravitational waves from the inspiral and merger of
binary black holes is the most likely source for detection by
gravitational wave observatories. The post-Newtonian regime
of the inspiral can be accurately modeled by the chirp waveforms obtained
by perturbation theory and the final ringdown waveform can be accurately modeled
by the known quasi-normal modes. This places special importance
on reliable waveforms for the nonlinear inspiral and merger waveform
which interpolates between these early and late time phases. Here CCE
plays the important role of providing an unambiguous waveform at
$\mathcal{I}^+$ which can be used to avoid the error introduced by
perturbative extraction techniques.

The application of CCE to binary  black hole simulations
was first carried out in~\cite{reis1,reis2} using an implementation of the PITT
code for the characteristic evolution. The Cauchy evolution
was carried out using a variant of the
BSSN formulation~\cite{mStN95,tBsS99}. Simulations of inspiral
and merger were carried out for equal mass non-spinning black holes and for
equal mass black holes with spins aligned with the orbital angular momentum.
For a binary of mass $M$, two separate choices of outer Cauchy boundary were located
at $R=3600M$ and $R=2000M$, with the corresponding
characteristic extraction worldtubes ranging from $R_E=100M$ to $R_E=250M$,
sufficient to causally isolate the  characteristic extraction from the
outer boundary during the simulation of 8 orbits prior to merger and ringdown.
The difference between CCE waveforms in this range
of extraction radii was found to be of comparable size to the numerical error.
In particular, for the grid resolutions used, the dominant numerical error was
due to the Cauchy evolution.

The CCE waveforms at $\mathcal{I}^+$ were 
also used to evaluate the quality of perturbative waveforms based upon
Weyl tensor extraction. In order to reduce finite extraction effects,
the perturbative waveforms were extrapolated to infinity by extraction at six
radii in the range $R=280M$ to $R=1000M$. It is notable that the results in~\cite{reis1}
indicate that the  systematic error in perturbative extraction had, previously,
been underestimated.

The lack of reflection symmetry in the spinning case leads to a recoil, or ``kick'', due to
the linear momentum carried off by the gravitational waves. The astrophysical consequence
of this kick to the evolution of a galactic core has accentuated the important role of CCE
waveforms to supply the energy, momentum and angular momentum radiated during binary black
hole inspirals. The radiated energy and momentum obtained from the $\psi_4$ Weyl component
obtained at $\mathcal{I}^+$ via CCE was compared to the corresponding value extracted at
finite radii and then extrapolated to infinity~\cite{reis2}. The extrapolated value was
found to be of comparable accuracy to the CCE result for the large extraction radii used.
For extraction at a single radius of $R=100M$, commonly used in numerical relativity, this
was no longer true and the error was 1 to 2 orders of magnitude larger. The CCE energy loss
obtained via  $\psi_4$ was also found to be consistent, within numerical error, to the
recoil computed from the news function. The work emphasizes the need for an accurate
description of the astrophysical consequences of gravitational radiation, which CCE  is
designed to provide.

In addition to the dominant oscillatory gravitational wave signals produced during binary
inspirals, there are also memory effects described by the long time scale change in the
strain $\Delta h=h(t,\theta,\phi)-h(-\infty,\theta,\phi)$. In a follow-up to the work
in~\cite{reis1,reis2}, these were studied by means of CCE~\cite{reis4} for the inspiral of
spinning black holes.  It was found that the memory effect was greatest for the case of
spins aligned with the orbital angular momentum, as might be expected since this case also
produces the strongest radiation. The largest spherical harmonic mode for the effect was
found to be the $(\ell=2,m=0)$ mode. Since CCE supplies either the news function or its
time derivative $\psi_4$, a major difficulty in measuring the memory is the proper setting
of the integration constants in determining the strain. This was done by matching the
numerical evolution to a post-Newtonian precursor. There is a slow monotonic growth of
$\Delta h$ during the inspiral followed by a rapid rise during the merger phase, which over
the time scale of the simulation leads to a step-like behavior modulated by the final
ringdown.  The simulations showed that the largest memory offset occurs for highly spinning
black holes, with an estimated value of 0.24 in the maximally spinning case. These results
are central to determining the detectability of the memory effect by observations of
gravitational waves. Since the size of the $(\ell=2,m=0)$ mode is small compared compared
to the dominant $(\ell=2,m=2)$ radiation mode, the memory effect is unlikely to be
observable in LIGO signals. However, the long period behavior of the effect might make it
more conducive to detection by proposed pulsar timing arrays designed to measure the
residual times-of-arrival caused by intervening gravitational waves.

Another application of CCE has been to the study of gravitational waves from precessing
binary black holes with spins aligned or anti-aligned to the  orbital angular
momentum~\cite{ccespin}. It was found that binaries with spin aligned with the orbital
angular  momentum are more powerful sources than the corresponding binaries with
anti-aligned spins. The results were confirmed by comparing the waveforms obtained using
perturbative extraction at finite radius  to those obtained using CCE. The comparisons
showed that the difference between the two approaches was within the numerical error of the
simulation.

\subsection{Application of CCE to stellar collapse}
\label{sec:stelcoll}
CCE has also been recently applied to study the waveform from the fully 3-dimensional
simulation  of the collapse and core bounce
of a massive rotating star~\cite{reis3}. After nuclear energy
generation has ceased, dissipative processes eventually push the core over its effective
Chandrasekhar mass. Radial instability then drives the inner core to nuclear densities
at which time the stiffened equation of state leads to a core bounce with tremendous
acceleration. The asymmetry of this bounce due to a rotating core potentially gives rise
to  a detectable source of gravitational quadrupole radiation which can be used to
probe the nuclear equation of state and the mass and angular momentum of the
star. Simulations were carried out for three choices of initial star parameters.
The gravitational waves emitted in the core bounce phase were compared using four
independent extraction techniques:

\begin{itemize}

\item The simplest technique was via the quadrupole formula, which estimates the 
waveform in terms of the second time derivative of the mass quadrupole tensor but does not
take into account  the effects of curvature or relativistic motion. 

\item Extraction at a finite radius via  the Newman--Penrose $\psi_4$ component.

\item  Perturbative extraction at a finite radius based upon the
Regge--Wheeler--Zerilli--Moncrief  (RWZM) formalism.

\item CCE, utilizing the PITT code, which avoids the near field or perturbative
approximations of the above techniques..

\end{itemize}

Historically, the quadrupole formula, which is computed in the inner region where
the numerical grid  is most accurate, has been the predominant extraction tool
used in stellar collapse. The metric or  curvature based methods suffer from numerical
error in extracting a signal which is many orders of magnitude weaker than that
from a binary inspiral.from the numerical noise. This is especially pertinent to RWZM and
$\psi_4$ extraction where the signal must be extracted in the far field.
In addition, the radiation is dominant in the $(\ell=2,m=0)$ spherical harmonic
mode, in which the memory effect complicates the relationship between
$\psi_4$ and the strain at low frequencies.

CCE was used as the benchmark in comparing the various extraction techniques. For all
three choices of initial stellar configurations, extraction via RWZM yielded the largest
discrepancy and showed a large spurious spike at core bounce and other  spurious
high frequency contributions. Quadrupole and $\psi_4$ extraction only led to
small differences with CCE. It was surprising that the quadrupole technique
gave such good agreement, given its simplistic assumptions.  
Overall, quadrupole extraction performed slightly better than $\psi_4$
extraction when compared to CCE. One reason is that the double time integration of $\psi_4$
to produce the strain introduces low frequency errors. Also,  $\psi_4$ extraction led
to larger peak amplitudes compared to either quadrupole extraction or CCE.

Several important observations emerged from this study.
(i) $\psi_4$ extraction and CCE  converge properly with extraction worldtube radius.
RWZM produces spurious high frequency effects which no other method reproduces.
(ii) Waveforms from
CCE, $\psi_4$ extraction and quadrupole
extraction agree well in phase. The high frequency contamination of RWZM makes
phase comparisons meaningless. (iii) Compared to CCE, the maximum amplitudes at
core bounce differ by $\approx$~1 to 7\%, depending on initial stellar parameters,
for $\psi_4$ extraction and by $\approx$~5 to 11\% for quadrupole extraction. (iv) Only
quadrupole extraction is free of low frequency errors. (v) For use in gravitational
wave data analysis, except for RWZM,
the three other extraction techniques yield results which are equivalent
up to the uncertainties intrinsic to matched-filter searches.

Certain technical issues cloud the above observations. CCE,  $\psi_4$ and RWZM
extraction are based upon vacuum solutions at the extraction worldtube, which is not
the case for these simulations in which the star extends over the entire computational grid.
This could be remedied by the inclusion of matter terms in the CCE technique,
which might also improve the low frequency behavior. In any case, this work represents
a milestone in showing that CCE has important relevance to waveform extraction
from astrophysically realistic collapse models.

The above study~\cite{reis3} employed a sufficiently stiff equation of state to
produce core bounce after collapse. In subsequent work, CCE was utilized
to study the gravitational radiation from a collapsar model~\cite{collapsar},
in which a rotating star collapses to form a black hole with accretion disk.
The simulations tracked the initial collapse and
bounce, followed by a post bounce phase leading to black hole formation.
At bounce, there is a burst of gravitational waves similar to the above study,
followed by a turbulent post bounce with weak gravitational radiation in which
an unstable proto-neutron star forms. 
Collapse to a black hole then leads to another pronounced spike in the waveform,
followed by ringdown to a Kerr black hole. The ensuing accretion flow does not lead to any
further radiation of appreciable size. The distinctive signature of the gravitational
waves observed in these simulations would enable a LIGO detection
to distinguish between core collapse leading to bounce and supernova and one leading to
black hole formation.

\subsection{LIGO accuracy standards}

The strong emission of gravitational waves from the inspiral and merger of
binary black holes has been a dominant motivation for the construction of
the LIGO and Virgo gravitational wave observatories.
The precise detail of the waveform obtained from numerical simulation is a
key tool to enhance detection and allow useful scientific interpretation of the gravitational
signal. The first  derivation~\cite{flanhugh} of
the accuracy required for numerically generated black hole waveforms 
to be useful as templates for gravitational wave data analysis was carried out in
the frequency domain. Proper accuracy standards must take into account
the power spectral density of the detector noise $S_n(f)$, which is calibrated
with respect to the frequency domain strain $\hat h(f)$.  Consequently, the
primary accuracy standards must be formulated in the frequency domain in order to
take detector sensitivity into account. 
See~\cite{lindbo} for a recent review.

It has been emphasized~\cite{linabuse} that the direct use of time domain errors
obtained in numerical simulations can be deceptive in assessing the accuracy
standards for  model waveforms to be suitable for gravitational wave data analysis.
For this reason, the frequency domain accuracy requirements have been translated
into requirements on the time domain $L_2$ error norms, so that they can
be readily enforced in practice~\cite{lindbo,lindob,lind}. 

There are two distinct criteria for waveform accuracy: (i) Insufficient accuracy
can lead to an unacceptable fraction of
signals to pass undetected through the corresponding matched-filter; (ii) the
accuracy affects whether a detected waveform can be used to measure the physical properties
of the source, e.g. mass and spin, to a level commensurate with the accuracy of
the observational data. Accuracy standards for model waveforms have been
formulated to prevent these potential losses in the detection of gravitational waves
and the measurement of their scientific content.

For a numerical waveform with strain component $h(t)$, the time domain error is
measured by
\begin{equation}
     {\cal E}_0 = \frac {|| \delta h||}{|| h ||},
      \label{eq:strainerr}
\end{equation}
where $\delta h$ is the error in the numerical approximation and $||F||^2=\int dt
|F(t)|^2$, i.e.\ $||F||$ is the $L_2$ norm, which in principle should be integrated
over the complete time domain of the model waveform obtained by splicing a
perturbative chirp waveform to a numerical waveform for the inspiral and merger.

The error can also be measured in terms of time derivatives of the strain. 
The first time derivative corresponds to the error in the news
\begin{equation}
        {\cal E}_1(Re N) = \frac {||\delta Re N ||}{|| Re N ||}  \, , \quad
        {\cal E}_1(Im N) = \frac {||\delta Im N ||}{|| Im N ||}
        \label{eq:newserr}
\end{equation}
and the second time derivative corresponds to the Weyl component error
\begin{equation}
        {\cal E}_2 (Re \Psi) = \frac {||\delta Re \Psi ||}{|| Re \Psi ||}  \, , \quad
        {\cal E}_2 (Im \Psi) = \frac {||\delta Im \Psi ||}{|| Im \Psi ||}  .
         \label{eq:weylerr}
\end{equation}

In~\cite{lindbo}, it was shown that sufficient conditions to satisfy data analysis
criteria for detection and measurement can be formulated in terms of any of the
error norms ${\cal E}_k=({\cal E}_0, {\cal E}_1,{\cal E}_2)$, i.e.\ in terms of the
strain, the news or the Weyl component. The accuracy requirement
for detection is
\begin{equation}
    {\cal E}_k \le C_k \sqrt{2\epsilon_{\mathrm{max}}},
    \label{eq:Ndet}
\end{equation}
and the requirement for measurement is
\begin{equation}
    {\cal E}_k \le C_k \frac {\eta_c}{\rho}.
    \label{eq:Nmeas}
\end{equation}
Here $\rho$ is the optimal signal-to-noise ratio of the detector,
defined by
\begin{equation}
         \rho^2 = \int_0^\infty \frac{4|\hat h(f)|^2}{S_n(f)} df ;
\end{equation}
$C_k$ are dimensionless factors introduced in~\cite{lindbo}
to rescale the traditional signal-to-noise ratio $\rho$ in making the transition
from frequency domain standards to time domain standards;
$\epsilon_{\mathrm{max}}$ determines the fraction of detections lost due to
template mismatch, cf.\ Eq.~(14) of~\cite{lindob}; and
 $\eta_c\le 1$ corrects for error introduced in detector calibration.
These requirements  for detection and measurement, for either $k=0, 1,2$
conservatively overstate the basic frequency domain requirements by
replacing $S_n(f)$ by its minimum value in transforming to the time domain.

The values of $C_k$ for the inspiral and merger of non-spinning equal-mass black
holes have been calculated in~\cite{lindbo} for the advanced LIGO noise spectrum.
As the total mass of the binary varies from $0 \rightarrow \infty$, $C_0$ varies
between $.65 > C_0 >0$, $C_1$ varies between $.24 < C_1 < .8$ and  $C_2$ varies
between $0 < C_2 < 1$. Thus only the error ${\cal E}_1$  in the news  can satisfy
the criteria over the entire mass range. The error in the strain ${\cal E}_0$
provides the easiest way to satisfy the criteria in the low mass case
$M<<M_{\odot}$ and the  error in the Weyl component  ${\cal E}_2$ provides the
easiest way to satisfy the criteria in the high mass case $M>>M_{\odot}$.

\subsection{A community CCE tool}

The importance of accurate waveforms has prompted development of a 
newly designed  CCE tool~\cite{ccetool} which meets the advanced LIGO
accuracy standards. Preliminary progress was reported in~\cite{babwz}.
The CCE tool is available for use by the numerical relativity community
under a general public license
as part of the Einstein Toolkit~\cite{einstool}.  It  can be applied
to a generic Cauchy code with extraction radius as small as $r=20M$, which
provides flexibility for many applications besides binary black holes,
such as waveform extraction from stellar collapse.

The matching interface was streamlined by introducing a pseudospectral
decomposition of the Cauchy metric in the neighborhood of the extraction worldtube.
This provides economical storage of the boundary data for the
characteristic code so that the waveform at $\mathcal{I}^+$ can be obtained
in post-processing with a small computational burden compared to the Cauchy evolution.
The new version incorporates stereographic grids with
circular patch boundaries~\cite{strateg}, which eliminates the large error
from the corners of the square patches
used previously. The finite difference accuracy of the
angular derivatives was increased to
4th order.  Bugs were eliminated that had been introduced in the process of
parallelizing the code using the Cactus framework~\cite{cactus}.
In addition,  the worldtube module, which supplies the inner boundary
data for the characteristic evolution, was  revamped so that it provides
a consistent, second order accurate start-up algorithm for numerically
generated Cauchy data. The prior module required differentiable
Cauchy data, as provided by analytic testbeds, to be consistent with convergence.

These changes led to clean second order convergence of all
evolved quantities at finite locations. Because 
some of the hypersurface equations become degenerate at $\mathcal{I}^+$,
certain asymptotic quantities, in particular the Bondi news function, are only
first order accurate.
However, the clean first order convergence allows the application
of Richardson extrapolation, based upon three characteristic grid sizes,
to extract waveforms with third order accuracy.

The error  norm for the extracted news function,
${\cal E}_1(N)$ as defined in (\ref{eq:newserr}),
has been measured for the simulation
of the inspiral of equal mass, non-spinning black holes obtained via
a BSSN simulation~\cite{ccetool}.  The  advanced LIGO criterion for detection
(\ref{eq:Ndet}), was satisfied for $\epsilon_{\mathrm{max}} =.005$ (which
corresponds to  less than a 10\% signal loss) and for 
values of $C_1$ throughout the entire binary mass range.
The criterion (\ref{eq:Nmeas}) for measurement
is more stringent. For the expected lower bound of the calibration  factor
$\eta_{\mathrm{min}} =0.4$, for $C_1=.24$ (corresponding to the most
demanding small mass limit)
and for  the most optimistic advanced LIGO signal-to-noise ratio
$\rho = 100$, the requirement for
measurement is $ {\cal E}_1(N) \le 9.6 \times 10^{-4}$.  This
measurement criterion was satisfied throughout the entire binary mass range by the
numerical truncation error $ {\cal E}_1(N)$ in the CCE waveform.

These detection and measurement criteria were satisfied for
a range of extraction worldtubes extending from $R=20M$ to $R=100M$. 
The ${\cal E}_1(N)$ error norm decreased
with larger extraction radius, as expected since the error
introduced by characteristic evolution depends upon the size of the
integration region between the extraction worldtube and $\mathcal{I}^+$.
However, the modeling error corresponding to the difference in waveforms
obtained with extraction at $R=50M$
as compared to $R_E=100M$ only satisfied
the measurement criterion for signal-to-noise ratios  $\rho<25$
(which would still cover the most likely advanced LIGO events).
This modeling error results from the different initial data which correspond
to different extraction radii. This error would be
smaller for longer simulations with a higher number of orbits.
The results suggest that the choice of extraction radius should be
balanced between  a sufficiently large radius to reduce initialization effects and
a sufficiently small radius where the Cauchy grid is more highly refined and outer
boundary effects are better isolated.

\subsection{Initial characteristic data for CCE}
\label{sec:initial}

Data on the initial null hypersurface must be prescribed
to begin the characteristic evolution. This data consist of the conformal
2-metric $h_{AB}$ of the null hypersurface. Because of the determinant
condition (\ref{eq:qdet}), this data can be formulated in terms of the
spin-weight 2 variable $J$ given in (\ref{eq:J}).
In the first applications of CCE, it was expedient to set
$J=0$ on the initial hypersurface outside some radius.
This necessitated a transition region to obtain continuity with the initial Cauchy
data, which requires non-zero initial characteristic data at the extraction
worldtube.

In~\cite{ccetool} the initialization was changed by requiring that the Newman--Penrose 
component of the Weyl tensor intrinsic to the initial null hypersurface vanish,
i.e.\ by setting $\psi_0 =0$. This approach is dual to the technique of
using $\psi_4$ to extract outgoing gravitational waves.
For a linear perturbation of the Schwarzschild metric, this $\psi_0$
condition eliminates incoming radiation crossing the initial null hypersurface.
Since $\psi_0$ consists of a second radial derivative of the characteristic data,
the  condition allows both continuity of $J$ at the extraction worldtube and the desired
asymptotic  falloff of $J$ at infinity. In the linearized limit, setting
$\psi_0=0$ reduces to $(\partial_\ell)^2 J=0$, in terms of the
compactified radial coordinate $\ell =1/r$. In terms of the compactified
grid coordinate $x=r/(R_E+r)$ (where $R_E$ is the
Cartesian radius of the extraction worldtube
defined by the Cauchy coordinates),  the corresponding solution is
\begin{equation}
    J=\frac{J|_{x_E}(1-x)x_E}{(1-x_E)x},
\end{equation} 
where $J|_{x_E}$ is determined by Cauchy data at the extraction
worldtube.  Since this solution also implies $J|_{\mathcal{I}^+}=J|_{x=1}=0$,
${\mathcal{I}^+}$ has unit sphere geometry so that the conformal gauge effects
discussed in Section~\ref{sec:conf} are minimized at the outset of the evolution.

Besides the extraneous radiation content in the characteristic initial data
there is also extraneous ``junk'' radiation in the initial Cauchy data for the binary black
hole simulation. Practical experience indicates that the effect of
this  ``junk'' radiation on the waveform is transient and becomes negligible
by the onset of the plunge and merger stage. However, another source of
waveform error with potentially longer time consequences can arise from a
 \emph{mismatch} between the initial characteristic and Cauchy data.
This mismatch arises because the characteristic data is given on the
outgoing null hypersurface emanating from the intersection
of the extraction worldtube and the initial Cauchy hypersurface.
Since in CCE the extraction worldtube cannot be located at the outer
Cauchy boundary, part of the initial null hypersurface
lies in the domain of dependence of the initial Cauchy data. Thus a free
prescription of the characteristic data can be inconsistent with the Cauchy
data.

The initial characteristic data $\psi_0 =0$  implies the absence of
radiation on the assumption that the geometry of the initial
null hypersurface is close to Schwarzschild. This assumption becomes valid as the
extraction radius becomes large and the exterior Cauchy data can be approximated
by Schwarzschild data. Thus this mismatch could in principle be reduced
by a sufficiently large choice of extraction worldtube. However, that approach is
counter productive to the savings that CCE can provide.

An alternative approach developed in~\cite{bishinit}
attempts to alleviate this problem
by constructing a solution linearized about Minkowski space.
The linearized solution is modeled upon binary black hole 
initial Cauchy data. By evaluating the solution on the initial characteristic
null hypersurface, this solves the compatibility issue up to curved space effects.
A comparison study based upon this approach
shows that the choice of $J=0$ initial data does affect the waveform
for time scales which extend long after
the burst of junk radiation has passed. Although this study
is restricted to CCE extraction radii $R>100M$ and does not explore the
additional benefits of the more gauge invariant $\psi_0 =0$ initial data
implemented in~\cite{ccetool}, it
emphasizes the need to control potential long terms effects
which might result from a mismatch between the Cauchy and characteristic
initial data.

Ideally, this mismatch could be eliminated by placing the extraction worldtube
at the artificial outer boundary of the Cauchy evolution by means of
a transparent interface with the outer characteristic evolution.
This is the ultimate goal of CCM, although a formidable
amount of work remains to develop a stable implementation.



\section{Numerical Hydrodynamics on Null Cones}
\label{sec:grace}

Numerical evolution of relativistic hydrodynamics has been traditionally
carried out on spacelike Cauchy hypersurfaces. Although the Bondi--Sachs
evolution algorithm can easily be extended to include matter~\cite{isaac}, the
advantage of a light cone approach for treating fluids is not as apparent as
for a massless field whose physical characteristics lie on the light cone.
However, results from recent studies of  relativistic stars and of fluid
sources moving in the vicinity of a black hole indicate that this approach can
provide accurate simulations of  astrophysical relevance such as supernova
collapse to a black hole, mass accretion, and the production of gravitational
waves.


\subsection{Spherically symmetric hydrodynamic codes}
\label{sec:shydro}

The earliest fully general relativistic simulations of fluids were carried out
in spherical symmetry. The first major work was a study of gravitational
collapse by May and White~\cite{may}. Most of this early work was carried out
using Cauchy evolution~\cite{tonirev}. Miller and Mota~\cite{miller} performed
the first simulations of spherically symmetric gravitational collapse using a
null foliation. Baumgarte, Shapiro and Teukolsky subsequently used a null
slicing to study supernovae~\cite{bst1} and the collapse of neutron stars to
form black holes~\cite{bst2}. The use of a null slicing allowed them to evolve
the exterior spacetime while avoiding the region of singularity formation.

Barreto's group in Venezuela applied characteristic
methods to study the self-similar collapse of spherical matter and charge
distributions~\cite{barreto96,barreto98,barreto99}. The assumption of
self-similarity reduces the problem to a system of ODE's, subject to boundary
conditions determined by matching to an exterior Reissner--Nordstr\"om--Vaidya
solution. Heat flow in the internal fluid is balanced at the surface by the
Vaidya radiation. Their simulations illustrate how a nonzero total charge can
halt gravitational collapse and produce a final stable
equilibrium~\cite{barreto99}. It is interesting that the pressure vanishes in
the final equilibrium state so that hydrostatic support is completely supplied
by Coulomb repulsion. In subsequent work~\cite{barreto09,barreto10}, they applied
their characteristic code to the evolution of a polytropic fluid sphere coupled to a
scalar radiation field to study the central equation of state, conservation of the
Newman--Penrose constant, the scattering of the scalar radiation off the polytrope
and its late time decay. The work illustrates how characteristic evolution can be
used to simulate radiation from a matter source in the simple
context of spherical symmetry. 

Font and Papadopoulos~\cite{toniphi} have given a state-of-the-art treatment
of relativistic fluids which is applicable to either spacelike or null
foliations. Their approach is based upon a high-resolution shock-capturing
(HRSC) version of relativistic hydrodynamics in flux conservative form, which
was developed by the Valencia group (for a review see~\cite{tonirev}).
In the HRSC scheme, the hydrodynamic equations are written in flux
conservative, hyperbolic form. In each computational cell, the system of
equations is diagonalized to determine the characteristic fields and
velocities, and the local Riemann problem is solved to obtain a solution
consistent with physical discontinuities. This allows a finite differencing
scheme along the characteristics of the fluid that preserves the conserved
physical quantities and leads to a stable and accurate treatment of shocks.
Because the general relativistic system of hydrodynamical equations is
formulated in covariant form, it can equally well be applied to spacelike or
null foliations of the spacetime. The null formulation gave remarkable
performance in the standard Riemann shock tube test carried out in a Minkowski
background. The code was successfully implemented first in the case of
spherical symmetry, using a version of the Bondi--Sachs formalism adapted to
describe gravity coupled to matter with a worldtube boundary~\cite{tam}. They
verified second order convergence in curved space tests based upon
Tolman--Oppenheimer--Volkoff equilibrium solutions for spherical fluids. In the
dynamic self-gravitating case, simulations of spherical accretion of a fluid
onto a black hole were stable and free of numerical problems. Accretion was
successfully carried out in the regime where the mass of the black hole
doubled. Subsequently the code was used to study how accretion modulates both
the decay rates and oscillation frequencies of the quasi-normal modes of the
interior black hole~\cite{imprints}.

The characteristic hydrodynamic approach of Font and Papadopoulos was first
applied to spherical symmetric problems of astrophysical interest. Linke, Font,
Janka, M{\"{u}}ller, and Papadopoulos~\cite{linke} simulated the spherical
collapse of supermassive stars, using an equation of state that included the
effects due to radiation, electron-positron pair formation, and neutrino
emission. They were able to follow the collapse from the onset of instability to
black hole formation. The simulations showed that collapse of a star with mass
greater than $5\times 10^5$ solar masses does not produce enough radiation to
account for the gamma ray bursts observed at cosmological redshifts.

Next, Siebel, Font, and Papadopoulos~\cite{sieb2002} studied the interaction of
a massless scalar field with a neutron star by means of the coupled
Klein--Gordon--Einstein-hydrodynamic equations. They analyzed the nonlinear
scattering of a compact ingoing scalar pulse incident on a spherical neutron
star in an initial equilibrium state obeying the null version of the
Tolman--Oppenheimer--Volkoff equations. Depending upon the initial mass and
radius of the star, the scalar field either excites radial pulsation modes or
triggers collapse to a black hole. The transfer of scalar energy to the star was
found to increase with the compactness of the star. The approach included a
compactification of null infinity, where the scalar radiation was computed. The
scalar waveform showed quasi-normal oscillations before settling down to a late
time power law decay in good agreement with the $t^{-3}$ dependence predicted by
linear theory. Global energy balance between the star's relativistic mass and
the scalar energy radiated to infinity was confirmed.


\subsection{Axisymmetric characteristic hydrodynamic simulations}
\label{sec:ahydro}

The approach initiated by Font and Papadopoulos has been applied in axisymmetry
to pioneering studies of gravitational waves from relativistic stars. The
gravitational field is treated by the original Bondi formalism using the
axisymmetric code developed by Papadopoulos~\cite{papath,papa}.
Because of the twist-free property of the axisymmetry in the original Bondi
formalism, the fluid motion cannot have a rotational component about the axis
of symmetry, i.e.\ the fluid velocity is constrained to the $(r,\theta)$ plane.
In his thesis work, Siebel~\cite{Siebel} extensively tested the combined
hydrodynamic-gravity code in the nonlinear, relativistic regime and demonstrated that
it accurately and stably maintained the equilibrium of a neutron star.

As a first application of the code, Siebel, Font, M{\"{u}}ller, and
Papadopoulos~\cite{sieb2002a} studied axisymmetric pulsations of neutron stars,
which were initiated by perturbing the density and $\theta$-component of
velocity of a spherically symmetric equilibrium configuration. The frequencies
measured for the radial and non-radial oscillation modes of the star were
found to be in good agreement with the results from linearized perturbation
studies. The Bondi news function was computed and its amplitude found to be in
rough agreement with the value given by the Einstein quadrupole formula. Both
computations involve numerical subtleties: The computation of the news
involves large terms which partially cancel to give a small result, and the
quadrupole formula requires computing three time derivatives of the fluid
variables. These sources of computational error, coupled with ambiguity in the
radiation content in the initial data, prevented any definitive conclusions.
The total radiated mass loss was approximately 10\super{-9} of the total mass.

Next, the code was applied to the simulation of axisymmetric supernova core
collapse~\cite{papadop2003}. A hybrid equation of state was used to mimic
stiffening at collapse to nuclear densities and shock heating during the
bounce. The initial equilibrium state of the core was modeled by a polytrope
with index $\Gamma=4/3$. Collapse was initiated by reducing the  polytropic
index to 1.3. In order to break spherical symmetry, small perturbations were
introduced into the $\theta$-component of the fluid velocity. During the
collapse phase, the central density increased by 5 orders of magnitude. At
this stage the inner core bounced at supra-nuclear densities, producing an
expanding shock wave which heated the outer layers. The collapse phase was
well approximated by spherical symmetry but non-spherical oscillations were
generated by the bounce. The resulting gravitational waves at null infinity
were computed by the compactified code. After the bounce, the Bondi news
function went through an oscillatory build up and then decayed in an $\ell=2$
quadrupole mode. However, a comparison with the results predicted by the
Einstein quadrupole formula no longer gave the decent agreement found in the
case of neutron star pulsations. This discrepancy was speculated to be due to
the relativistic velocities of $\approx 0.2 c$ reached in the core collapse as
opposed to $10^{-4}c$ for the pulsations. However, gauge effects and numerical
errors also make important contributions which cloud any definitive
interpretation. This is the first study of gravitational wave production by
the gravitational collapse of a relativistic star carried out with a
characteristic code. It is clearly a remarkable piece of work which offers up
a whole new approach to the study of gravitational waves from astrophysical
sources.


\subsection{Three-dimensional characteristic hydrodynamic simulations}
\label{sec:3dhydro}

The PITT code has been coupled with a rudimentary matter source to carry out
three-dimensional characteristic simulations of a relativistic star orbiting a
black hole. A naive numerical
treatment of the Einstein-hydrodynamic system for a perfect fluid was incorporated
into the code~\cite{matter}, but a more accurate HRSC hydrodynamic algorithm has not
yet been implemented. The fully nonlinear matter-gravity null code was tested
for stability and accuracy to verify that nothing breaks down as long as the
fluid remains well behaved, e.g., hydrodynamic shocks do not form. The code
was used to simulate a localized blob of matter falling into a black hole,
verifying that the motion of the center of the blob approximates a geodesic and
determining the waveform of the emitted gravitational radiation at ${\cal
I}^+$. This simulation was a prototype of a neutron star orbiting a black
hole, although it would be unrealistic to expect that this naive treatment of the fluid
could reliably evolve a compact star for several orbits.
A 3D HRSC characteristic hydrodynamic code would open the way to
explore this important astrophysical problem.

Short term issues were explored with the code in subsequent
work~\cite{matter2}. The code was applied to the problem of determining
realistic initial data for a star in circular orbit about a black hole. In
either a Cauchy or characteristic approach to this initial data problem, a
serious source of physical ambiguity is the presence of spurious gravitational
radiation in the gravitational data. Because the characteristic approach is
based upon a retarded time foliation, the resulting spurious outgoing waves
can be computed by carrying out a short time evolution. Two very different
methods were used to prescribe initial gravitational null data:
\begin{enumerate}
\item a Newtonian correspondence method, which guarantees that the
  Einstein quadrupole formula is satisfied in the Newtonian
  limit~\cite{newt}, and
  \label{method_1}
\item setting the shear of the initial null hypersurface to zero.
  \label{method_2}
\end{enumerate}
Both methods are
mathematically consistent but suffer from physical shortcomings. Method~\ref{method_1}
has only approximate validity in the relativistic regime of a star in close
orbit about a black hole while Method~\ref{method_2} completely ignores the
gravitational lensing effect of the star. It was found that, independently of
the choice of initial gravitational data, the spurious waves quickly radiate
away, and that the system relaxes to a quasi-equilibrium state with an
approximate helical symmetry corresponding to the circular orbit of the star.
The results provide justification of recent approaches for initializing the
Cauchy problem which are based on imposing an initial helical symmetry, as
well as providing a relaxation scheme for obtaining realistic characteristic
data.


\subsubsection{Massive particle orbiting a black hole}
\label{sec:part}

One attractive way to avoid the computational expense of hydrodynamics in
treating a star orbiting a massive black hole is to treat the star as a
particle. This has been attempted using the PITT code to model a star of mass
$m$ orbiting a black hole of much larger mass, say $1000\,m$~\cite{particle}.
The particle was described by the perfect fluid energy-momentum tensor of a
rigid Newtonian polytrope in spherical equilibrium of a fixed size in its
local proper rest frame, with its center following a geodesic. The validity of
the model requires that the radius of the polytrope be large enough so that
the assumption of Newtonian  equilibrium is valid but small enough so that the
assumption of rigidity is consistent with the tidal forces produced by the
black hole. Characteristic initial gravitational data for a double null
initial value problem were taken to be Schwarzschild data for the black hole.
The system was then evolved using a fully nonlinear characteristic code. The
evolution equations for the particle were arranged to  take  computational
advantage of the energy and angular momentum conservation laws which would
hold in the test body approximation.

The evolution was robust and could track the particle for two orbits as it
spiraled into the black hole. Unfortunately, the computed rate of inspiral was
much too large to be physically realistic: the energy loss was $\approx$~10\super{3}
greater than the value expected from perturbation theory. This discrepancy
might have a physical origin, due to the choice of initial  gravitational data
that ignores the particle or  due to a breakdown of the rigidity assumption,
or a numerical origin due to improper resolution of the particle. It is a
problem whose resolution would require the characteristic AMR techniques being
developed~\cite{pretlehn}.


These sources of error can be further aggravated by the introduction of matter
fields, as encountered in trying to make definitive comparisons between the
Bondi news and the Einstein quadrupole formula in the axisymmetric studies of
supernova collapse~\cite{papadop2003} described in Section~\ref{sec:ahydro}. In
the three-dimensional characteristic simulations of a star orbiting a black
hole~\cite{matter2,particle}, the lack of resolution introduced by a localized
star makes an accurate calculation of the news highly problematical. There
exists no good testbed for validating the news calculation in the presence of
a fluid source. A perturbation analysis in Bondi coordinates of the
oscillations of an infinitesimal fluid shell in a Schwarzschild
background~\cite{bishlin} might prove useful for testing constraint
propagation in the presence of a fluid. However, the underlying Fourier mode
decomposition requires the gravitational field to be periodic so that the
solution cannot be used to test computation of mass loss or radiation reaction
effects.



\section*{Acknowledgments}
\label{acknow}

This work was supported by National Science Foundation grant
PHY-0854623 to the University of Pittsburgh. I want to thank the many
people who have supplied me with material. Please keep me updated.

\newpage


\bibliography{refs}

\end{document}